\documentclass[iop,revtex4,12pt,preprint]{emulateapj_recent}

\usepackage[utf8]{inputenc}

\usepackage{graphics,graphicx}
\usepackage{helvet}
\usepackage{subfigure}
\usepackage[utf8]{inputenc}
\usepackage[T1]{fontenc}
\usepackage{soul} 
\usepackage[colorlinks,urlcolor=blue,citecolor=blue,linkcolor=blue,pdfusetitle]{hyperref}
\usepackage{appendix}
\usepackage{amsmath, amssymb, gensymb}
\usepackage{morefloats}
\usepackage{float}
\usepackage{ulem}
\usepackage{url}
\usepackage{natbib}
\usepackage{microtype}
\usepackage{multirow}
\usepackage{morefloats}
\usepackage{rotating}
\usepackage{here}
\usepackage{xspace}
\usepackage{color}
\usepackage{lipsum}
\usepackage{xspace}
\usepackage{float}
\usepackage{stmaryrd}

\def\Pf{\mathcal{P_{\rm{frac}}}}
\def\S{\mathcal{S}}
\def\StimesP{\S\times\Pf}

\def\eg {e.g.,\xspace} %e.g.,
\def\ie {i.e.,\xspace} %i.e.,
\def\etal {\textit{et al.}\xspace}
\def\ThOriC {$\theta^1$ Ori C\xspace}
\def\krat {$k$-RAT\xspace}
\def\brat {$B$-RAT\xspace}
\def\krats {$k$-RATs\xspace}
\def\brats {$B$-RATs\xspace}

\def\Td {$T_\textrm{d}$\xspace}
\def\NHt {$N_{\textrm{H}_2}$\xspace}
\def\NHtc {$N_{\textrm{H}_{2},}\;_{\textrm{from}}\;_{\,\theta^{1}\, \textrm{Ori}\,\textrm{C}}$\xspace}
\def\NHtsqT {$N_{\textrm{H}_2}\times\sqrt{T_\textrm{d}}$\xspace}

\begin{document}

% \linenumbers

\title{The origin of dust polarization in the Orion Bar}
\shorttitle{The origin of dust polarization in the Orion Bar}

\author{Valentin J. M. Le Gouellec\altaffilmark{1}}, \author{B-G Andersson\altaffilmark{1}}, \author{Archana Soam\altaffilmark{2,1}}, \author{Thiébaut Schirmer\altaffilmark{3}}, \author{Joseph M. Michail\altaffilmark{4,5}}, \author{Enrique Lopez-Rodriguez\altaffilmark{6}}, \author{Sophia Flores\altaffilmark{7}}, \author{David T. Chuss\altaffilmark{8}}, \author{John E. Vaillancourt\altaffilmark{9}}, \author{Thiem Hoang\altaffilmark{10,11}}, \author{Alex Lazarian\altaffilmark{12}}

\affiliation{\altaffilmark{1}SOFIA Science Center, Universities Space Research Association, NASA Ames Research Center, Moffett Field, California 94035, USA}
\affiliation{\altaffilmark{2}Indian Institute of Astrophysics (IIA), Sarjapur Road, Kormangala, Bangalore 560034, India}
\affiliation{\altaffilmark{3}Department of Space, Earth and Environment, Chalmers University of Technology, Onsala Space Observatory, 439 92 Onsala, Sweden}
\affiliation{\altaffilmark{4}Department of Physics and Astronomy, Northwestern University, 2145 Sheridan Rd., Evanston, IL 60208, USA}
\affiliation{\altaffilmark{5}Center for Interdisciplinary Exploration and Research in Astrophysics (CIERA), Northwestern University, 1800 Sherman Ave., Evanston, IL 60201, USA}
\affiliation{\altaffilmark{6}Kavli Institute for Particle Astrophysics \& Cosmology (KIPAC), Stanford University, Stanford, CA 94305, USA}
\affiliation{\altaffilmark{7}Physics Department, Santa Clara University, 500 El Camino Real, Santa Clara, CA 95053, USA}
\affiliation{\altaffilmark{8}Department of Physics, Villanova University, 800 E. Lancaster Ave., Villanova, PA 19085, USA}
\affiliation{\altaffilmark{9}Lincoln Laboratory, Massachusetts Institute of Technology, 244 Wood St., Lexington, MA 02421-6426}
\affiliation{\altaffilmark{10}Korea Astronomy and Space Science Institute, Daejeon 34055, Republic of Korea}
\affiliation{\altaffilmark{11}Korea University of Science and Technology, 217 Gajeong-ro, Yuseong-gu, Daejeon, 34113, Republic of Korea}
\affiliation{\altaffilmark{12}Department of Astronomy, University of Wisconsin-Madison, USA}

\shortauthors{Le Gouellec \etal}
\email{valentin.j.legouellec@nasa.gov}

\slugcomment{Accepted for publication in ApJ on 04/22/2023}

\begin{abstract}

The linear polarization of thermal dust emission provides a powerful tool to probe interstellar and circumstellar magnetic fields, because aspherical grains tend to align themselves with magnetic field lines. While the Radiative Alignment Torque (RAT) mechanism provides a theoretical framework to this phenomenon, some aspects of this alignment mechanism still need to be quantitatively tested. One such aspect is the possibility that the reference alignment  direction changes from the magnetic field ("\brat") to the radiation field k-vector ("\krat") in areas of strong radiation fields.
We investigate this transition toward the Orion Bar PDR, using multi-wavelength SOFIA HAWC+ dust polarization observations.
The polarization angle maps show that the radiation field direction is on average not the preferred grain alignment axis.
We constrain the grain sizes for which the transition from \brat to \krat occur in the Orion Bar (grains $\geq$ 0.1 $\mu$m toward the most irradiated locations), and explore the radiatively driven rotational disruption that may take place in the high-radiation environment of the Bar for large grains. 
While the grains susceptible to rotational disruption should be in supra-thermal rotation and aligned with the magnetic field, \krat aligned grains would rotate at thermal velocities.
We find that the grain size at which the alignment shifts from \brat to \krat corresponds to grains too large to survive the rotational disruption. 
Therefore, we expect a large fraction of grains to be aligned at supra-thermal rotation with the magnetic field, and potentially be subject to rotational disruption depending on their tensile strength.
\end{abstract}

% \vspace{0.5cm}

% \maketitle

\section{Introduction}
\label{sec:intro}

Photodissociation, or photon-dominated, regions (PDRs) designate the regions of the interstellar medium (ISM) intensively affected by the radiative energy produced by close-by massive stars \citep{Tielens1985a}. A PDR has a layered structure consisting of the transition between a hot plasma gas to the molecular region of the parental cloud. Located at the edges of high-mass star formation regions, PDRs harbor a variety of radiation driven chemical and physical processes (for a recent review see \citealt{Wolfire2022}, and references therein), in which interstellar dust grains have major roles. Small and large grains absorb part of the intense far-UV (FUV) radiation from O and B stars and re-radiate it as infra-red (IR) continuum emission. Very small grains and large molecules also heat the gas via the photoelectric effect \citep{Bakes1994,Weingartner2001b}. In addition, H$_2$ formation on dust grain surfaces are activated in such irradiated regions \citep{LeBourlot2012,Andersson2013,Bron2014,JonesA2015}. 

Understanding the physical processes acting on dust grains in PDRs enable us to constrain how the dust properties vary in such environments, which in turn affect the efficiency of the mechanisms dictating the evolution of PDRs. Multi-wavelength photometric studies have already constrained the formation and destruction of nano-grains in PDRs \citep{Arab2012,VanDePutte2020,Schirmer2022}. However, the properties of large dust grains (\ie $\geq\,0.1\,$nm) are harder to constrain using only the total intensity of the dust thermal emission. Analyzing its linear polarization is a powerful tool to study those large grains. The focus of this paper is thus to investigate the polarized dust emission toward a well-studied PDR, in order to study the radiation-driven mechanisms acting on dust grains.

It is well established that interstellar continuum polarization is due to elongated dust grains aligned, generally, with the magnetic field \citep{Hiltner1949,Andersson2015}. A number of physical processes are involved in the required grain alignment.  Grains must both achieve "internal alignment", whereby their rotation axis is aligned with a grain symmetry axis (ensuring a constant projection of the individual grain shape) and "external alignment", whereby the ensemble of grains align along a common, external, direction - usually the local magnetic field.  Both of these processes rely on paramagnetic effects in the grain bulk, in which the rotation will flip some of the free spins (this is known as the "Barnett effect"; \citealt{Purcell1979}). This is the inverse of the Einstein-de Haas effect, well-known in laboratory physics \citep{Einstein1915}.  An asymmetric grain rotating around a non-symmetry axis will experience nutation.  If this nutation is rapid enough the Barnett effect will not achieve a steady state but will cause energy dissipation \citep{Purcell1979}.  Under angular momentum conservation the lowest energy state of a rotating grain occurs when the angular momentum axis is parallel to the grain's axis of maximum inertia (smallest axis).  Hence Barnett relaxation leads to efficient internal alignment of paramagnetic grains\footnote{Nuclear and inelastic relaxation effects also play an important role in the alignment of dust grains \citep{Lazarian1999a,Lazarian1999c}}.  Because quantum spins carry both angular momenta and magnetic moments, in steady state the Barnett effect causes magnetization of these grains.  The interaction of this induced magnetization and an external field then leads to the external alignment.  

Both of these effects rely on the rapid rotation of the grain.  This is now understood to be accomplished through the interaction of the grain with an anisotropic radiation field, described by the Radiative Alignment Torque (RAT) theory \citep{Dolginov1976, Draine1996,Draine1997,LazarianHoang2007}, in which the right- and left-hand circular polarization components of the radiation field scatter differentially on an aspherical grain.  In most ISM cases the grains align and cause polarization relative to the magnetic field direction, so called ``\brat''.
However, in the case of a strong anisotropic radiation field, RAT theory predicts that the alignment axis can change from the magnetic field to the radiation field \citep[also called ``\krat'';][]{LazarianHoang2007,Tazaki2017,Hull2022}. 
This effect is stronger for large grains and can affect large internally aligned grains located close to a strong radiation source, if rotating at thermal velocities.
However, an intense radiation field can also trigger the RAdiative Torque Disruption mechanism (RATD, \citealt{Hoang2019NatAs}), which results in the fragmentation of grains. This occurs when the RAT-induced grain spin increases to such rotation speeds that the rotational energy overcomes its cohesion, or tensile strength, causing the grain to fragment. 
Within PDRs, our interests are two fold. First, we aim to predict the preferred axis of alignment in order to determine whether polarized dust emission preferentially traces the orientation of the magnetic field or the radiation field. 
Second, we study the effect of the intense irradiation on the population of aligned grains to ascertain whether the radiation contributes significantly to the evolution of dust toward PDRs.

Our study focuses on the Orion Bar, a PDR illuminated over a very broad spectrum, including the FUV, by the O7-type star \ThOriC, the most massive and luminous member of the Trapezium young stellar cluster \citep{ODell2001}. Located at 390 pc from us \citep{Kounkel2017,Kounkel2018}, the Orion Veil nebula forming the near-side shell around the \ion{H}{2} region is strongly shaped by the intense ionizing radiation and strong winds from \ThOriC that expands into the background Orion Molecular Cloud (OMC) \citep{Gudel2008,Pabst2019,Kavak2022}. The Bar forms a denser part of the foreground edge of the Veil. The edge of the PDR is located at the ionization front (IF), across which the gas converts from fully ionized to fully neutral. The gas density in the atomic gas rises to $n_{\rm{H}}\,=\,4-5\,\times\,10^{4}$ cm$^{-3}$, as indicated by the [OI] and [CII] forbidden line emission \citep{Tielens1985b,Tielens1993,Hogerheijde1995}, which are the main cooling agents of the gas phase \citep{Herrmann1997,Ossenkopf2013}. The location of the dissociation front (DF), \ie where the H/H$_2$ transition takes place, depends on the attenuation of the dissociating FUV photons, which in turn depends on the dust FUV absorption cross-section. In the Bar, the DF is located at $\sim$\,10-15$^{\prime\prime}$ from the IF (see Figure 1 of \citealt{Habart2022}), at $A_V\,\sim\,0.5-2$ mag \citep{Allers2005}. 
Mid-infrared (MIR) photometry using SOFIA/FORCAST \citep{Salgado2016} suggests that the UV and infrared dust opacities in the region are low by a factor of 5 to 10 compared to the diffuse ISM. Grain growth through coagulation may be responsible for this decrease. More recently, near-infrared (NIR) and MIR observations of emission by polycyclic aromatic hydrocarbons (PAH) pointed toward an efficient destruction of small PAHs in this type of highly illuminated PDRs \citep{Murga2022}. \citet{Schirmer2022} also concluded that nano-grains would suffer strong depletion in the Bar, proposing a scenario where these grains are formed via the fragmentation of large grains due to radiative pressure-induced collisions, and then be destroyed by photo-dissociation processes. Far-infrared (FIR) photometry is primarily sensitive to larger and colder dust, and can also probe dust grain evolution, such as coagulation \citep{Arab2012}. Interpreting polarized dust emission as tracing the plane of the sky component of the magnetic field, FIR dust polarization studies concluded that the support by magnetic field energy \citep{Chuss2019,Guerra2021} against the thermal expansion and mechanical feedback from \ThOriC \citep{Pellegrini2009,Pabst2020} is such that it can play a role in the Bar dynamics.

This paper is structured as follows. In Section \ref{sec:obs}, we present the multi-wavelength SOFIA/HAWC+ polarization observations of the Orion Bar and the data reduction steps. Results are presented in Section \ref{sec:results}, where we investigate variations of the polarization quantities as function of wavelength, spatial location, and environmental conditions.
In Section \ref{sec:modeling}, we perform a grain alignment timescale analysis using the available constraints on the environmental conditions toward the Orion Bar. We estimate for what grain size the \brat to \krat transition happens, the grain size parameter space potentially affected by RATD.
Finally, in Section \ref{sec:disc}, we discuss the characteristics of dust grains populating the Orion Bar PDR in light of the physical processes acting on them, \ie radiative alignment torques and rotational disruption. We draw our conclusions in Section \ref{sec:ccl}.

\section{SOFIA/HAWC+ polarization observations}
\label{sec:obs}

\begin{table*}[!tbph]
% \nolinenumbers
\centering
\small
\caption[]{Polarization observation summary}
\label{t.obs}
\setlength{\tabcolsep}{0.5em} %\vspace*{0.1in}
\begin{tabular}{p{0.13\linewidth}ccccccc}
\hline \hline \noalign{\smallskip}
HAWC+ Band & Band Center & Field of View & FWHM beam size & pixel size & Total exposure time & SOFIA archival ID 
\vspace{0in}\\
& ($\mu$m) & ($^{\prime}$) & ($^{\prime\prime}$) & ($^{\prime\prime}$) & (s) & \\
\noalign{\smallskip}  \hline
\noalign{\smallskip}
A & \phantom{0}53 &1.4 $\times$ 1.7 & \phantom{0}4.9 & 1.2 & 445 & 08\_0209 \& 09\_0037 \\
\noalign{\smallskip}
C & \phantom{0}89 & 2.1 $\times$ 2.7 & \phantom{0}7.8 & 2.0 & 3549 & 09\_0107 \\
\noalign{\smallskip}
D & 154 & 3.7 $\times$ 4.6 & 13.6 & 3.4 & 254 & 08\_0209 \\
\noalign{\smallskip}
E & 214 & 4.2 $\times$ 6.2 & 18.2 & 4.6 & 127 & 08\_0209 \\
\noalign{\smallskip}
\hline
\smallskip
\end{tabular}
% \vspace*{-0.1in}
% \tablecomments{\small }
\end{table*}

\begin{figure*}[!tbh]
\centering
\subfigure{
\includegraphics[scale=0.9,clip,trim= 0.2cm 5.35cm 10.1cm 4.5cm]{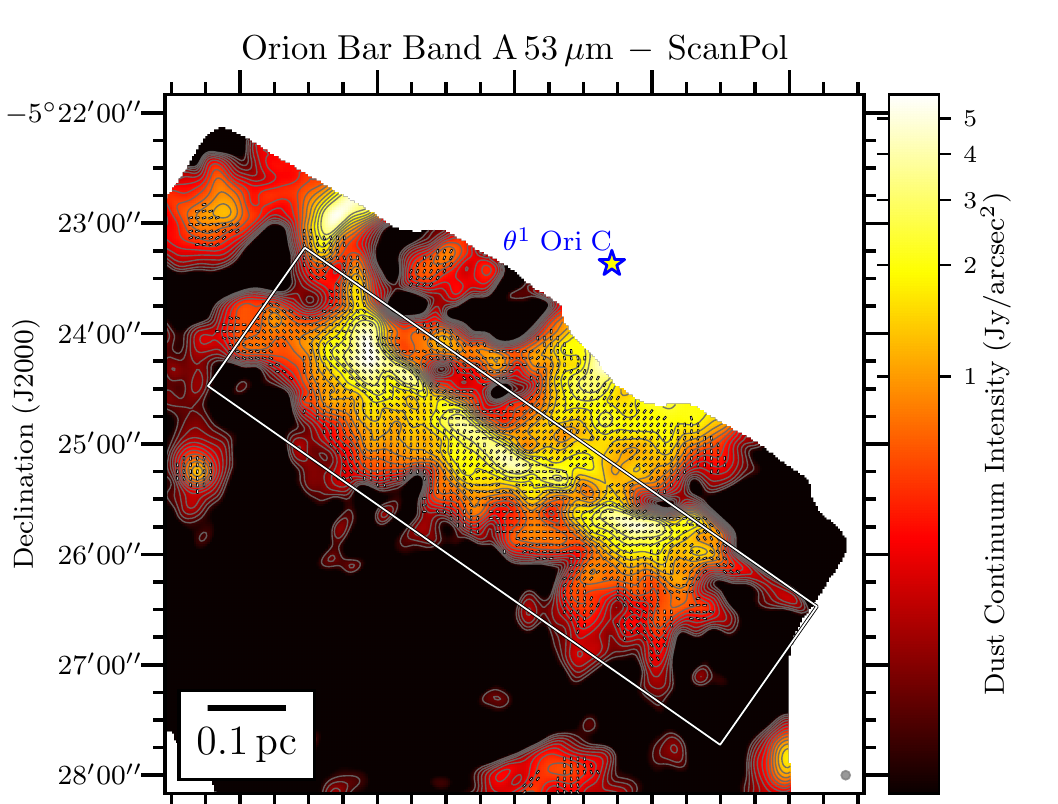}}
\subfigure{
\includegraphics[scale=0.9,clip,trim= 2cm 5.35cm 10.1cm 4.5cm]{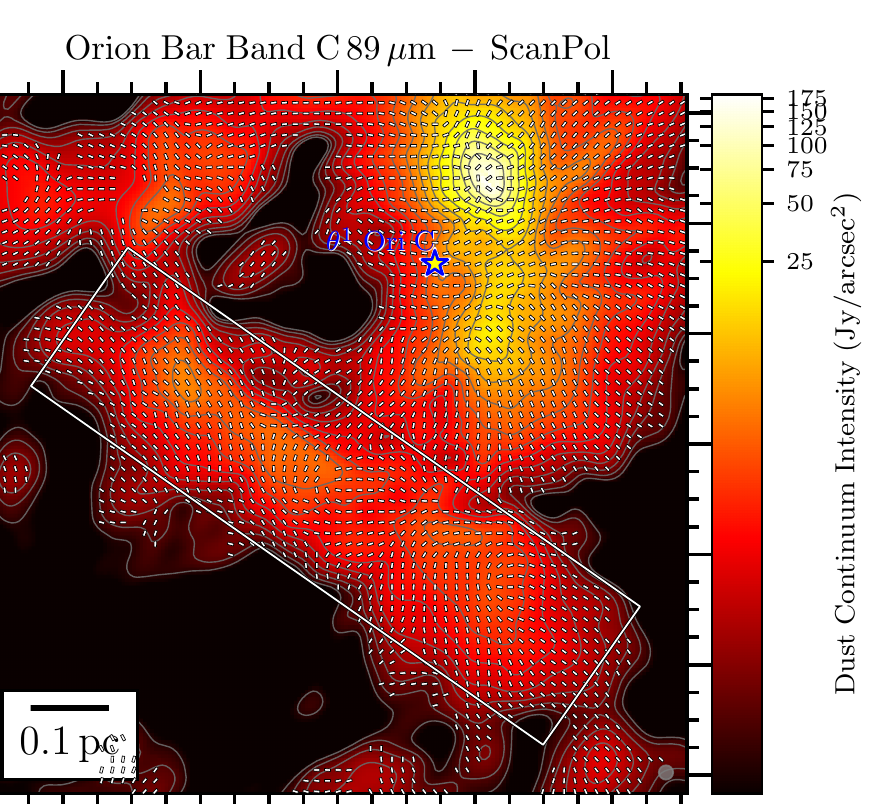}}
\subfigure{
\includegraphics[scale=0.9,clip,trim= 0.2cm 4.5cm 10.1cm 4.5cm]{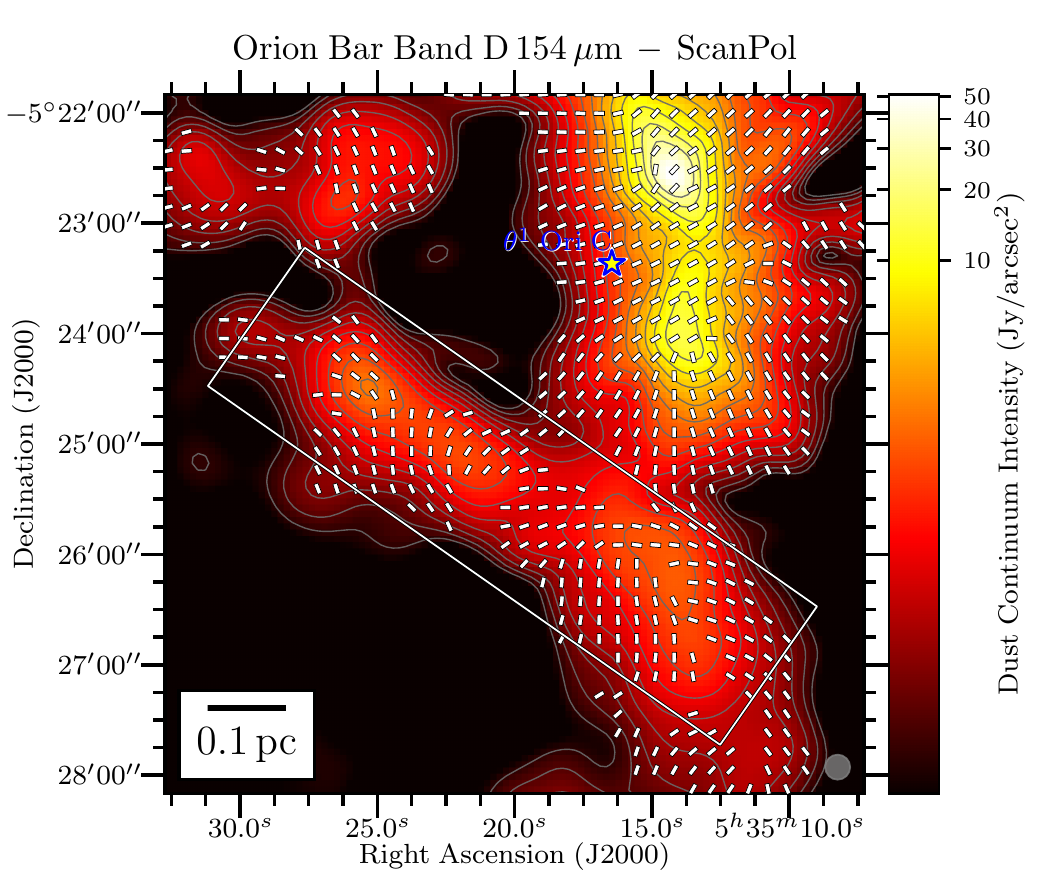}}
\subfigure{
\includegraphics[scale=0.9,clip,trim= 2cm 4.5cm 10.1cm 4.5cm]{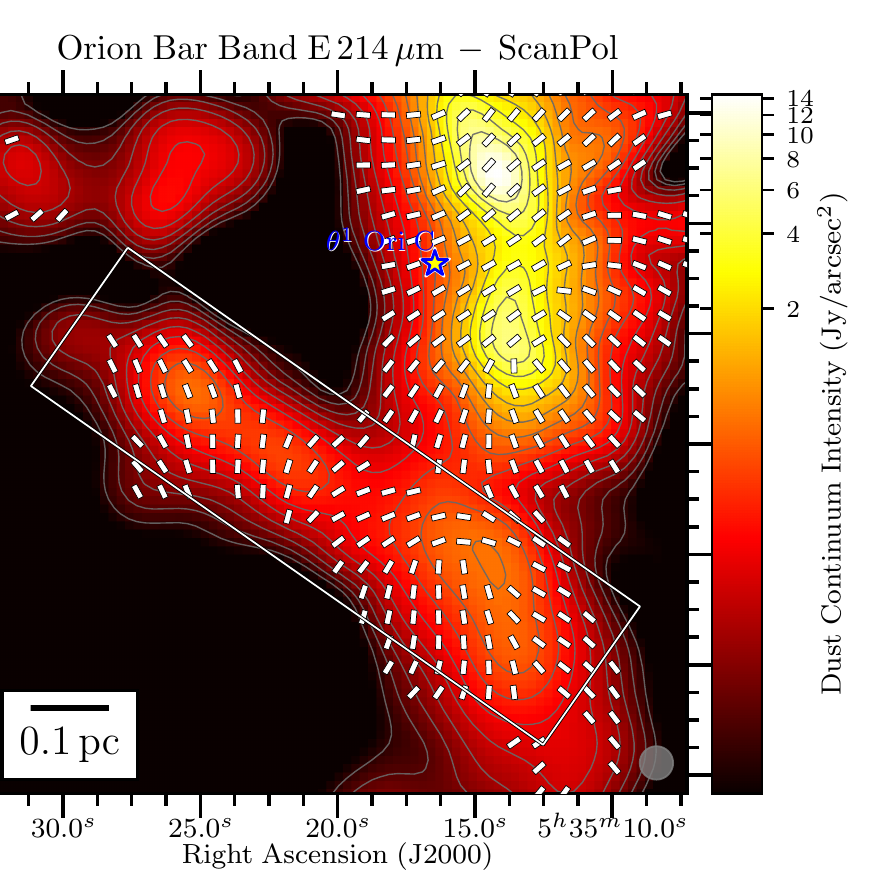}}
% \vspace{-0.1cm}
\caption{\small Polarization maps of the Orion Bar, observed with SOFIA HAWC+ using the OTFMAP polarimetric mode at 53 $\mu$m (Band A, top left), 89 $\mu$m (Band C, top right), 154 $\mu$m (Band D, bottom left), and 214 $\mu$m (Band E, bottom right). Lines segments represent the magnetic field orientation, rotated by 90$^{\circ}$ from the $E$-vector polarization angle maps. Vectors are plotted if $I/\sigma_I\;\geq\;100$ and $P/\sigma_P\;\geq\;5$. The length of the vectors does not represent any quantity. The color scale is the total intensity (Stokes $I$) of the thermal dust emission, shown from 35 $\sigma_I$. The black contours trace the dust continuum emission. The beam size is shown in the bottom right corner of each field (see Table \ref{t.obs}). The position of the $\theta^1\,$Ori\,C star, the most luminous star of the Trapezium cluster, is indicated. The rectangular grey box shows the location of the Orion Bar and denotes the region that this paper is focusing on.}
\label{fig:obs_bar_4bands}
\vspace{0.2cm}
\end{figure*}

Polarization data of the Orion Bar were obtained using the HAWC+ camera on board SOFIA \citep{Vaillancourt2007,Dowell2010,Harper2018} in two observing modes under two separate programs. Standard chop-nod (C2N) observations of the full Orion Molecular Cloud-1 (OMC1) region have been presented in \citet{Chuss2019}, gathering data acquired on flights 354, 355, 442, 444, 447, 450, 454, from December 2016 to September 2018, under the HAWC+ GTO programs 88\_0005 and 07\_0509  (PI: C.D. Dowell).
To evaluate the possible impact of off-beam contamination of the polarization, and gain additional signal-to-noise in the Orion Bar region, we acquired additional observations of the Orion Bar, using the on-the-fly-map (OTFMAP) polarimetric mode (also called scan-pol mode throughout this paper), on 2020 September 11-12, Flight 686, and 2022 September 26-27, flight 919, under program 08\_0209 and 09\_0037 (PI: B-G Andersson).  We also utilize archival OTFMAP observations from program 09\_0107 (PI: A. Tielens).

Quantitative comparison and discussion between the polarization quantities obtained with the C2N mode and those obtained with the OTFMAP mode (using different data reduction techniques) are presented in Appendix \ref{app:c2n_vs_otfmap}. However, throughout the body of this paper we only use the OTFMAP polarimetric maps for our results and analysis.
Here, we briefly present the mapping and reduction methods of the OTFMAP observations used in this study.  The general characteristics of the observations are presented in Table \ref{t.obs}. We use the standard pixel size of the pipeline given the large dynamic range of OMC-1. We note that lower-resolution using Nyquist sampling during the map-making algorithm is a more optimal approach for fainter objects \citep{LiPS2022,LopezRodriguez2022}. 

We reduced the data using the SOFIA data reduction pipeline\footnote{\href{https://github.com/SOFIA-USRA/sofia_redux}{https://github.com/SOFIA-USRA/sofia\_redux}} v.1.2.3 that integrates the CRUSH algorithm \citep{Kovacs2006PhD,Kovacs2008}. The OTFMAP polarimetric mode performs successive scans over a specific region (generally no larger than a few fields-of-view) parametrized with a Lissajous pattern. For each half-wave plate position angles (5.0$^{\circ}$, 27.5$^{\circ}$, 50.0$^{\circ}$, and 72.5$^{\circ}$), a scan is taken and reflective and transmissive data are collected as time series. Each set of four scans are reduced by the CRUSH algorithm, which estimates and removes the correlated atmospheric noise, removes the instrumental signals, performs gain estimations and noise weighting in an iterated framework. 

While the main advantage of the OTFMAP mode over the C2N mode is the optimization of the total observing time to reach a given level of signal-to-noise ratio (SNR), it has some difficulties in recovering large-scale, diffuse, and faint emission (see \citealt{LiPS2022,LopezRodriguez2022} for a description of the OTFMAP mode of HAWC+). Indeed, a comparison by eye between the C2N data of OMC-1 \citep{Chuss2019} with the OTFMAP data we present here reveals that the C2N mode is more efficient at recovering the diffuse polarized emission of the background cloud, between the Orion-KL and the Orion Bar for example. In order to improve the SNR of the the faint polarized dust emission of the Orion Bar we applied different numerical filters and filtering options of the pipeline \eg \texttt{faint, extended, deep}, and varied the number of iterations, while paying careful attention to the different polarization quantities produced by the full pipeline reduction algorithm. For the Band A, D, and E data, comparing with the nominal reduction pipeline configuration, we find that the \texttt{extended} and \texttt{deep} filters produce artificial polarization signals. For example, we retrieve significant rotation of polarization position angles, and modification of the structure in Stokes $I$ and high polarization fractions in regions of SNR in total intensity of $\geq\,200$, compared to the nominal reduction. We also retrieve uniform polarization position angles toward some regions of faint emission, not recovered by the nominal reduction. This is likely because the corresponding integration times are not long enough to ensure most of the faint emission to be at a reasonable level of SNR. However, the \texttt{faint} filter worked reasonably well on these data, offering a slight increase in SNR in the polarized dust emission. The Band C data are of much higher SNR due to longer integration time (see Table \ref{t.obs}), and the recovery of the extended and diffuse emission improves significantly by using the \texttt{faint} and \texttt{deep} filters. However, the Orion Bar is already completely detected with a reasonable SNR criteria by the standard pipeline using the nominal configuration for CRUSH. Therefore, for consistency, no specific filters were ultimately used in the four OTFMAP datasets we use here. In addition, we explored a range of iterations \texttt{rounds} for the reduction of each dataset. Increasing \texttt{rounds} systematically increases the SNR of the polarized dust emission in regions of SNR $\geq\,200$ in total intensity, up to a point where artificial polarization increases significantly. Applying a conservative approach, and use \texttt{rounds} of 30, 15, 10, 10, for Band A, C, D, E, respectively.

An important step in the reduction of HAWC+ data is the estimation of the zero-level background of the observations. Because the instrument is subject to both the variable atmosphere and the emission from the astrophysical source, the reduction algorithm may produce areas of negative flux where the emission from the source and the atmosphere are at similar levels. This can, in turn, potentially cause a loss of flux, but which can be corrected by the reduction algorithm. In order to correct for the zero-level background, we have determined a region of faint emission using \textit{Herschel} archival images at 70, 100, 160 and 250 $\mu$m, covering the HAWC+ field for each observation wavelength (see the corresponding regions highlighted in Figure \ref{fig:obs_bar_4bands_zero_lvl_zone}). We require the pipeline to estimate the mean of this zero-level region and to add this value to the entire map in each scan. Following the method presented in \citet{LiPS2022} (see their Section 2.3), we find that the zero-level background correction contributes a median of 2.5 $\pm$ 0.6\,\%, 0.55 $\pm$ 0.07\,\%, 1.5 $\pm$ 0.5\,\%, 0.3 $\pm$ 0.3\,\%, in the Band A, C, D, and E data, respectively.

The properties of the linear polarization of thermal dust emission are expressed by the Stokes parameters $Q$ and $U$. Stokes $I$ represents the total intensity of the emission. $\sigma_I$, $\sigma_Q$, $\sigma_U$ are the error of Stokes $I$, $Q$, and $U$, respectively. For linear polarization, the polarized intensity is defined as $P\,=\,\sqrt{Q^2 +U^2}$, which we have systematically debiased using the expression in \citet{WardleKronberg1974,Vaillancourt2006,Hull2015b}. The polarization fraction is the part of the total intensity which is linearly polarized, and is defined as $\Pf\,=\,P/I$. Finally, the polarization position angle $E$-vector is defined as $\phi\,=\,0.5\;\textrm{arctan}\left(U/Q\right)$. The corresponding uncertainties $\sigma_P$, $\sigma_{\Pf}$, and $\sigma_\phi$, are derived following \citet{Gordon2018}. We apply three conservative cuts in SNR on the HAWC+ polarization quantities throughout this paper which are: $I/\sigma_I\;\geq\;100$ (we use sometimes a SNR value of 200 for more conservative quantification), $P/\sigma_P\;\geq\;5$, and $\Pf\;\leq\;30\,\%$. After correction for instrumental polarization, HAWC+ has an absolute error of 3$^\circ$ in polarization position angle, and 0.4\% in polarization fraction \citep{Harper2018}.

\section{Results}
\label{sec:results}

In this Section, we analyze the spatial variation of the polarization quantities along and across the Orion Bar, as well as how these quantities vary as a function of wavelength. We are particularly interested in analyzing how the polarization varies along the minor axis of the Bar, transitioning from the line-of-sights (LOSs) toward the irradiated side to the colder region exposed to much lower UV intensity, but also along the major axis of the Bar, along which the Bar also experiences a gradient in radiation field strength. 

A transition in the grain alignment mechanism from $B$- to \krats would change the alignment axis of the grain, but detecting such transition highly depends on the projection of the relative orientation of the 3D magnetic and radiation fields, on the plane-of-the-sky (POS). The detectability of the $B$- to \krat transition is maximal when the plane defined by ($\vec{k};\vec{B}$) (where $\vec{k}$ and $\vec{B}$ are the radiative and magnetic field vectors, respectively) is parallel to the POS, in which case we could measure a change in polarization angle of $\angle(\vec{k},\vec{B})$ passing from $B$- to \krats. If the plane defined by ($\vec{k};\vec{B}$) and the POS are orthogonal to one another, any such transition would not be detectable in dual polarization. In practice, radiative transfer effects must also be taken into account to predict the apparent change of polarization angle, because the entire population of grains susceptible of RAT alignment, \ie both $k$-aligned and $B$-aligned grain population, will contribute to the observed polarized dust emission. In addition, the relative level of grain alignment efficiency of $B$- and \krats, as well as the potential differences between the $\angle(\vec{k},\overrightarrow{LOS})$ and $\angle(\vec{B},\overrightarrow{LOS})$ angles, will also affect the resulting dust polarization signal.

\citet{Salgado2016} estimated an inclination of the Bar of 4$^\circ$, which means that the anisotropic component of the radiation field received by the Orion Bar is roughly in the POS. The magnetic field from the Orion KL nebula of OMC1, which is located behind the Orion nebula \citep{Genzel1989}, is mostly perpendicular to the major axis of the dense filament \citep{Hough1986,Rao1998,Schleuning1998,Chrysostomou1994,Tang2010,Hull2014,Pattle2017,WardThompson2017,Chuss2019,Cortes2021}. From the observations presented here, we can see that the component of the OMC1 magnetic field projected on the POS, has roughly the same orientation of the radiation field vector from the Trapezium cluster, for the medium between the Bar and the Trapezium. As a consequence, if all the aligned grains of the Orion Bar were aligned via \krats, or via \brats with respect to the initial large scale magnetic field of OMC1, the polarization $B$-vectors would be uniform and roughly aligned the minor axis of the Bar, \ie pointing toward \ThOriC. This is clearly not the case (Figure \ref{fig:obs_bar_4bands}). Therefore, the magnetic field of the Bar is likely more complex, and a precise analysis of the polarized dust emission and the grain alignment conditions (dependent on the dust characteristics and the environmental conditions of the Bar, \eg the gas density and dust temperature) is required to investigate whether the aligned dust grains are susceptible to the \krats mechanism in the Bar.

\subsection{Multi-wavelength analysis of the polarization quantities in the Orion Bar}

\begin{figure*}[!tbh]
\centering
\subfigure{
\includegraphics[scale=0.36,clip,trim= 1.6cm 15.8cm 2cm 0.8cm]{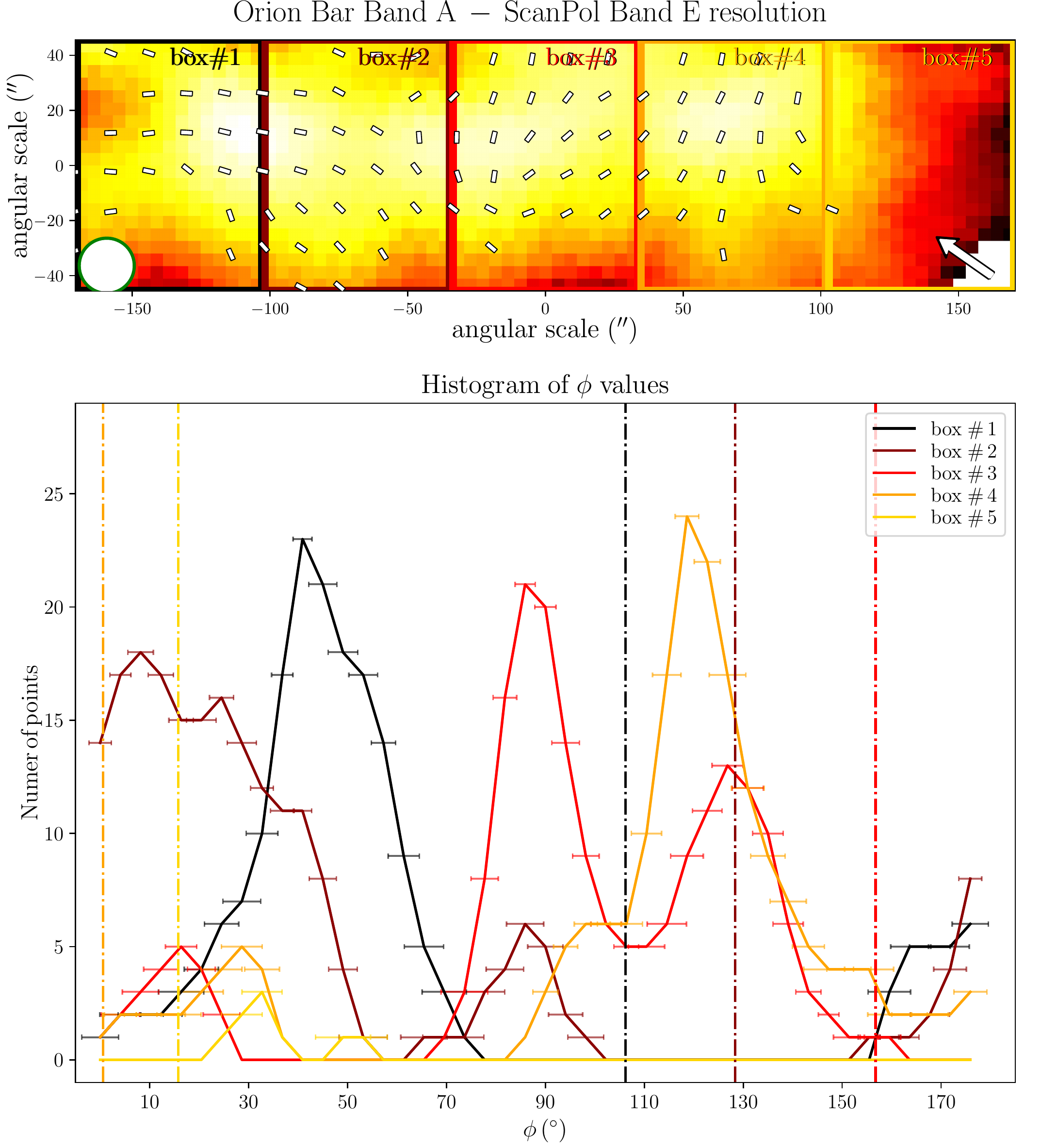}}
\subfigure{
\includegraphics[scale=0.36,clip,trim= 1.6cm 15.8cm 2cm 0.8cm]{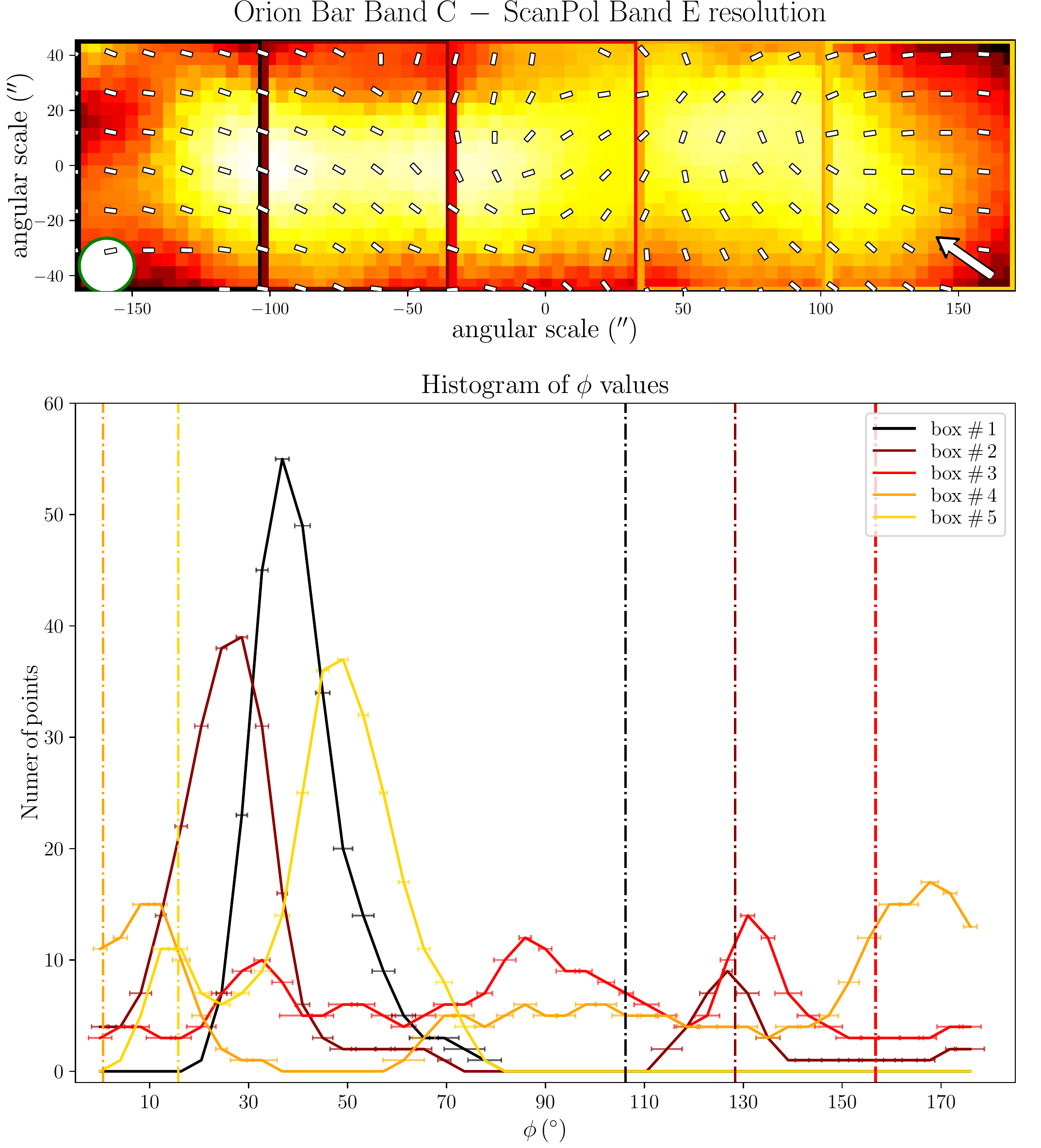}}
\subfigure{
\includegraphics[scale=0.36,clip,trim= 1.6cm 15.8cm 2cm 0.8cm]{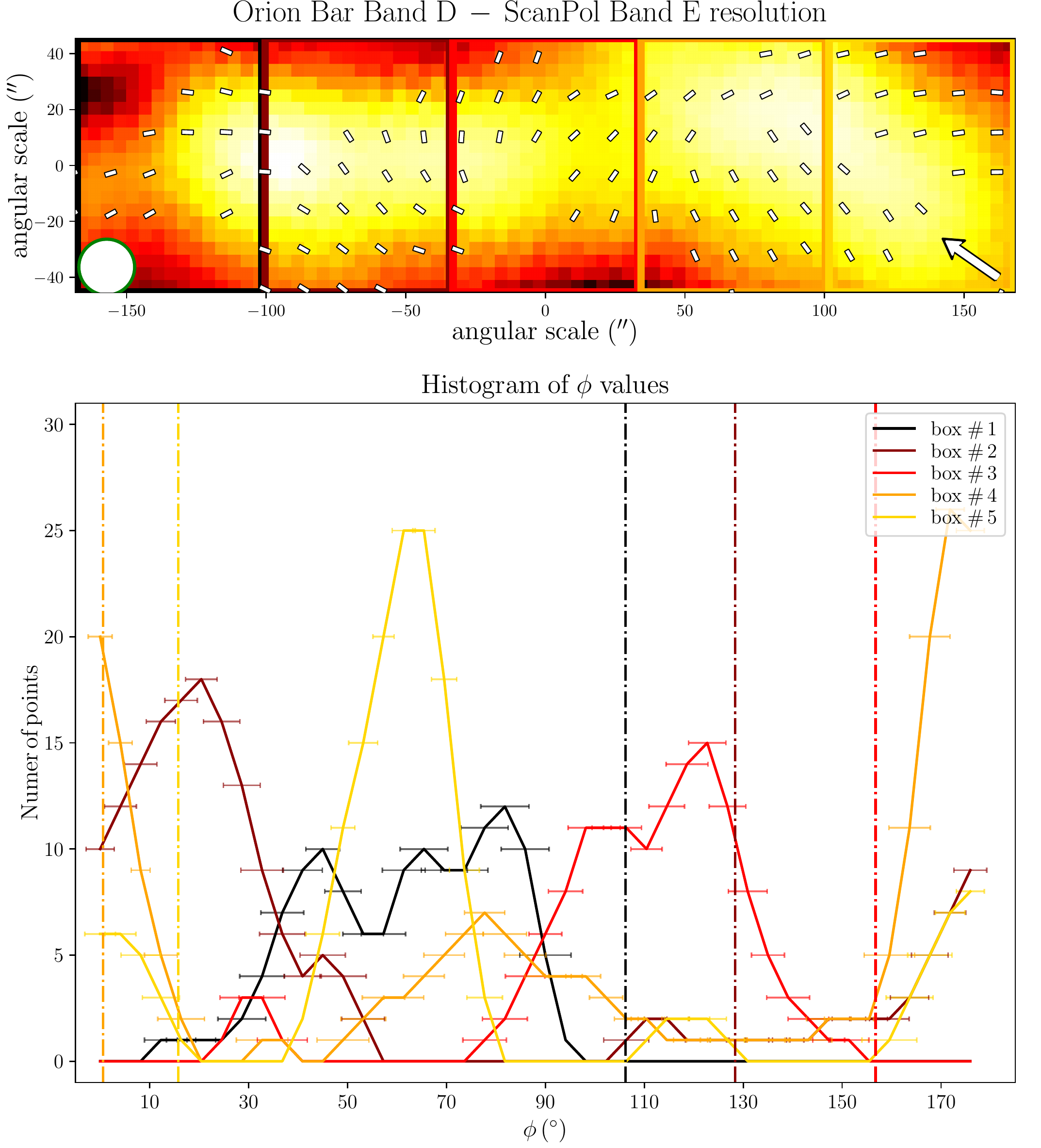}}
\subfigure{
\includegraphics[scale=0.36,clip,trim= 1.6cm 15.8cm 2cm 0.8cm]{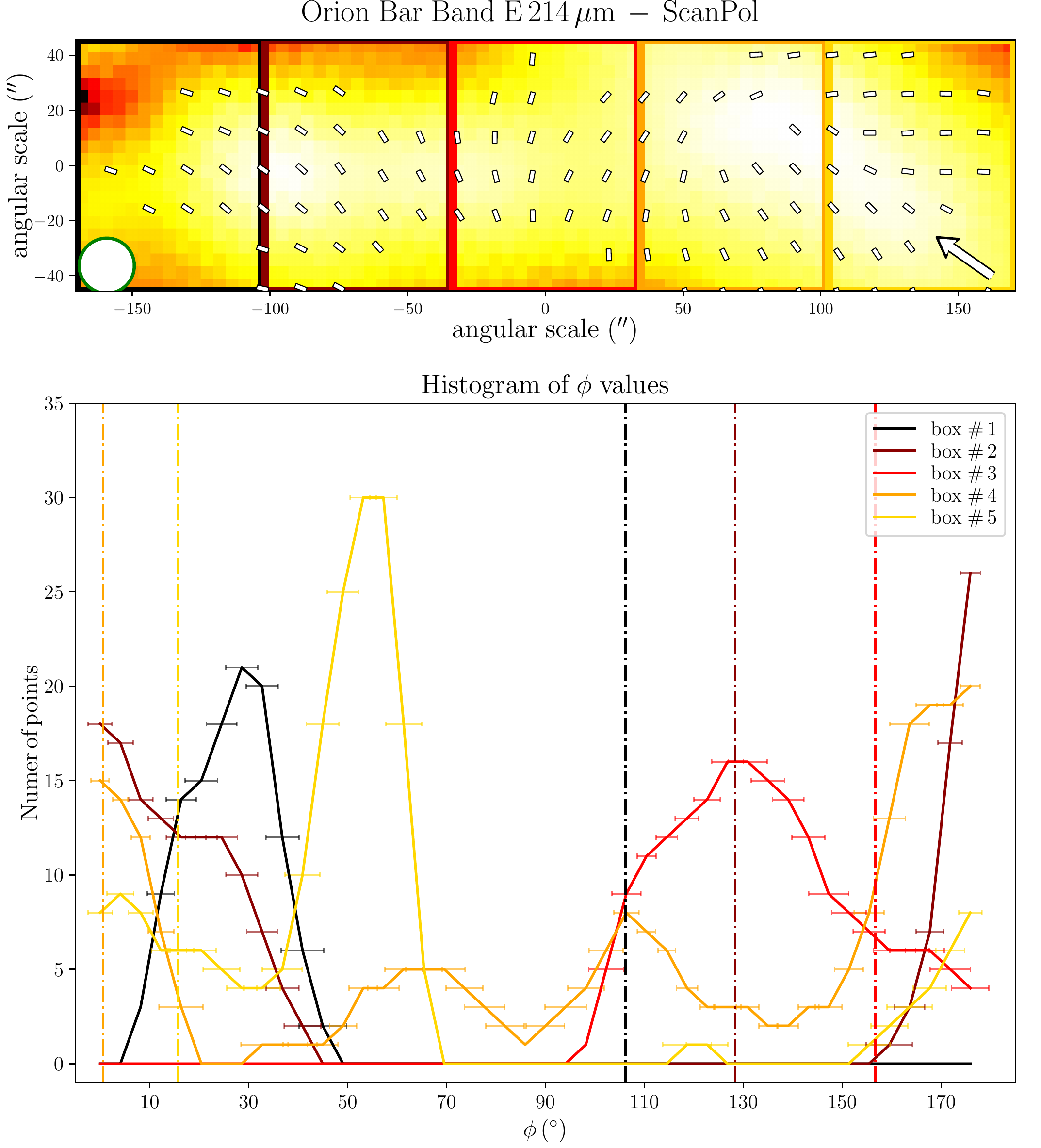}}
\vspace{-0.4cm}
\caption{\small Histograms of polarization position angles in the Orion Bar in the four HAWC+ bandpasses. The polarization maps from each bandpass have been smoothed and regridded to the Band E resolution and pixel size (see Table \ref{t.obs}). 
In each panel, the upper figure shows the total intensity in color scale and polarization position angles in the form of line segments, for the region highlighted in Figure \ref{fig:obs_bar_4bands} with grey rectangles. We split the Orion Bar into five regions along its major axis. The five sub regions are indicated with colored boxes. The polarization angle $B$-vectors are shown with respect to North, that is indicated by a white arrow. The white circles is the Band E beam resolution element.
The lower figure presents the histograms of polarization position angle $B$-vectors taken from North. Each colored line represent the histograms of the polarization angles within each of the five sub regions. The orientations of the radiation field emanating from \ThOriC and pointing to the center of the five boxes are indicated by the vertical dot-dashed lines of the corresponding colors. 
The SNR criteria for the polarization angles are SNR($I$)$\;\geq\;100$, SNR($P$)$\;\geq\;5$, $\Pf\;\leq\;30\,\%$. The error bars correspond to the mean of the uncertainty in the values of $\phi$ within each bin of the histograms. The histogram lines have been smoothed with a 1D-Gaussian kernel of a size of 4$^{\circ}$ for clarification.
Grains aligned via $k$-RAT would produce $B$-vectors polarization angles parallel to the radiation field. However, the peaks of the different histograms do not correspond to the vertical lines, that indicate the average radiation field direction in each box.}
\label{fig:obs_bar_hist5boxes}
\vspace{0.2cm}
\end{figure*}

\begin{figure*}[!tbh]
\centering
\subfigure{
\includegraphics[scale=0.36,clip,trim= 1cm 1cm 1cm 0cm]{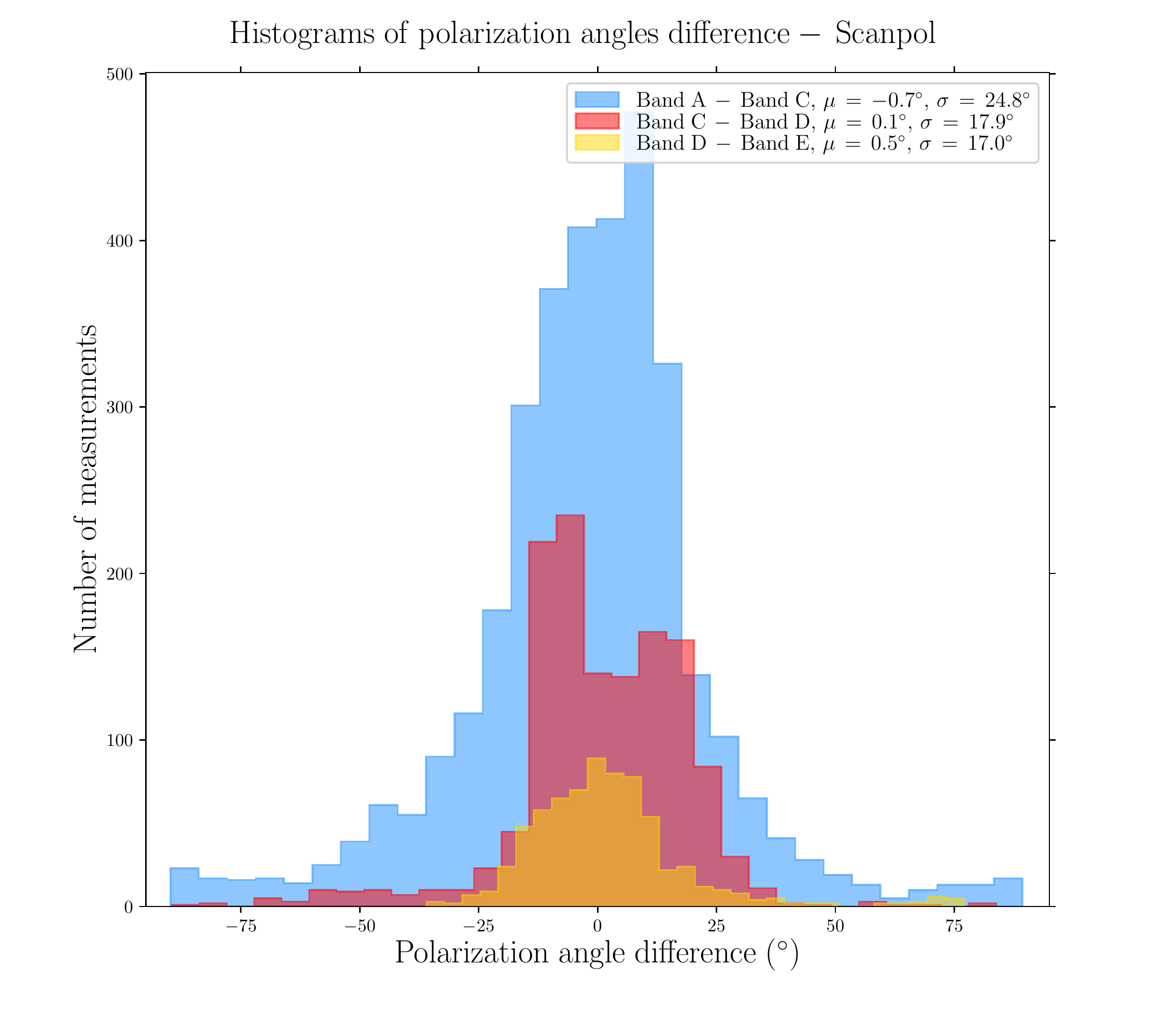}}
\subfigure{
\includegraphics[scale=0.36,clip,trim= 1cm 1cm 1cm 0cm]{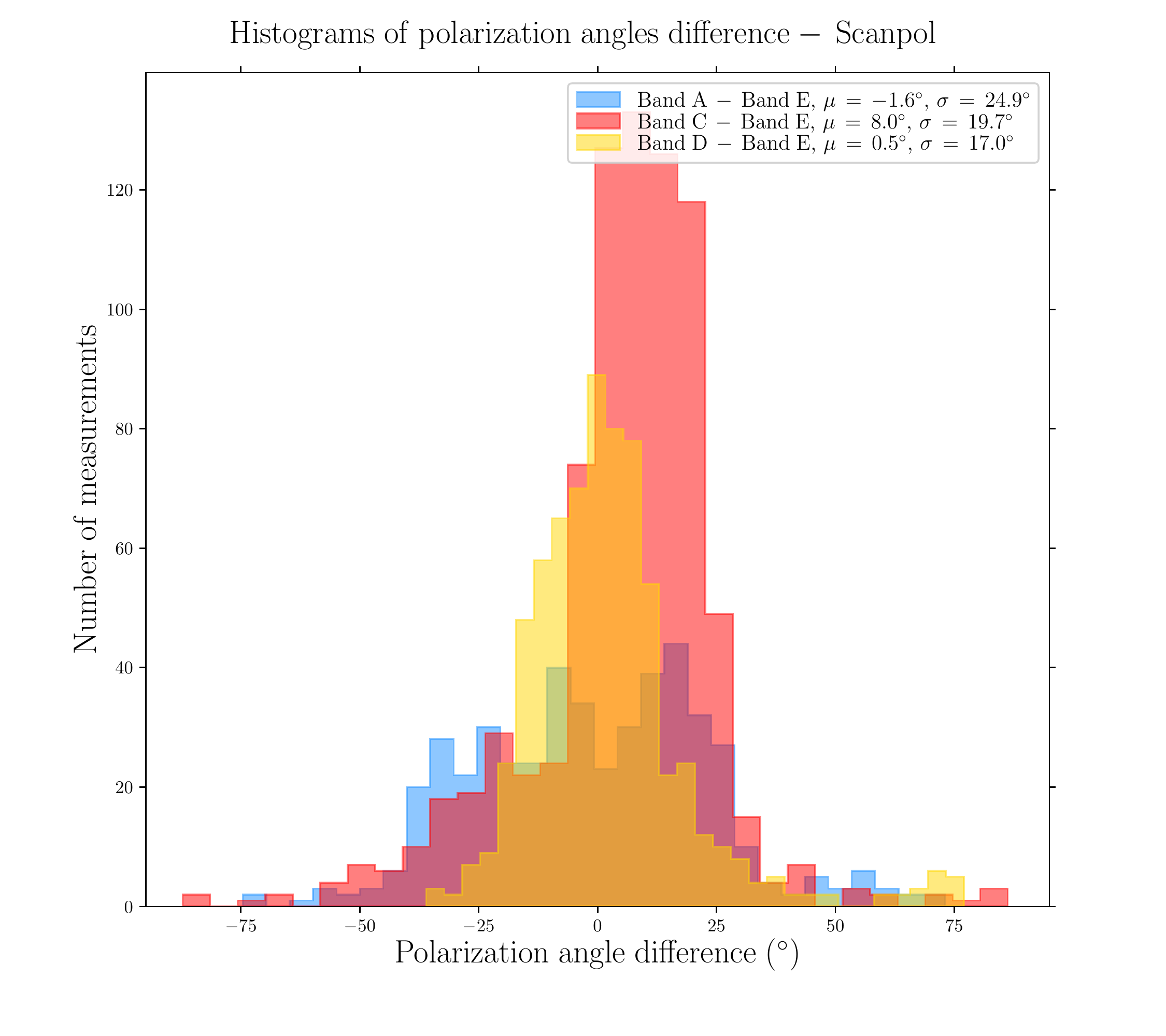}}
\subfigure{
\includegraphics[scale=0.36,clip,trim= 1cm 1cm 1cm 0cm]{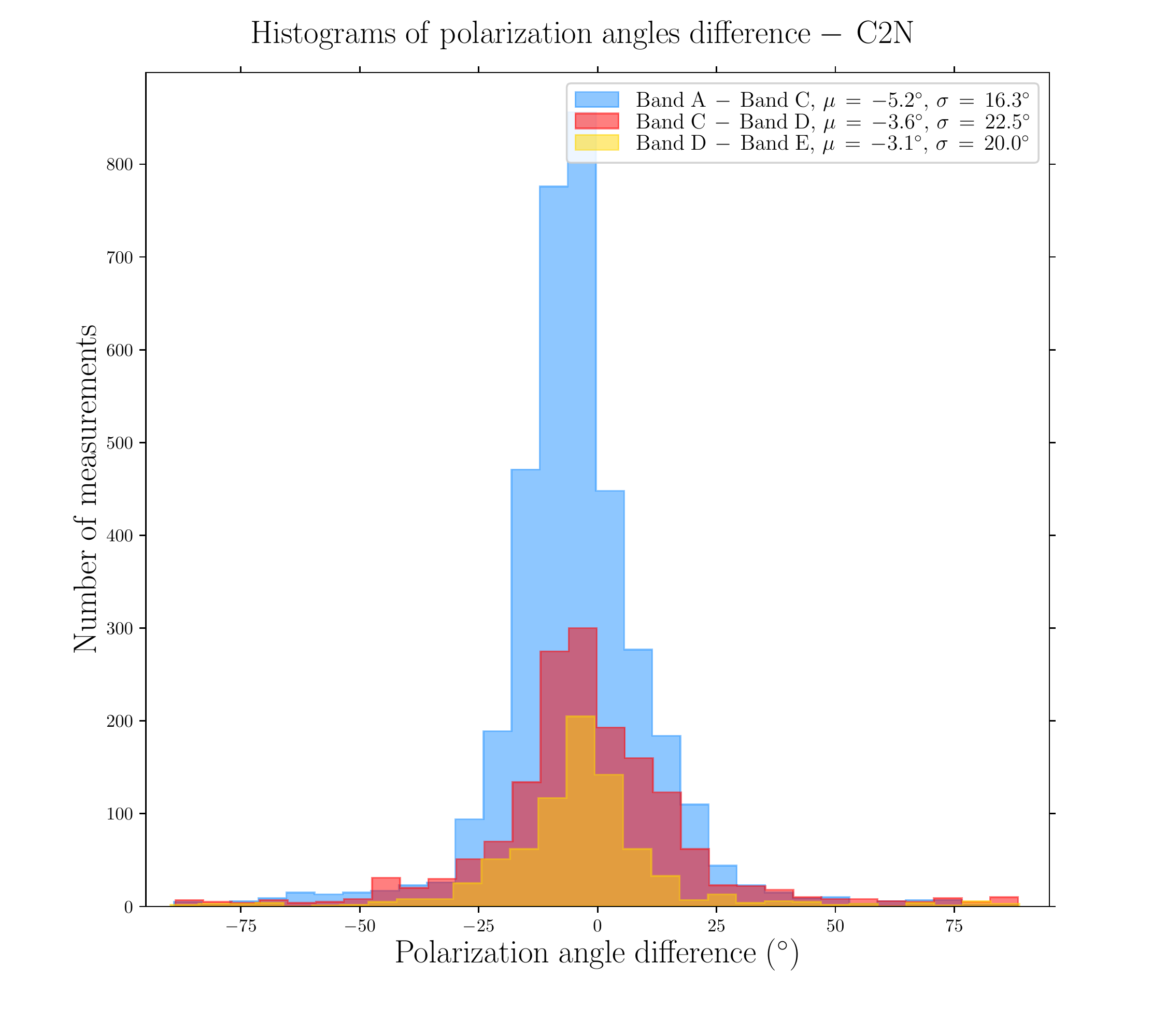}}
\subfigure{
\includegraphics[scale=0.36,clip,trim= 1cm 1cm 1cm 0cm]{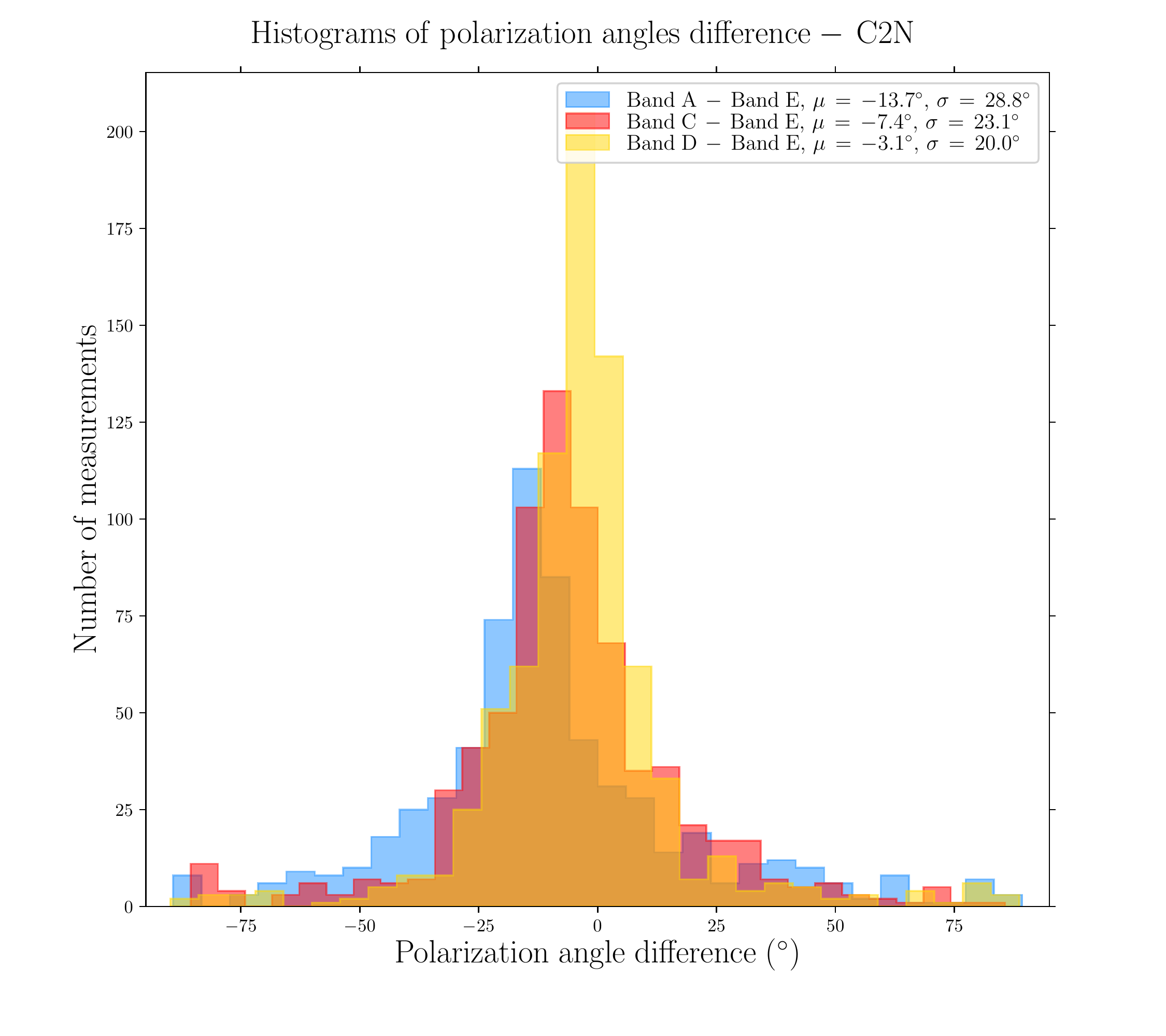}}
\vspace{-0.4cm}
\caption{\small Histograms of polarization angle difference in the Orion Bar between various pair of observations at different wavelengths, from the OTFMAP mode (top panels) and the C2N mode (bottom panels). For each pair of observations, the observations of highest angular resolution have been smoothed and regridded to the lowest resolution observation. The mean and standard deviation of each distribution is indicated as $\mu$, and $\sigma$, respectively. The data points come exclusively from the Orion Bar region denoted in Figure \ref{fig:obs_bar_4bands}. 
In order to calculate an angle difference to plot in the histograms for each distribution, at any specific point in the map, we require that the 2 differenced values both have SNR($I$)$\;\geq\;200$, SNR($P$)$\;\geq\;5$, $\Pf\;\leq\;30\,\%$.}
\label{fig:hist_comp_over_bands}
\end{figure*}

\begin{figure*}[!tbh]
\centering
\includegraphics[scale=0.6,clip,trim= 2cm 2cm 1cm 0cm]{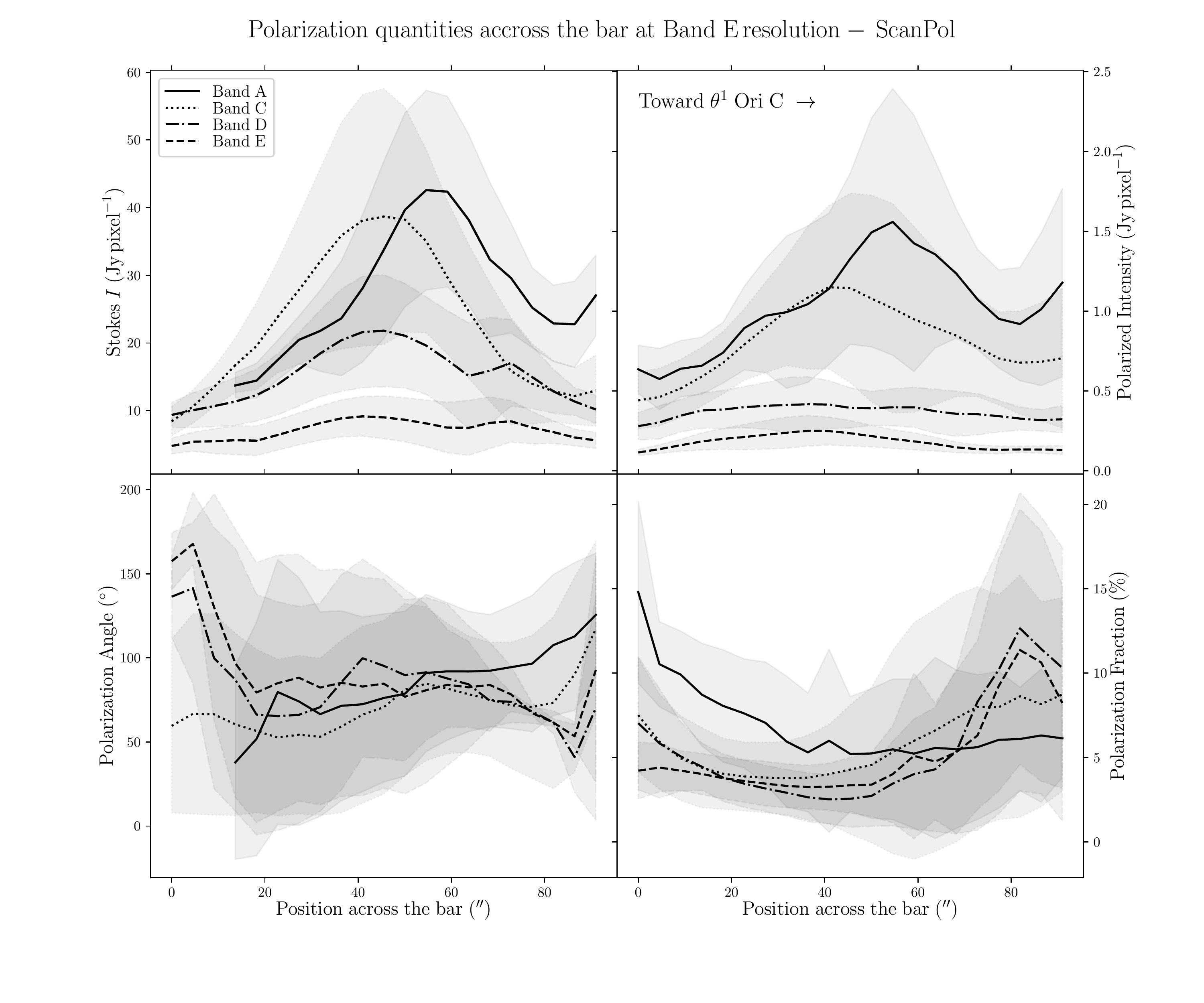}
% \vspace{-0.4cm}
\caption{\small Polarization quantities across the Orion Bar at the resolution of Band E (18.2 $^{\prime\prime}$) from the OTFMAP mode observations. The total intensity (top left panel), polarized intensity (top right panel), polarization position angle (bottom left panel), and polarization fraction (bottom right panel) profiles have been obtained by averaging the SNR-selected data along the major axis of the Orion Bar (long axis of the rectangle shown in Figure \ref{fig:obs_bar_4bands}). The profiles thus represent average quantities across the Orion Bar, as a function of position along its minor axis, which is directed toward the irradiation front. The shaded areas represent $\pm$ the standard deviation of each bin of points. The SNR criteria for the total intensity, polarized intensity, polarization position angle, and polarization fraction data are SNR($I$)$\;\geq\;200$, SNR($P$)$\;\geq\;5$, SNR($I$)$\;\geq\;200$ \& SNR($P$)$\;\geq\;5$ \& $\Pf\;\leq\;30\,\%$, and SNR($\Pf$)$\;\geq\;5$ \& $\Pf\;\leq\;30\,\%$, respectively.}
\label{fig:obs_bar_4bands_accross}
\end{figure*}

We present in Figure \ref{fig:obs_bar_4bands} the maps of the flux density in the Orion Bar at each of the 4 HAWC+ bandpasses overlaid with the polarization position angles showing the apparent magnetic field lines (rotating the polarization $E$-vectors by 90$^\circ$), adopting a sampling pattern of 1-2 vectors per beam resolution element. 
The rectangular grey box shows the location of the Orion Bar and denotes the region discussed in detail in the following Sections. The box's major axis has a position angle with respect to North of 55$^{\circ}$, and its dimensions are $90^{\prime\prime}\times340^{\prime\prime}$, centered on the FIR emission.
In the four plots of Figure \ref{fig:obs_bar_4bands}, we conserve the original resolution of the observations. 
However, later on in the analysis, we do regrid and smooth the Band A, C, and D observations to the Band E gridding and resolution\footnote{Smoothing polarization and their covariance map must account for the fact that the polarization reference direction frame on the celestial sphere varies over a projected map, and thus within the smoothing 2D-kernel. However, this effect is ignored because of the relatively small size of the HAWC+ maps. Therefore, the Stokes maps and their uncertainty are smoothed independently, and we rebuild the polarization quantities from the smoothed and regridded maps.}, to ensure accurate and quantitative comparisons across different wavelength (using the \texttt{reproject} python package of astropy; see Appendix \ref{app:add_pol_plots}).
The magnetic field exhibits a complex morphology, with several sub regions having homogeneous polarization patterns, but different from one another. Indeed, different organized components of the magnetic field co-exist, as seen in the South West of the Bar in the Band C, D and E observations, where two patches of polarized dust emission exhibit orthogonal magnetic field orientations. Band A observations, which have the highest spatial resolution, retrieve several structures in Stokes $I$, not observed in the lower resolution Band C, D and E maps. Different wavelengths can preferentially probe different regions if the distribution of dust temperature is not uniform on the line of sight. If those regions have different magnetic field geometries, they will contribute to polarization angles disparities across wavelengths.

To quantitatively estimate whether any regions of the Bar can exhibit dust polarization consistent with $k$-RATs, Figure \ref{fig:obs_bar_hist5boxes} presents histograms of polarization position angles ($B$-vectors) in five sub-regions, obtained by splitting the Orion Bar along its major axis. The three central regions experience a stronger radiation field given their higher proximity to the Trapezium cluster. Data at the four wavelengths, all with a gridding and resolution corresponding to those of the Band E, are presented. 
In each histogram, the orientations of the anisotropic radiation field emanating from \ThOriC and pointing to the center of the five boxes are indicated by vertical dot-dashed lines. Then, if a given region exhibits $B$-vector angles similar to the radiation field, those vectors would be consistent with the $k$-RAT mechanism.
In each box, and at all four observation wavelength, none of the main peak of the different histogram components correspond to the vertical lines. This indicates that the ambient radiation field direction does not dictate the direction of grain alignment in the Bar. 
However, two small areas in Figure \ref{fig:obs_bar_hist5boxes} exhibit $B$-vector polarization angles consistent with the radiation field: the center of box \# 3 in the Band E data (toward one equivalent beam surface area), and the bottom left of box \# 4 in the Band C, D, and E data (toward one equivalent beam surface area at Band C, two at Band E and D). This latter one corresponds to the cold and dense dust component in the South West of the Bar mentioned above, not recovered by the Band A observations (see below).

Toward the northern and middle regions (boxes \# 1, 2, 3) of the Bar, the polarization angles across the four wavelengths are consistent within 20$^{\circ}$ (based on the differences across wavelength of the peaks of the histogram components).
However, the southern region (histogram in box \# 4) shows clear differences between the Band A observations, and the observations taken at Band C, D, and E, because the band C, D, and E data have picked up the emission from a dense ($N_{\textrm{H}_2}\;\geq\;5\,\times\,10^{22}$ cm$^{-2}$) and cold ($T_{\textrm{d}}\;\leq\;30$ K) region, that is visible in total intensity (the $N_{\textrm{H}_2}$ and $T_{\textrm{d}}$ maps are from \citealt{Chuss2019}, see their Figure 3., and the Section \ref{sec:Pf_vs_Td_NH2} of this paper). In contrast, the emission from the Band A observations, sensitive to warmer dust, corresponds to regions closer to the irradiated front of $T_{\textrm{d}}\;\geq\;40$ K. The $B$-vectors in box \# 4 of Band A are mainly parallel to the minor axis of the Bar, which is also a pattern of emission seen with the three others wavelengths, alongside this component of colder dust. The polarization angles of box \# 5 are consistent across the Band C, D, and E observations within 10$^{\circ}$.
Finally, the polarization angles exhibited by the HAWC+ Band E observations are also consistent with the 850 $\mu$m dust polarization observations obtained with POL-2 on the JCMT \citep{WardThompson2017,Pattle2017}. The standard deviation of the distribution of polarization angles differences between the HAWC+ Band E and POL-2 data is $\sim$ 20$^{\circ}$, similar to what is obtained among the HAWC+ bands (see Figure \ref{fig:hist_comp_over_bands}). 

In order to go further in quantitatively comparing polarization angles across wavelength, Figure \ref{fig:hist_comp_over_bands} presents comparisons (as pair-wise differences) of the polarization angles between the various pairs of observations at different wavelengths, using the OTFMAP mode maps, and the C2N mode data presented in \citet{Chuss2019}. The distributions peak around 0$^\circ$ and the standard deviations are within 15-25$^\circ$, while $\sigma_\phi\,\leq\,5^{\circ}$ in the regions of Bar where SNR($I$)$\;\geq\;200$, SNR($P$)$\;\geq\;5$. Comparing the difference between the mean and the standard deviation of the polarization angle difference distributions, it is not possible to separate them with statistical significance, even if we increase the SNR criteria on the polarized intensity and Stokes $I$, in which case we also deselect too many data points. The distribution of the polarization difference between Band A and Band C, and between Band A and Band E, show a significant number of points outside the central peak of the distribution, corresponding to typical difference of +30 $^{\circ}$ and -30 $^{\circ}$. Those data points come from the south-east side of the Bar, where the SNR of Band A detections are the lowest. However, if we weight the polarization difference by their relative uncertainty (\ie computing $(\phi_{\lambda_1}-\phi_{\lambda_2})/\sqrt{(\delta\phi_{\lambda_1}^2+\delta\phi_{\lambda_2}^2)}$), we find that the histograms using the Band A still exhibit larger standard deviation compared to the others.
We thus suspect those data points may be artifacts from the data reduction. 

Figure \ref{fig:obs_bar_4bands_accross} presents the variation of Stokes $I$, $P$, $\phi$, and $\Pf$ for the four wavelengths OTFMAP observations as a function of the position along the minor axis of the Bar (\ie the minor axis of the gray rectangle used as a reference so far that is shown in Figure \ref{fig:obs_bar_4bands}), which is directed toward the irradiation front. We highlight that the data points of Figure \ref{fig:obs_bar_4bands_accross} are SNR-selected independently across bands. The goal is to determine how important the effects of the environmental conditions are for the emission. Indeed, our region of interest, defined by the rectangle highlighted in Figure \ref{fig:obs_bar_4bands}, shows a clear dust temperature gradient (see \citealt{Chuss2019}). 
The peak in Stokes $I$ and $P$ for Band A is clearly closer by 15$^{\prime\prime}$ to the irradiation front compared to the Stokes $I$ and $P$ peaks of Band C, D, and E observations, which is likely a dust temperature effect. As suggested above, Band A is more sensitive to the hot dust layer located at the irradiated edge of the Bar, compared to the dust emission emanating from the colder region not directly exposed to the irradiation of the Trapezium cluster.

The $B$-vector map exhibits several components, which explains why the variations of the mean polarization angle across the Bar in Figure \ref{fig:obs_bar_4bands_accross} shows significant discrepancies between the bands. The Band A observations at Band E resolution do not have a large number of independent polarization detections. This explains the marked rotations at the beginning and end of the profile of $\phi$ as a function of position across the Bar for this band, which are not statistically significant. Band C data, having a much higher signal to noise than the other bands, trace more reliably the underlying polarized dust emission emanating from the more tenuous outer part of the Bar, and/or from the background OMC1 cloud. Band D and E polarization angle profiles also show significant differences (up to 80$^{\circ}$ between the mean polarization angles) compared to Band A and C for positions $\leq$ 20$^{\prime\prime}$ from the cold side of the PDR. Again, lack of uniform detections toward these faint regions explains the retrieved discrepancies. Finally, the polarization fraction profiles follow opposite trends compared to the total intensity profiles. Effects due to both dust temperature and gas density must play a role in the resulting $\Pf$ profiles.  We will investigate these in Section \ref{sec:Pf_vs_Td_NH2}.

In this data analysis, no variations of the polarization position angles as a function of wavelength are detected with statistical significance. 
In all our observations, the polarization angles analyzed alongside the radiation field direction along the Bar show no evidence for $k$-RAT aligned grains.
However, the analysis of the dust polarization observational data alone is limited by the complexity of the apparent magnetic field lines structure, the observational wavelengths probing different environmental conditions across the Orion Bar, and the observation data quality.

\subsection{Variations of the polarization degree as a function of environmental conditions}
\label{sec:Pf_vs_Td_NH2}

The effects of the radiation field and the gas density govern the radiative torques and the efficiency of de-alignment by gaseous collisions, respectively. This in turn regulates the polarization level of the dust emission. We use the work of \citet{Chuss2019} (see also \citealt{Arab2012}), who gathered HAWC+, \textit{Herschel} PACS \& SPIRE \citep{Andre2010,Abergel2010,Andre2011,Bendo2013}, JCMT/SCUBA-2 \citep{Mairs2016}, GBT and VLA \citep{Dicker2009}  observations of the Orion nebula and fitted modified blackbody spectra for each pixel.  These fits yielded column density, dust temperature and emissivity maps with 18.7$^{\prime\prime}$ angular resolution and a 3.7$^{\prime\prime}$ square pixel size. To ensure statistical independency, we regrid these maps to a Nyquist sampled map of 4 pixels per beam area, resulting in a pixel size of 8.3$^{\prime\prime}$. We then smooth and regrid the maps of the four bands of HAWC+ observations to the resolution and pixel sampling of the dust temperature and gas column density maps, \ie to 18.7$^{\prime\prime}$ angular resolution with 8.3$^{\prime\prime}$ square pixel size.

We compare our polarization results with three physical quantities describing the local environmental conditions: the line of sight dust temperature \& column density, and the derived gas column density between the Bar and the Trapezium cluster \NHtc.  That is, the column density ``seen'' towards the Trapezium by each location we map in the Orion Bar. The dust temperature will serve as a proxy for the radiation field intensity, a major parameter that govern the efficiency of radiative torques, and we use \NHtsqT as a proxy for the efficiency of de-alignment by gaseous collisions because the gaseous collisional rate is proportional to this quantity \citep{Draine1996}. 

In order to compute \NHtc, we first estimate the gas column density of the Bar itself \NHt, by subtracting 2$\times$10$^{21}$ cm$^{-2}$ from the derived column densities, corresponding to the background cloud OMC1 based on the SED fitting of the maps. We then estimate the gas volume density in the Bar $n_{\rm{H}}$ by dividing \NHt by 0.28 pc, the size of the Bar along the LOS derived by \citet{Salgado2016}, who used an estimation of the absorbing area at the surface of the PDR. We finally sum the gas density values along the lines separating each pixel in the Bar from the Trapezium cluster to derive \NHtc.

\begin{figure}[!tbh]
\centering
\includegraphics[scale=0.62,clip,trim= 0.cm 0.5cm 1cm 1.5cm]{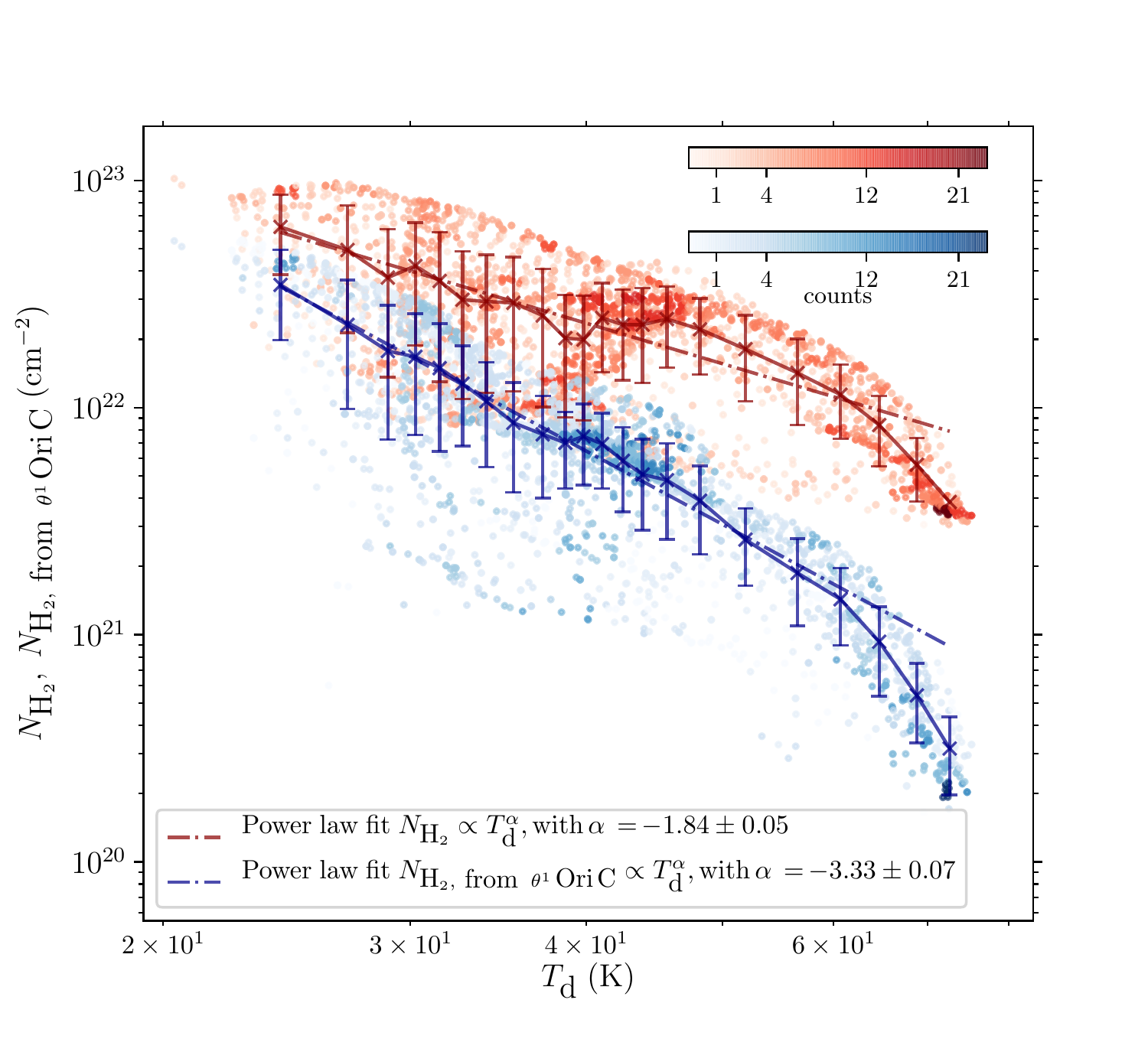}
\caption{\small Gas column densities \NHt (blue) and \NHtc (red) as a function of the dust temperature obtained from the grey body SED fitting of \citet{Chuss2019}. The solid lines are the running mean, and the error bars represent the standard deviation of each bin. Power law fits are shown with the dot dashed lines.
}
\label{fig:Td_NH}
\end{figure}

Figure \ref{fig:Td_NH} shows the two-dimensional histogram of the gas
column density after the correction for the background OMC1 material as a function of dust temperature toward the Orion Bar. The gas column density is inversely proportional to the dust temperature for the complete range of \Td values probed here. 
We note that \citet{Chuss2019} performed Markov Chain Monte Carlo tests to evaluate the systematic covariance between the SED fitting parameters (\ie \Td, \NHt, and $\beta$ the dust emissivity spectral index, see their Section 3.1.5). While there is some covariance between \Td and \NHt, the width of the likelihood functions are such that we can consider that the trends of Figure \ref{fig:Td_NH} have a physical origin.
The RAT alignment theory predicts that the minimum size of aligned dust grains decreases with increasing radiation field, and with decreasing volume density. We note this point because \citet{Tram2021c} flagged values below a specific dust temperature, below which gas column density was positively proportional to the dust temperature. In our case, we expect the polarization degree to increase (decrease) as a function of \Td (\NHtsqT), for the full range of our \Td (\NHtsqT) values. A departure from this trend can be explained by an evolution of dust properties as a function of the environmental conditions. In this context, \citet{Tram2021,Tram2021b,Tram2021c} proposed that the decrease of $\Pf$ as a function of \Td observed at high dust temperature ($T_\textrm{d}\,\gtrsim\,40-50$ K) with HAWC+ toward star forming clouds/dense cores can be caused by RATD. In addition, we note that \NHtsqT and \NHtc are used here to quantify the effect of two different physical quantity on the efficiency of radiative torques. While we use \NHtsqT as a proxy for the efficiency of collisional gaseous de-alignment, we use \NHtc as a proxy for the reddening of the radiation field emanated from the Trapezium cluster. Because the radiative torque efficiency strongly decreases for grains smaller than the wavelength of impinging photons, the grain alignment efficiency induced by RATs should be affected by the reddening \citep{LazarianHoang2007}.

\begin{figure*}[!tbh]
\centering
\includegraphics[scale=0.43,clip,trim= 4.5cm 4cm 3cm 0.5cm]{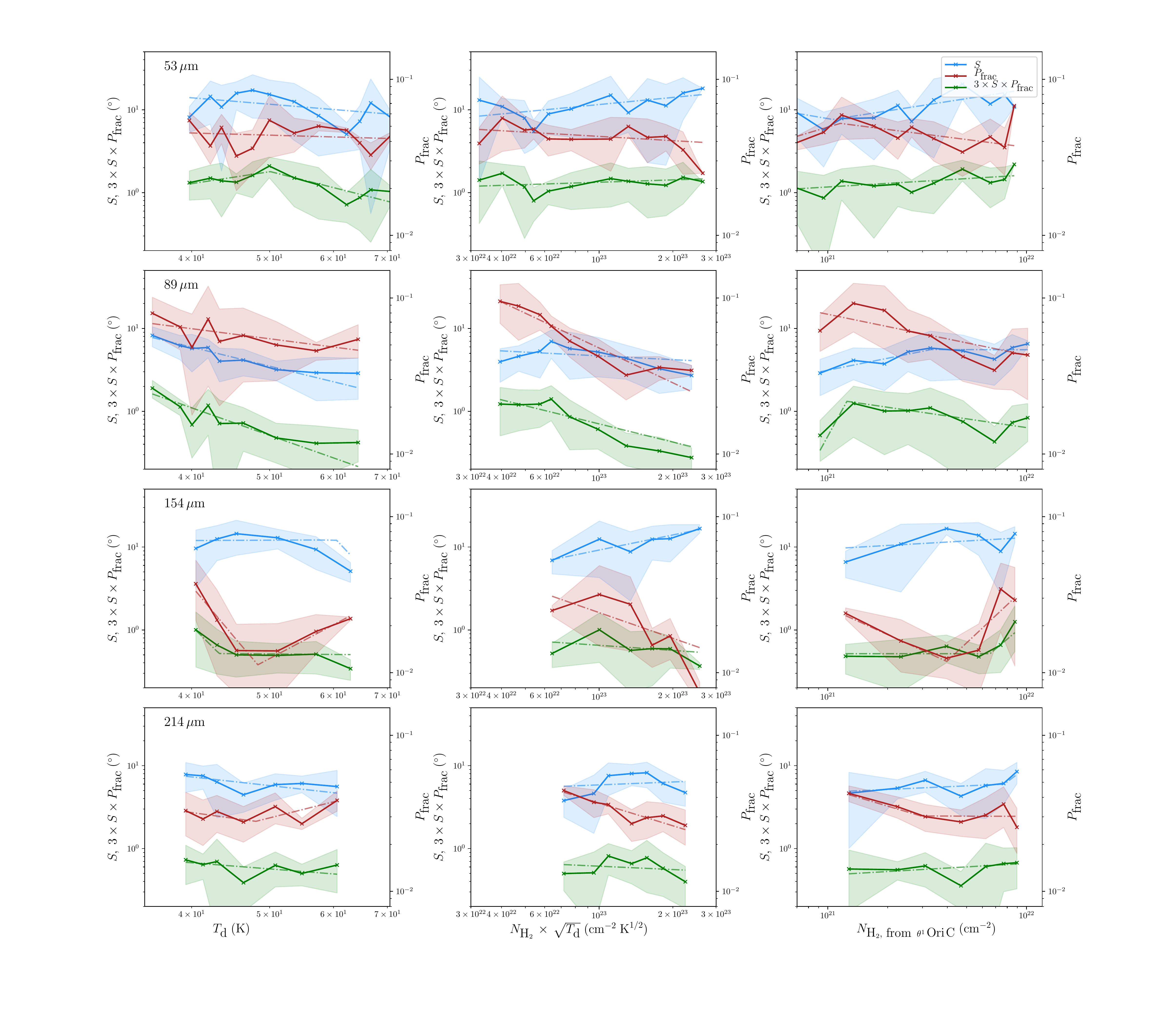}
\caption{\small Function of the mean of $\S$ (blue lines), $\Pf$ (red lines), and $\StimesP$ (green lines), as a function of the dust temperature $T_{\textrm{d}}$ (left column), \NHtsqT (middle column), and the plane-of-the-sky gas column density between \ThOriC and the Bar \NHtc (right column), obtained with the HAWC+ polarization data observed at 53 $\mu$m (top row), 89 $\mu$m (top-middle row), 154 $\mu$m (bottom-middle row), and 214 $\mu$m (bottom row). For each distribution, the dotted-dash line is a power law fit to the data. A broken power law is used when it fits the distribution with more accuracy (based on a $\chi^2$ criteria). The shaded area represents $\pm$ the standard deviation of each bin of data points.
At each wavelength, the maps have been regridded to 4 pixels per beam area to ensure statistical independence. 
}
\label{fig:multi_S_Pf}
\end{figure*}

The disorganized component of the magnetic field can be a source of depolarization due to cancellation of the polarization signals on the line of sight, as shown by \citet{Jones1992}.  It can also cause a decrease in the observed polarization when the typical scale of the POS magnetic field fluctuations are smaller or comparable to the spatial resolution. Therefore, alongside the polarization fraction $\Pf$, we also consider the dispersion of polarization angles in the POS $\S$, which allows us to quantify the level of disorganization of the apparent magnetic field lines.  Loss of grain alignment also directly impacts the fractional polarization. To disentangle between these effects, one needs to analyze $\S$ and $\Pf$ together, using the disorganized component of the magnetic field provided by $\S$ to determine whether changes in $\Pf$ can be attributed to changes in the grain alignment efficiency. Several studies have revealed that $\Pf$ and $\S$ are correlated, showing the role of depolarization due to magnetic field disorganization \citep{Fissel2016,Planck2018XII,Chuss2019,LeGouellec2020}. We use the nearest neighbors approach  to derive $\S$ \citep{LeGouellec2020}, given the pixel sampling of our maps. To obtain $\S$, we debias the maps of measured dispersion $\S_m$ by subtracting its uncertainty: $\S\,=\,\sqrt{\S_{\textrm{m}}^2-\sigma_{\S}^2}$, following the relation of \citet{Alina2016}. We then use $\StimesP$ as a proxy for the polarization degree corrected for the depolarization effects induced by the disorganized component of the magnetic field. We use SNR($\Pf$)$\,\geq\,$5 and SNR($I$)$\,\geq\,$200 for our data selection criteria.

Figure \ref{fig:multi_S_Pf} presents the evolution of $\Pf$ (right-hand axes), $\S$, and $\StimesP$ (left-hand axes) as functions of \Td, \NHtsqT, and \NHtc, for the four wavelength observations (see Section \ref{sec:required_RAT} for the polarization fraction spectra specifically). With this approach, we want to determine whether the evolution of polarization degree (tentatively corrected from depolarization caused by the disorganized component of the magnetic field) as a function of environmental conditions suggests varying alignment conditions. Given the limited sensitivity of the Band A observations, and the large dynamic range between the Band A initial resolution and the angular scale at which we smooth the maps to, \ie 18.7$^{\prime\prime}$, the corresponding distribution shown in Figure \ref{fig:multi_S_Pf} are not conclusive and must be interpreted with caution.  However, Band C, D, and E, present clearer systematic trends. 

%%%%%%%%%%%%

\subsubsection{Polarization quantities versus \Td}

The slope of the $\Pf$ trends as a function of \Td appears to be on average slightly decreasing or flat at 53, 89, and 154 $\mu$m. A small increase is seen at 214, and 154 $\mu$m, for \Td\,$\geq$\,50\,K. At the four wavelengths of observation, $\S$ is flat or decreasing with increasing \Td. Given the anti-correlation of \NHt with \Td, this decrease of $\S$ with \Td can be explained by an increase of magnetic field apparent disorganization, or structure complexity, with the line-of-sight depth in the Bar, which is know to harbors high density substructures \citep{Habart2022}. 
$\StimesP$ systematically decreases with \Td, or exhibits a flat trend.
As $\StimesP$ is proposed to trace grain alignment corrected for depolarization by the B-field structure, the grain alignment efficiency is on average found to slightly decrease with \Td. These trends are not able to be explained by the basic RAT theory alone, and suggest an evolution of dust grain properties as a function of temperature, which could be caused by RATD.

A line of sight with a given derived value of dust temperature may actually correspond to multiple dust components of different inherent dust temperatures. This can explains the different distributions of $\Pf$ at the different wavelengths, for the same range of dust temperatures. Shorter wavelength observations are more sensitive to hotter dust, more affected by RATD. This can explain why the decrease of $\StimesP$ versus \Td at 89 $\mu$m is more important than at 154 and 214 $\mu$m. We note that if RATD were strongly efficient throughout the Orion Bar (see Section \ref{sec:RATD}), we would expect a clear and ubiquitous anti-correlation between polarization degree and dust temperature. In addition, the grain alignment efficiency within the hotter side, \ie where \Td$\gtrsim\,$ 50 K, remains high, with $\StimesP$ values around $0.15\,-\,0.3^{\circ}$, suggesting that the radiative feedback does not totally hinder grain alignment.

\subsubsection{Polarization quantities versus \NHtc}

In the four bands $\Pf$ decreases with \NHtc, expect in the $154\,\mu$m data, where $\Pf$ increasing with increasing \NHtc, for \NHtc$\,\geq\,4\,\times\,10^{21}$ cm$^{-2}$. $\S$ systematically increases with \NHtc. The situation of $\StimesP$ is less clear; it exhibits mostly flat trends and is potentially increasing at high \NHtc. While the complexity of the environment preclude us from drawing clear conclusions about the evolution of grain alignment versus \NHtc, the flat or increasing trends of grain alignment efficiency versus \NHtc is hard to explain with basic RAT theory, because the increasing of reddening is supposed to reduce the number of grains susceptible to alignment. The region where \Td$\lesssim\,$ 45 K and \NHtc$\gtrsim\,5\times$10$^{21}$ cm$^{-2}$ corresponds to the southern side of the Bar, which is nominally not exposed to significant heating. 
If deep into the cloud the effects of higher reddening on the grain alignment efficiency are compensated by the increase of the minimum grain size of aligned grains caused by the loss of RATD efficiency with increasing density, the potential increase of $\StimesP$ at high \NHtc can be explained by an evolution of dust properties across the Bar caused by RATD.
Hence, we may probe a significant gradient in the alignment properties of dust grains, from the PDR edge where the radiative feedback via photo-fragmentation of aggregates and/or RATD is effective, to the inner quiescent part of the Bar, where larger aggregates survive.

\subsubsection{Polarization quantities versus \NHtsqT}

The evolution of both $\Pf$ and $\StimesP$ versus \NHtsqT (a proxy for the gas-dust collision rate) show clear decreases at the four wavelength of observations. The trends of $\S$ are either flat or increasing with \NHtsqT, which suggest that a part of the decrease of $\Pf$ versus \NHtsqT is due to the magnetic field structure. From these results we conclude that the grain dis-alignment by gas pressure is effective in the Bar.

To summarize, we suggest that the evolution of the grain alignment efficiency with the local conditions (using the available constrains on the dust temperature and gas density) cannot be explained by basic RAT theory alone, without considering dust evolution. The evolution of dust properties throughout the Bar can explain the slight decrease (increase) of alignment efficiency with dust temperature (reddening), which might be caused by RATD. However, we note that RATD, if occurring, does not totally hinder grain alignment in the Bar. 
The dis-alignment by gas collisions efficiently decrease the grain alignment degree. We now confront the hypothesis raised above with the predictions provided by the computation of the different grain alignment timescales as a function of the characteristic grain sizes describing the alignment and disruption of grains.

\section{Modeling of grain Alignment timescales and characteristic grain sizes}
\label{sec:modeling}

\begin{figure*}[!tbh]
\centering
\includegraphics[scale=0.6,clip,trim= 3cm 1.5cm 3cm 0cm]{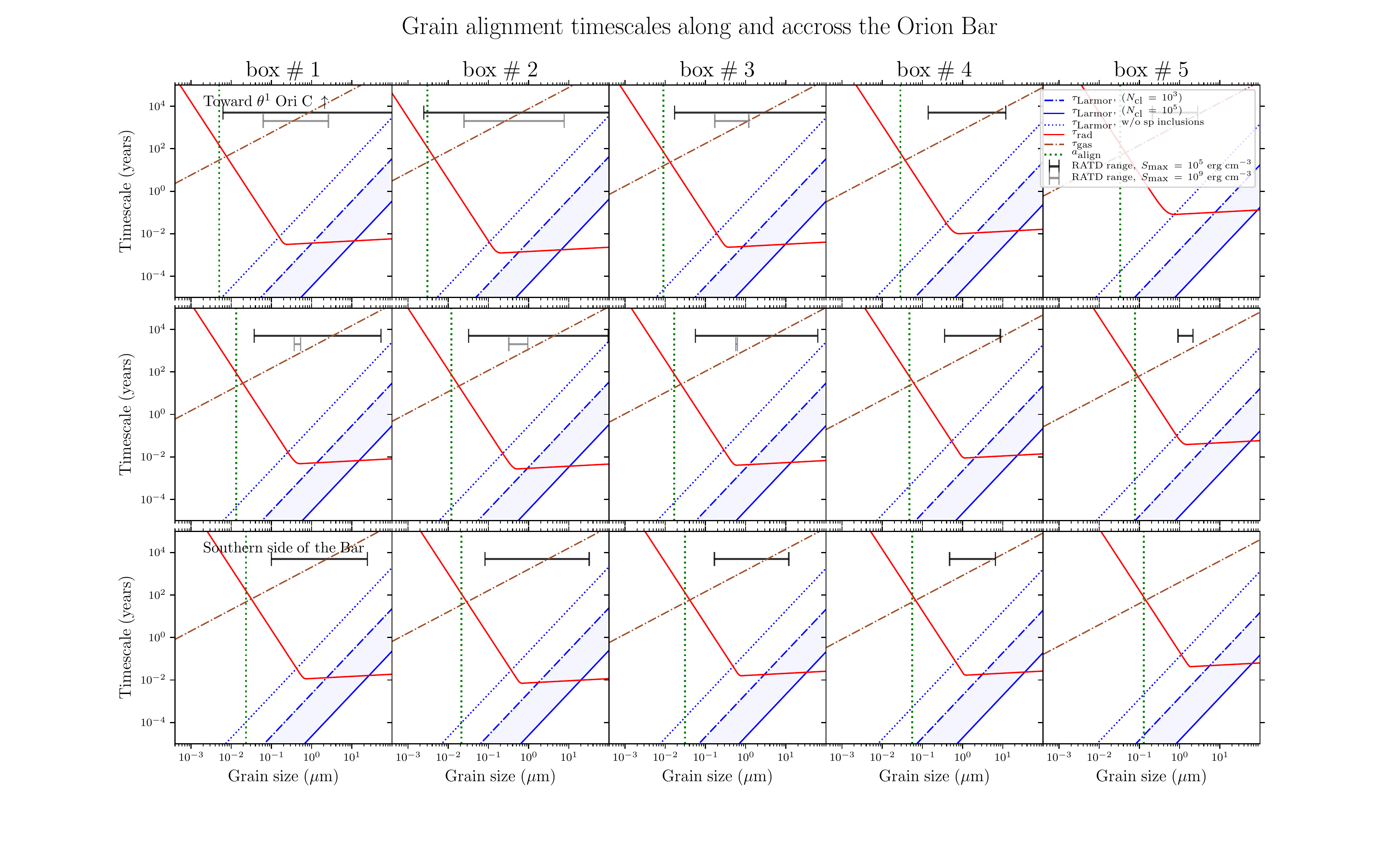}
\caption{\small Grain alignment timescales as a function of grain size along and across the Orion Bar. The results of each box have been obtained using the average environmental conditions of a specific spatial location in the Orion Bar. The Orion Bar (grey rectangle in Figure \ref{fig:obs_bar_4bands}) is divided in 5$\;\times\;$3 boxes, 5 boxes along its major axis such as box \#1 is the most northern one, and 3 boxes along its minor axis such as the top row corresponds to the irradiated edge of the Bar close to \ThOriC, and the bottom row corresponds to the southern, colder edge of the Bar. In each plot we show the Larmor precession timescale $\tau_{\textrm{Larmor}}$ (for a grain without super paramagnetic inclusions with the dotted line, and for two values of $N_{\textrm{cl}}$, being the number of atoms per cluster: 10$^3$ and 10$^5$, with the dot-dash and solid blue lines, respectively), the radiative precession timescale $\tau_{\textrm{rad}}$, and the collisional gaseous damping time of the grain $\tau_{\textrm{gas}}$ (dot-dash brown line). The minimum grain size of aligned grains $a_{\textrm{align}}$ is shown by the vertical dotted green line. The range of grain size affected by rotationally disruption $a_{\textrm{disr}}-a_{\textrm{disr},\textrm{max}}$), for $S_{\textrm{max}}\,=\,10^{5}$ and $10^{9}\,\textrm{erg}\,\textrm{cm}^{-3}$, are shown by the horizontal dark and grey lines, respectively. At a given grain size, the shortest precession timescale between $\tau_{\rm{Larmor}}$ and $\tau_{\textrm{rad}}$ dictates the direction of alignment, \ie with the magnetic field or the radiation field.}
\label{fig:timescales}
\end{figure*}

\begin{figure*}[!tbh]
\centering
\includegraphics[scale=0.6,clip,trim= 3cm 0cm 3cm 0cm]{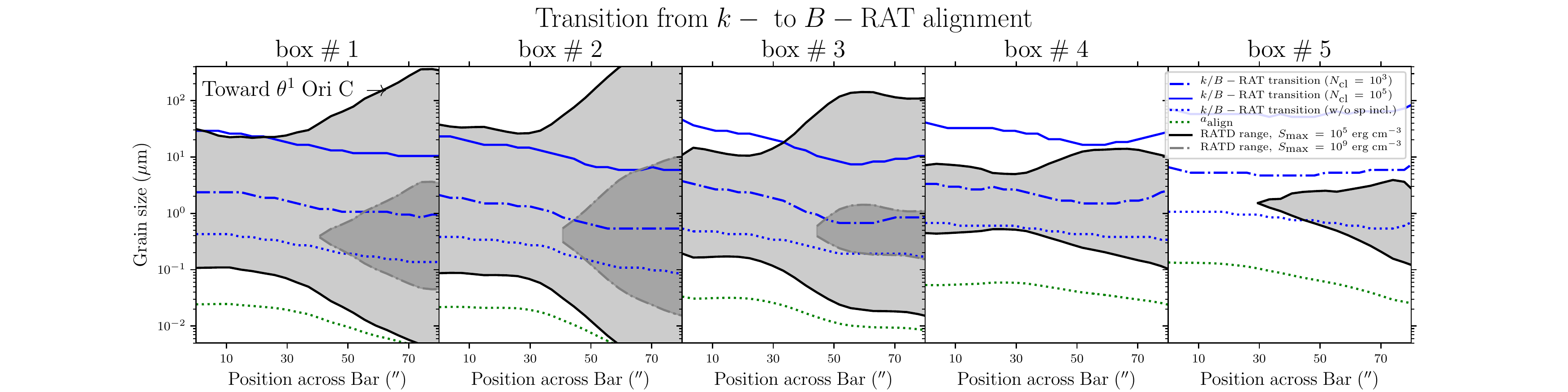}
\vspace{-0.1cm}
\caption{\small Grain size corresponding to the $k$- to \brat transition as function of the position across the Orion Bar. In this Figure, the Orion Bar is divided in 5 boxes along its major axis such as box \#1 is the most northern one. In each panel, the $x$-axis is the position along the minor axis of the Orion Bar (spanned by the three rows of Figure \ref{fig:timescales}), with increasing values corresponding to closer location from \ThOriC. For each position across the Orion Bar, the grain size at which the $k$- to \brat transition occurs is found by equating the radiative precession timescale with the Larmor precession timescale for a grain without super paramagnetic inclusions (blue dotted line), and for two values of $N_{\textrm{cl}}$, \ie 10$^3$ and 10$^5$ (blue dot-dash and solid blue lines, respectively). 
Grain sizes larger than this transition size $a_{\textrm{trans}}$ can be affected by \krat.
The dotted green line is the minimum grain size of aligned grains, $a_{\textrm{align}}$. 
The range of grain size affected by rotational disruption $a_{\textrm{disr}}-a_{\textrm{disr},\textrm{max}}$, for $S_{\textrm{max}}\,=\,10^{5}$ and $10^{9}\,\textrm{erg}\,\textrm{cm}^{-3}$, are shown by the shaded region encompassed by the dark and grey lines, respectively.
}
\label{fig:timescales_Krat_transit}
\end{figure*}

Studying the grain alignment timescales allows us to estimate for which grain sizes the \krat mechanism can be important, and in parallel, what grains are potentially affected by RATD. Ultimately, the goal is to determine whether the aligned grains that are susceptible to be aligned via the \krat alignment can survive the RATD phenomenon, which also affects in theory those aligned grains. We see below that the \krat and RATD phenomena affect the large end of the typical ISM dust grain size distribution, where the relative number of grains is low \citep{Mathis1977}. However, those grains are the origin of the FIR to (sub-)millimeter polarized dust emission.
The spatial resolution of this grain alignment timescale analysis corresponds to the resolution of the original dust temperature and column density map of \citet{Chuss2019}, \ie with 18.7$^{\prime\prime}$ angular resolution and 3.7 $^{\prime\prime}$ square pixel size.

\subsection{Grain size parameter space for \krat}
\label{sec:krat_study}

% \BG{[We need to note, somewhere, in this section that the alignment (and disruption) acts on an underlying - total - grain size distribution - i.e. the Mathis, Rumpl \& Nordsieck, 1977, n$\sim a^{-3.5}$]}
To estimate which grain alignment mechanism is dominating, we derive the different timescales that describe the efficiency of the different phenomena involved in the alignment of dust grains by radiative torques. RATs can be presented as a balance between gaseous de-alignment induced by collisions of gas particles onto dust grains, the precession speed of the grain's magnetic moment induced by grain rotation (for paramagnetic grains) around the magnetic field, and the efficiency of the radiative torques applied by the anisotropic radiation field impinging onto grains. We neglect the infrared emission damping and plasma drag effects \citep{Draine1998}.  The efficiency of these three processes can be compared to one another from comparing the relevant characteristic timescales, \ie the collisional gaseous damping timescale $\tau_{\rm{gas}}$, the Larmor precession timescale $\tau_{\rm{Larmor}}$, and the radiative precession time-scale $\tau_{\rm{rad}}$ \citep{LazarianHoang2007,Tazaki2017}. 
Ultimately, the minimum size of aligned grains $a_{\textrm{align}}$ denotes the grain size over which dust grains can be considered aligned \citep{Hoang2008}. 

The collisional gaseous damping time of the grain is estimated by:
\begin{equation}
\tau_{\textrm{gas}}\,=\,\frac{3}{4\sqrt{\pi}} \frac{I_1}{n_{\textrm{H}} \mu m_{\textrm{H}} v_{\textrm{th}} a^4}\,\,,
\end{equation}
where $v_{\textrm{th}}\,=\,\sqrt{2 k_{\textrm{B}} T_{\textrm{gas}} / \mu m_{\textrm{H}}}$ is the thermal velocity of gas atoms of mass $m_{\rm H}$, $n_{\rm H}$ is the gas density, $T_{\textrm{gas}}$ is the gas temperature (we will use \Td as a proxy for $T_{\textrm{gas}}$, which is not necessarily true in PDRs; see \citealt{Koumpia2015}),
$I_1$ is the principal grain moment of inertia (which scales as $\rho a^5$, where $\rho$ the dust grain density), $a$ is the grain effective size, and $\mu$ is the mean molecular weight per hydrogen molecule.

The Larmor precession timescale, describing the precession of the magnetized grain's angular momentum around an external magnetic field $B$ resulting from the interaction of the grain magnetic moment with $B$ is given by:
\begin{equation}
\tau_{\rm{Larmor}}\,\simeq\,1.3\,\hat{\rho}\hat{s}^{-2/3}a^{2}_{-5}\hat{B}^{-1}\hat{\chi}^{-1} \,\,\rm{years}\,\,,
\end{equation}
where $\hat{B}\,=\,B/5\,\mu$G is the magnetic field, $\hat{\rho}\,=\,\rho/3\,$g cm$^{-3}$, $\hat{s}\,=\,s/0.5$ the grains' aspect ratio, and $\hat{\chi}\,=\,\chi(0)/10^{-4}$ the grains' paramagnetic zero-frequency susceptibility. In our calculations we fix the magnetic field strength to $B\;=\;200\;\mu$G, given the results of \citet{Chuss2019,Guerra2021}. We adopt $\rho\,=\,3$\,g\,cm$^{-3}$ and s\,=\,0.5 \citep{Hildebrand1995}. The zero-frequency susceptibility of a dust grain depends on its paramagneticity. For a super-paramagnetic grain, we have from \citet{Morrish2001}:
\begin{equation}
\chi(0)\,=\,\chi{_\textrm{sup}}(0)\,=\,1.2\times10^{-2} N_{\textrm{cl}} \phi_{\textrm{sp}} \left( \frac{15\,\textrm{K}}{T_\textrm{d}} \right)\,\,,
\end{equation}
where $\phi_{\rm{sp}}$ is the fraction of atoms that are super-paramagnetic, and $N_{\rm{cl}}$ is the number of atoms per cluster. 
The GEMS measurements derived values  $\phi_{\rm{sp}}$ = 0.03 \citep{Bradley1994,Martin1995,Goodman1995a}, and $N_{\rm{cl}}$ is expected to be $N_{\rm{cl}}$ = 10$^3$-10$^5$ \citep{Kneller1963,Jones1967}. Lastly, we consider an ordinary paramagnetic grain, with:
\begin{equation}
\chi(0)\,=\,4.2\,\times\,10^{-2} f_p \left( \frac{15\,\textrm{K}}{T_\textrm{d}} \right)\,\,,
\end{equation}
where $f_p$ is the fraction of atoms in the grain that are paramagnetic, evaluated at 10\% \citep{Tazaki2017}. Results obtained from modeling and observations of dust polarization of the diffuse ISM and dense cores tend to favor the scenario where the efficiency of RATs can only be reproduced if grains' magnetic relaxation is sufficiently fast, \ie for super paramagnetic grains \citep{Hoang2016,Reissl2020,LeGouellec2020}. We also note that H$_2$ formation occurring at the grains' surface provide additional torque that also increase the grain rotational velocity, eventually bringing the grain to suprathermal rotation \citep{Purcell1979,Hoang2015}, especially toward PDRs \citep{LeBourlot2012,Andersson2013,Soam2021a}. These both effects of grains' super-paramagneticity and H$_2$ formation supplemental torque increase the fraction of grains with high angular momentum, \ie the fraction of grains at the the so-called high-$J$ attractor point \citep{LazarianHoang2007,Hoang2009a,Hoang2016}. While grains at high-$J$ can be considered perfectly aligned, grains with a low angular momentum (\ie at the the so-called low-$J$ attractor point) would be poorly aligned and produced, exhibiting a Rayleigh reduction factor of $\sim$ 0.1 \citep{Hoang2016}. The relative fraction of grains at high-$J$, \ie $f_\textrm{high-J}$, would increase with increasing grains' magnetic susceptibility and magnetic field strength, and decreasing grain size, gas density and temperature \citep{ChauGiang2022}.
%For reference, \citet{Draine1996} proposed $\chi(0)$ values in the range 4$\times10^{-5}-10^{-3}$ for ordinary paramagnetic grains.

Finally, the radiative precession time-scale $\tau_{\rm{rad}}$ is given by:
\begin{equation}
\begin{split}
\tau_{\rm{rad}}\,\simeq\, 110 \hat{\rho}^{-1/2}\hat{s}^{-1/3}a^{1/2}_{-5}\hat{T}_{\textrm{d}}^{1/2} \left( \frac{u_{\rm{rad}}}{u_{\rm{ISRF}}}\right)^{-1} \left( \frac{\bar{\lambda}}{1.2\,\mu\rm{m}}\right)^{-1} \\
\times \left( \frac{\bar{\gamma} |\bar{\mathcal{Q}_{\Gamma}}(\bar{\lambda})|}{0.01}\right)^{-1}\,\rm{years}\,\,,
 \end{split}
 \label{equ:tau_rad}
\end{equation}
where $a_{-5}\,=\,a/10^{-5}$ cm, $\hat{T}_{\textrm{d}}\,=\,T_{\textrm{d}}/100$ K,
$u_{\rm{rad}}$ is the radiation field intensity,
$\bar{\gamma}$ is the anisotropy of the mean radiation field (equation from \citealt{Tazaki2017}),
$\bar{\mathcal{Q}}_{\Gamma}(\bar{\lambda})$ is the RAT efficiency (see the relation 10 in \citealt{Hoang2020b}),
and $\bar{\lambda}\,=\,\frac{\int^{\infty}_{0}  \lambda u_{\rm{rad}} d\lambda }{u_{\rm{rad}}}$ is the mean wavelength of the radiation field.
% \footnote{The values with a bar above result from the integration over the radiation field spectrum of the corresponding wavelength-dependent value, weighted by the radiation field $u_\lambda$}.
We estimate the radiation field using the relation from \citet{Draine2011}:
\begin{equation}
u_{\rm{rad}}\,\approx\, \left( \frac{T_{\rm{d}}}{16.4} \left(\frac{a}{0.1\,\mu{\textrm{m}}} \right)^{1/15} \right)^{6}\,\,.
\end{equation}
To estimate $\bar{\lambda}$, \ie the reddening of the radiation field, we use the relation from \citet{Hoang2020b}, who, using an analytical model, derives the mean wavelength $\bar{\lambda}$ as a function of the blackbody temperature of a star and the gas column density integrated between the star and the region where $\bar{\lambda}$ is estimated. We use T$_\textrm{eff}\,=\,$39 000 K (corresponding to \ThOriC; \citealt{SimonDiaz2006}), and  the \NHtc value derived above with the SED fitting performed in \citet{Chuss2019} to estimate the reddening of the radiation field $\bar{\lambda}$ in every pixel of the Bar.

Finally, we note that line-of-sight integration of the background OMC1 cloud must affect the results from the SED fitting that we use, especially for the determination of the dust temperature. In addition, the coarse resolution of this SED modeling, \ie 18.7$^{\prime\prime}$, precludes us from resolving the hotter layer of dust.  This is more efficiently traced by the mid-infrared photometry study performed with FORCAST in \citealt{Salgado2016}, where a cold and a warm modified blackbody component were used. Therefore, the grain alignment timescale analysis does not probe physical scales corresponding to this hot dust layer where \Td$\,\geq\,$100\,K.
% the physical scales corresponding to the grain alignment timescale analysis do not exactly probe the edge of the PDR. 
However, the Orion Bar is resolved in its minor axis such that we can establish a clear gradient in dust temperature with a range of 35$-$75 K and in projected column density with a range of 10$^{21}-2\,\times\,10^{22}$\;cm$^{-2}$.

Equation \ref{equ:tau_rad} is only valid for grains aligned at low-$J$. High-$J$ aligned grains are most likely be aligned with the magnetic field, because the spin-up effect of RATs dominate over the radiative precession \citep{Hoang2022b}. We will thus compare the radiative precession timescale with the Larmor precession timescale of ordinary paramagnetic grains.

Figure \ref{fig:timescales} presents the grain alignment conditions of the Orion Bar, showing the different timescales discussed above. The Bar is divided into 3$\times$5 sub regions, \ie 5 along its major axis, and 3 along its minor axis (\ie each of the five boxes displayed in Figure \ref{fig:obs_bar_hist5boxes} are simply separated in three sub-boxes along the minor axis of the Bar). Dust grains larger than the typical size at which $\tau_{\rm{Larmor}}\;>\;\tau_{\rm{rad}}$ can be considered as aligned at low-$J$ with the radiation field \citep{LazarianHoang2007}. This transitional grain size is referred to as $a_{\textrm{trans}}$ hereafter. Toward the region probed by the box \#2, corresponding to the highest irradiation conditions that we probe, $a_{\textrm{trans}}$ ranges from $\sim$ 0.1 to 0.5\;$\mu$m. In the southern section of the region in box \#5, corresponding to the coldest and densest conditions that we probe, $a_{\textrm{trans}}$ would be $\sim$\,1\,$\mu$m.

The evolution of the transitional $a_{\textrm{trans}}$ values as a function of the proximity to the irradiated front is shown within the 5 boxes distributed along the major axis of the Bar in Figure \ref{fig:timescales_Krat_transit}. The southern boxes \#4 and 5 exhibit on average higher values of $a_{\textrm{trans}}$ and $a_{\textrm{align}}$ compared to boxes \#1$-$3. These latter ones must correspond to most of the radiation absorbing area at the surface of the PDR.

% Presentations of Figures \ref{fig:timescales} and \ref{fig:timescales_Krat_transit}.

\subsection{Constraints from grain alignment disruption}
\label{sec:RATD}

The constraints on the grain alignment timescales obtained above determine which grains are potentially aligned with the radiation field, depending on its strength. However, aligned dust grains subject to high irradiation can also trigger the so-called RAdiative Torque Disruption (RATD; \citealt{Hoang2019NatAs}) mechanism, which causes in the disruption of grains into fragments, occurring when the grain rotational energy induced by radiative torques exceed the grain cohesion forces. Comparing the angular velocity at which dust grains are rotationally disrupted with the rotation speed induced by RATs, allows us to constrain which grain sizes are affected by the RATD mechanism. 
Grains subject to RATD would thus be those with high angular momentum, \ie at high-$J$.
Given the two regimes of radiative torque efficiency, \ie when $a\;>\;\bar{\lambda}$ or $a\;<\;\bar{\lambda}$, it is possible to define the grain size interval $[a_{\rm{disr}}\,;\,a_{\rm{disr,max}}]$, inside which the aligned dust grains are rotationally disrupted. From \citet{Hoang2020b}, we have :
\begin{equation}
\label{equ:a_disr}
\begin{split}
a_{\rm{disr,min}}\,=\,\bar{\lambda}\left( \frac{ 0.8 n_{\rm{H}} \sqrt{2\pi m_{\rm{H}} k T_{\rm{gas}}} }{\gamma u_{\rm{rad}}} \right)^{1/2} \left(\frac{S_{\rm{max}}}{\rho}\right)^{1/4}
\\
\times\,(1+F_{\rm{IR}})^{1/2}\,\,, 
\end{split}
\end{equation}
\begin{equation}
\begin{split}
\label{equ:a_disr_max}
a_{\rm{disr,max}}\,=\, \frac{ \gamma u_{\rm{rad}} \bar{\lambda} }{ 16 n_{\rm{H}} \sqrt{2\pi m_{\rm{H}} k T_{\rm{gas}}} } \left(\frac{S_{\rm{max}}}{\rho}\right)^{-1/2} \\
\times (1+F_{\rm{IR}})^{-1}\,\,,
\end{split}
\end{equation}
where $S_{\rm{max}}$ is the tensile strength of dust grains, $F_{\rm{IR}}$ is the rotational damping coefficient due to the emission of infrared photons emitted by the grain which reduces the grain’s angular momentum (see \citealt{Draine1998}; we adopt the relation of \citealt{Hoang2020b}). The tensile strength of interstellar dust is uncertain, because it depends on both the grain structure, \ie compact versus composite, and the grain composition. Dust grain evolution in the ISM is ruled by fragmentation and coagulation
processes, as well as the formation of ice mantles. Dust grains with a
core-mantle structure \citep{Desert1990,Jones1990}, and dust composite models composed of silicate and carbon grain aggregates \citep{Mathis1989,Zubko2004,Kohler2015,Draine2021}, have been proposed. We explore the range $10^{5} - 10^{9}\,\rm{erg}\,\rm{cm}^{-3}$ for $S_{\rm{max}}$ in order to consider both the case of large aggregates, and the case of compact silicate core grains \citep{Hoang2019c}. Figures \ref{fig:timescales} and \ref{fig:timescales_Krat_transit} show $[a_{\rm{disr}}$ and $a_{\rm{disr,max}}]$ for the two extreme values of $S_{\rm{max}}$ that we consider here.

We find that for the most irradiated region of the Bar (box \#1, 2, and 3 columns in Figure \ref{fig:timescales}), the RATD mechanism can disrupt aggregates of tensile strength $S_{\rm{max}}\;=\;10^{5}\,\rm{erg}\,\rm{cm}^{-3}$ and size $a\;\gtrsim$ 0.01 $\mu$m toward the northern side exposed to the irradiation (top-row panels), and grains $\gtrsim$ 0.1 $\mu$m toward the southern, colder regions (bottom-row panels). Compact grains of tensile strength $S_{\rm{max}}\;=\;10^{9}\,\rm{erg}\,\rm{cm}^{-3}$ appear to be affected by RATD only toward the northern irradiated side of the Bar, where grains $\gtrsim$ 0.1 $\mu$m can be rotationally disrupted. 
Setting aside the considerations about $f_\textrm{high-J}$, that does not depend only on the grain's characteristics, Figure \ref{fig:timescales_Krat_transit} clearly shows that in box \#1, 2, and 3, grains potentially aligned with the radiation field, \ie larger than $a_{\textrm{trans}}$, are generally disrupted by RATD if they are aggregates. 
Compact grains of high tensile strength are also affected by RATD close to the irradiation front.
Large and compact grains can in theory be aligned via \krat in this region if $a\;\geq\;a_{\textrm{trans}}$ and $a\;\geq\;a_{\rm{disr,max}}$, \ie $a\;\gtrsim\;1-10\;\mu$m. 
However, such large grains, which are expected to be formed by collisions, are likely to be aggregates-type, and likely to be rotationally destroyed if they are efficiently aligned.
In addition, it is unlikely that such large grains dominate the polarized dust emission in the FIR.

\section{Discussion}
\label{sec:disc}

% The results presented in Section \ref{sec:results} show that (1) the HAWC+ polarimetric observations do not exhibit any evidence for the \krat mechanism to be the cause of the polarized dust emission, and (2) RATD can potentially affect the population of the largest aligned grains. Below we discuss further the physical conditions that could trigger \krat, the consequences of high RATD efficiency in removing the large aligned grains, and perspectives relative to grain alignment theories.

\subsection{Where could the \krat alignment operate?}
\label{sec:loc_kRAT_RATD}

We note that with its 18.7 $^{\prime\prime}$ (\ie 0.035 pc at 390 pc) angular resolution, our grain alignment timescale study does not resolve the hot dust layer close to the dissociation front, where the temperature can reach 400-700 K \citep{Goicoechea2011,Goicoechea2017,Parikka2017}. In addition, observations and models have suggested that the Orion Bar PDR actually consists of high density clumps $n_{\textrm{H}}\,=\,10^{6}-10^{7} \;\textrm{cm}^{-3}$ embedded in a lower density medium of $n_{\textrm{H}}\,\approx\,5\,\times\,10^{4} \;\textrm{cm}^{-3}$ mainly responsible for the extended PDR emission (\citealt{Lis2003,AndreeLabsch2017,Habart2022} and references therein). These substructures could have been induced by UV radiation driven
compression \citep{Gorti2002,Tremblin2012}, advecting the molecular gas through the atomic gas \citep{Goicoechea2016}. Based on the results of Sections \ref{sec:krat_study} and \ref{sec:RATD}, it is possible to predict if those specific environmental conditions, toward these denser and warmer substructures, could trigger \krat. Fixing all other parameters, we have : $a_{\textrm{disr,min}}\,\propto\,u_{\textrm{rad}}^{-1/2} n_{\textrm{H}}^{1/2} \bar{\lambda}$ and $a_{\textrm{trans}}\,\propto\,$ $u_{\textrm{rad}}^{-0.6} \bar{\lambda}^{-0.65}$. From \citet{Hoang2020b}, $\bar{\lambda}\,\propto\,A_{V,\star}^{\alpha}$, with $\alpha\,\approx\,0.6$, where $A_{V,\star}$ is the visual extinction measured from the illuminating star to a given location in the cloud.
Therefore, an increase in density of two orders of magnitude, with a stronger or equal apparent radiation field, could sufficiently decrease the ratio $a_{\textrm{trans}}/a_{\textrm{disr,min}}$ 
such that grains with radii of $0.1-1\,\mu$m can potentially subject to \krats, depending on their compactness.
In parallel, the fraction $f_{\textrm{high-J}}$ would also decrease with increasing dust temperature and density, enabling a larger fraction of $k$-RAT aligned grains to contribute to the polarization.

% Given the spatial resolution of our multi-wavelength dust polarization analysis presented here, we are thus unable to probe such $k$-RAT-driven polarization. 
The results from the HAWC+ polarimetric observations presented in the histograms of Figure \ref{fig:obs_bar_hist5boxes} does not show that grains are aligned with the radiation field emanating from the Trapezium cluster. The low-$J$ aligned grains cannot dominate the FIR dust polarized emission because the polarization fraction would be much lower (\ie $\leq\,1\,\%$) than what is observed. In addition, if large low-$J$ aligned grains were to dominate the polarization, they would generate a rotation of the polarization angle with increasing wavelength, where those large grains contribute more to the total emission, which is not observed. 
This latter effect would also induce a clear decrease of polarization fraction with wavelength, which is not the case in the Bar (see Section \ref{sec:required_RAT} and Figure \ref{fig:pfrac_spectra}). 
Finally, given the transition sizes of $\sim\,0.1\,-\,0.5\,\mu$m probed in Section \ref{sec:krat_study} and the maximum size of the dust size distribution ($\leq\,0.5\,\mu$m; see \citealt{Schirmer2022}, who proposed that fragmentation of large grains due to collisions caused by radiative pressure is an efficient mechanism in the Bar), the low-$J$ $k$-RAT aligned grains are not expected to dominate the FIR emission.
Therefore, while there may be a population of $k$-RAT aligned grains, there is no evidence that they are the dominant cause of the FIR dust polarization observations. FIR polarization thus probe grains efficiently aligned at high-$J$ with the magnetic field. 
% Fractions of grains efficiently aligned at high-J with the magnetic field is thus not negligible.
We also note that mechanically aligned grains can produce the same polarization pattern that $k$-RAT aligned grains. Because such pattern are not favored in our analysis, we do not consider this mechanism further in our discussion. However, Appendix \ref{app:METs} present an analytic exploration of this grain alignment mechanism, where we find that it should not dominate the origin of the polarized dust emission.

\subsection{On the depletion of large silicates}
\label{sec:deple_silic}

\begin{figure}[!tbh]
\centering
\includegraphics[scale=0.42,clip,trim= 0.4cm 1cm 2cm 2cm]{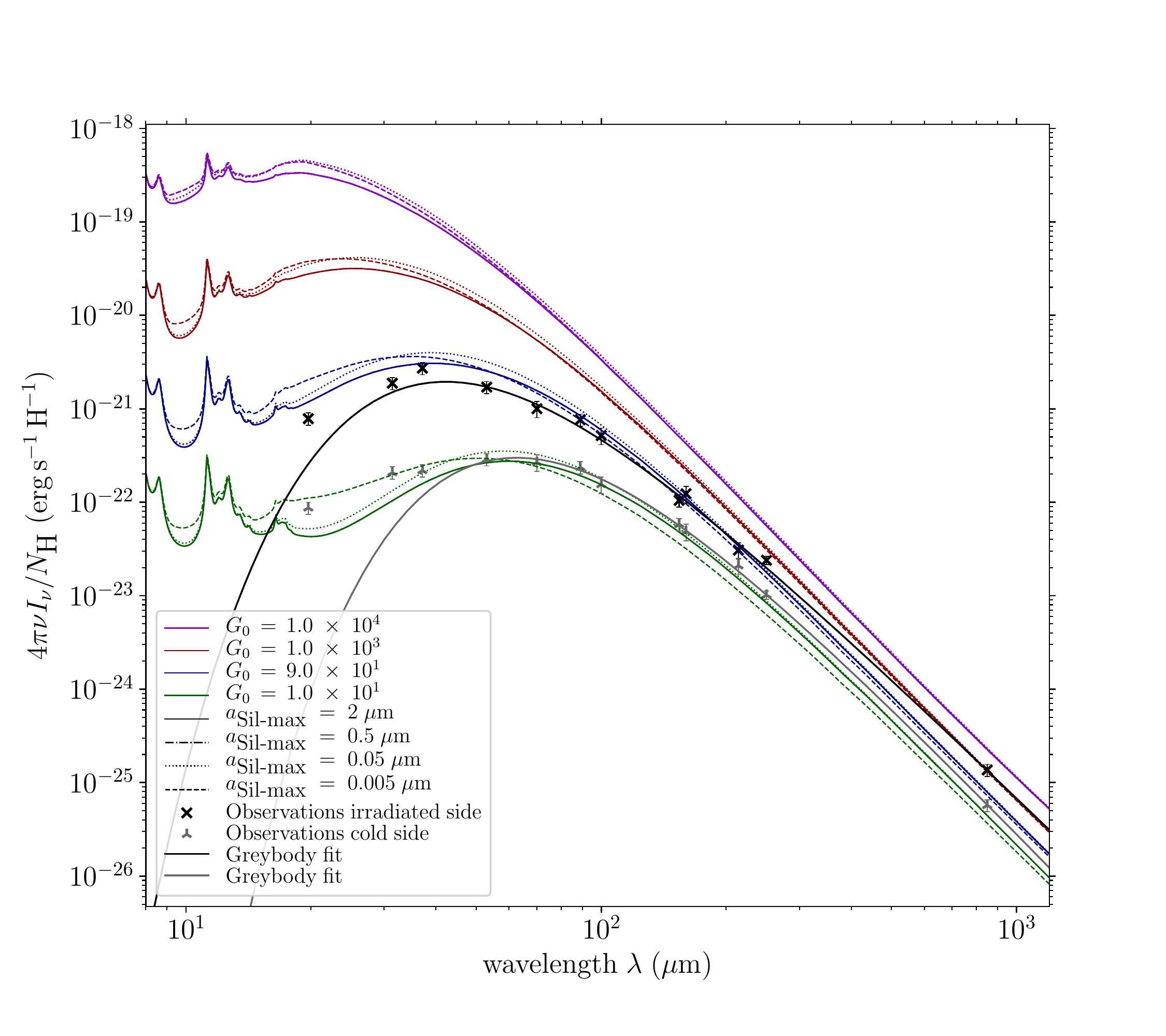}
\caption{\small DustEM models using \citet{Compiegne2011} ISM dust models, for different levels of irradiation field strength $G_0$, and different maximum grain size of the silicate power-law distribution $a_{\textrm{Sil-max}}$ (2, 0.5, 0.05, and 0.005 $\mu$m). Two sets of flux measurements, taken on the irradiated side of the PDR (\Td$\,\sim\,$65 K, black points) and on the cold embedded (\Td$\,\sim\,$40 K, grey points) side of the PDR, are represented. The corresponding grey-body fits, of \citet{Chuss2019} are shown with the black and grey solid lines.
% \ie fitting the FIR side of the SED using the temperature, dust emissivity and optical depth
}
\label{fig:SED}
\end{figure}

% mention also that high alignment efficiency is necessary to ensure high RATD efficiency.
We find that a population of high-$J$ aligned grains should contribute significantly to the dust polarization because the $k$-RAT non-detection implies that the population of low-$J$ aligned grains is negligible. 
However, the analysis of Figure \ref{fig:multi_S_Pf} showed that the level of grain alignment efficiency toward the irradiated side (\ie $\Pf$$\,\sim\,2-10$\,\%) of the Bar suggests that a significant fraction of grains is still aligned.
% However, the analysis of Figure \ref{fig:multi_S_Pf} suggested that the efficiency of the RATD mechanism is limited because no strong anti-correlation between polarization degree and dust temperature is detected.
We now discuss the efficiency of RATD for those efficiently aligned grains. RATD is a mechanism that is relatively hard to precisely constrain observationally from emission data, as the abundance and size of the largest aligned grains is difficult to measure. Indeed, as shown in Section \ref{sec:RATD}, RATD can be efficient at fragmenting all the large grains of $0.1-10\,\mu$m in size if they are aggregates of low tensile strength. However, we also note that RATD cannot be so efficient that it would deplete all grains $\geq\,0.1\,\mu$m in size, because we expect the grain alignment efficiency to be limited for small grains \citep{LazarianHoang2007,Andersson2015}, \ie $\leq\,0.01\,\mu$m in size. Besides, the polarized dust we observe with HAWC+ does not indicate a scenario where only small poorly aligned grains are present. Therefore, one needs to determine whether the efficiency of this mechanism is limited, and/or to characterize the structure of grains in such environment. 

\begin{figure}[!tbh]
\centering
\includegraphics[scale=0.65,clip,trim= 0.4cm 0cm 1.2cm 0.8cm]{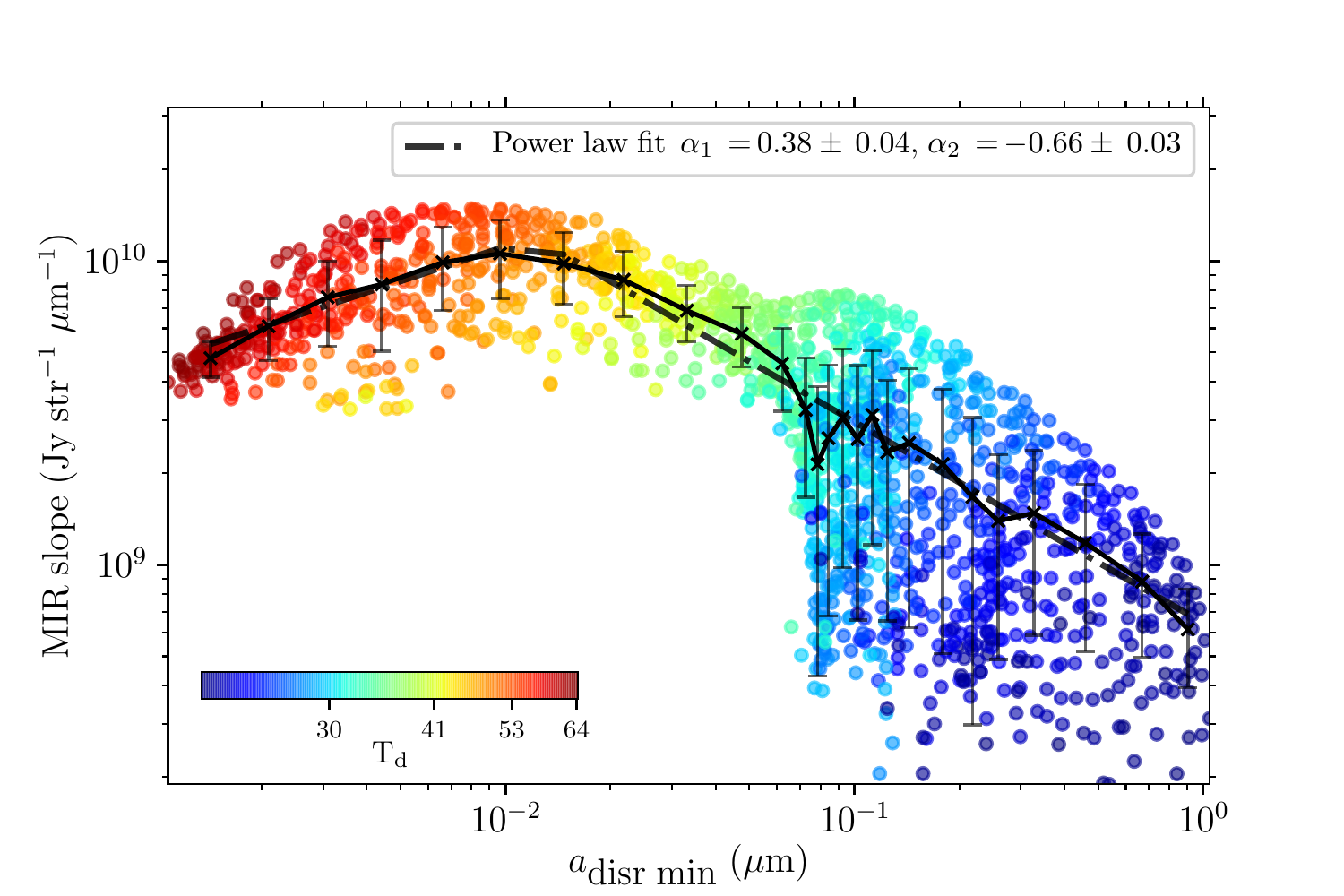}
\caption{\small 2D-distribution of MIR-to-FIR slope (using the FORCAST data at 19.4, 31.7, and 37.1 $\mu$m from \citealt{Salgado2016}) versus $a_{\rm{disr,min}}$ (derived in Section \ref{sec:loc_kRAT_RATD}). The color scale indicates the dust temperature. The black line is the running mean, and the associated error bars represents the standard deviation of each bin of data. The dot-dashed line is a broken power-law fit, whose indexes are shown in the legend.}
\label{fig:slopeMIR}
\end{figure}

To investigate the hypothesis of high RATD efficiency precluding the survival of $\gtrsim\;0.1\;\mu$m size grains, we  investigate the impact of such a depletion of large silicate grains on the SED of the Orion Bar PDR. To do this, we use the photometry files used in \citet{Chuss2019} for their SED fitting,  as well as the SOFIA FORCAST MIR observations (19.4, 31.7, and 37.1 $\mu$m) from \citet{Salgado2016}, that we smooth and regrid to 18.7 $^{\prime\prime}$ resolution and 3.7 $^{\prime\prime}$ pixel size with a flux conserving algorithm. In parallel, we use DustEM \citep{Compiegne2011} to simulate the effects of both the irradiation level $G_0$ and the diminution of the maximum size of silicates on the SED of a standard ISM dust grain population, using compositions of \citet{Compiegne2011}. Results are shown in Figure \ref{fig:SED}. For each level of $G_0$ that we implement, we vary the maximum size of silicates $a_{\textrm{Sil-max}}$ among the following values of 2, 0.5, 0.05, and 0.005 $\mu$m. The impact of removing large silicates on the SED is to increase the emission in the MIR, thus reducing the SED slope between the MIR and FIR (the MIR-to-FIR slope we are referring to here is the slope of the flux versus wavelength, calculated with the FORCAST data between 19.4 and 37.1 $\mu$m)\footnote{For reference, we also did this analysis using the THEMIS dust model \citep{Jones2013,Jones2017}, and also noticed a evolution of the MIR-to-FIR slope when decreasing the maximum size of the silicates grains. However, this evolution is different that of using the \citet{Compiegne2011} model. Precise dust modeling would be required, but this goes beyond the scope of this paper.}. Indeed, as the mass of dust is kept constant, cutting out the distribution toward the largest grains redistribute the mass toward small grains, whose emission is thus increased. 
% However, one would need to investigate whether those small silicates are still at thermal equilibrium, or if stochastic heating is dominant. While the emission of large grains can be considered as successive black bodies, a temperature distribution is required to model the emission of small grains heated stochastically. 
We also note that the spectral signature of the $\sim$ 15 $\mu$m band can also help to identify a scenario of a higher abundance of small silicates.

Also shown in Figure \ref{fig:SED} are two sets of flux measurements, taken from the irradiated side of the PDR (\Td$\,\sim\,$65 K; star symbols) and on the cold embedded (\Td$\,\sim\,$40 K; plus symbols) side of the PDR \footnote{We do not expect to fit the SED with the DustEM models.  We only illustrate the impacts of the depletion of large silicates on the SED.}. These data points correspond to relatively low $G_0$, $\sim\,10-100$ in Habing units, compared to the known FUV radiation field incident on the Orion Bar PDR, \ie $G_0$\,$\sim\,1-4\,\times\,10^{4}$ \citep{Marconi1998}, because of the low angular resolution of our photometric maps. The two SEDs shown in Figure \ref{fig:SED} exhibit a shallower MIR-to-FIR slope than the model with no large silicate depletion, \ie with $a_{\textrm{Sil-max}}\,=\,2\,\mu$m. 

In order to investigate whether the evolution in the MIR-to-FIR slope in the observational data can be caused by a depletion of silicates, we show in Figure \ref{fig:slopeMIR} the slope versus the minimum size of rotationally disrupted grains $a_{\textrm{disr,min}}$ derived in Section \ref{sec:RATD} for $S_{\rm{max}}\;=\;10^{5}\,\rm{erg}\,\rm{cm}^{-3}$. We find a clear bimodal change in the MIR-FIR slope as a function of $a_{\textrm{disr,min}}$, where it initially increases with $a_{\textrm{disr,min}}$, up to $a_{\textrm{disr,min}}\approx\,$1$\times$10$^{-2} \mu$m, after which it decreases.  The different slopes are mirrored in the dust temperature, such that positive slopes correspond to larger dust temperatures (\Td>\,50 K) with negative slope corresponding to cooler dust temperatures. 
This decrease of the MIR-to-FIR slope slope, for $a_{\textrm{disr,min}}<\,$1$\times$10$^{-2} \mu$m, for decreasing (increasing) $a_{\textrm{disr,min}}$ (\Td) is consistent with RATD affecting the size distribution of the silicate grains by increasing the abundance of small silicates, assuming that the MIR-to-FIR slope is sensitive to such effects.
For larger $a_{\textrm{disr,min}}$, the decrease of the MIR slope for increasing (decreasing) $a_{\textrm{disr,min}}$ (\Td) is likely due to a dust temperature effect.

The interpretation that RATD is responsible for the decrease of the MIR slope for decreasing $a_{\textrm{disr,min}}$ at \Td>\,40 K, is, however, subject to several caveats. For example, in a PDR the NIR to FIR flux ratio is smaller than in the diffuse ISM because of the intense dust evolution processes at the PDR irradiation front \citep{Goicoechea2007}. While nano-grains are efficiently photo-destroyed, large grains fragment due to collisions caused by radiative pressure, in addition to the RATD effect, which in turn can re-form nano-grains via sticking collisions at higher extinction \citep{Schirmer2020,Schirmer2022}. This also can have an impact on the MIR slope of the SED\footnote{Using the PDR-constrained THEMIS parameters derived in \citet{Schirmer2022}, we find that, however, this only affects marginally the MIR-to-FIR slope.}. In addition, the characteristic timescale of the photo-fragmentation of large grains, \ie the collision timescale $\tau_{\rm{coll}}$, is of the same order of magnitude of the rotational disruption timescale $\tau_{\rm{disr}}$. Indeed, \citet{Schirmer2022} derived $\tau_{\rm{coll}}\,\approx\, 5\,\times\,10^{1}-10^{2}$ years for the Orion Bar, while we find $\tau_{\rm{disr}}\,\approx\,10^{0}-10^{2}$ years (using \citealt{Hoang2019NatAs,Lazarian2021}), for grains of 0.1$-$1\,$\mu$m. Therefore, RATD is not the only process responsible for the fragmentation of the large grains in the Orion Bar. These two timescales are grain size dependent, and the effects of RATD is more important for large grains \citep{Hoang2019c}, while the photo-fragmentation of grains is more efficient for small grains. Therefore, the idea of using the MIR-to-FIR slope to probe the depletion of the largest aligned grains caused by RATD remains to be explored.

\begin{figure*}[!tbh]
\centering
\includegraphics[scale=0.5,clip,trim= 1.5cm 1.3cm 2.4cm 0.1cm]{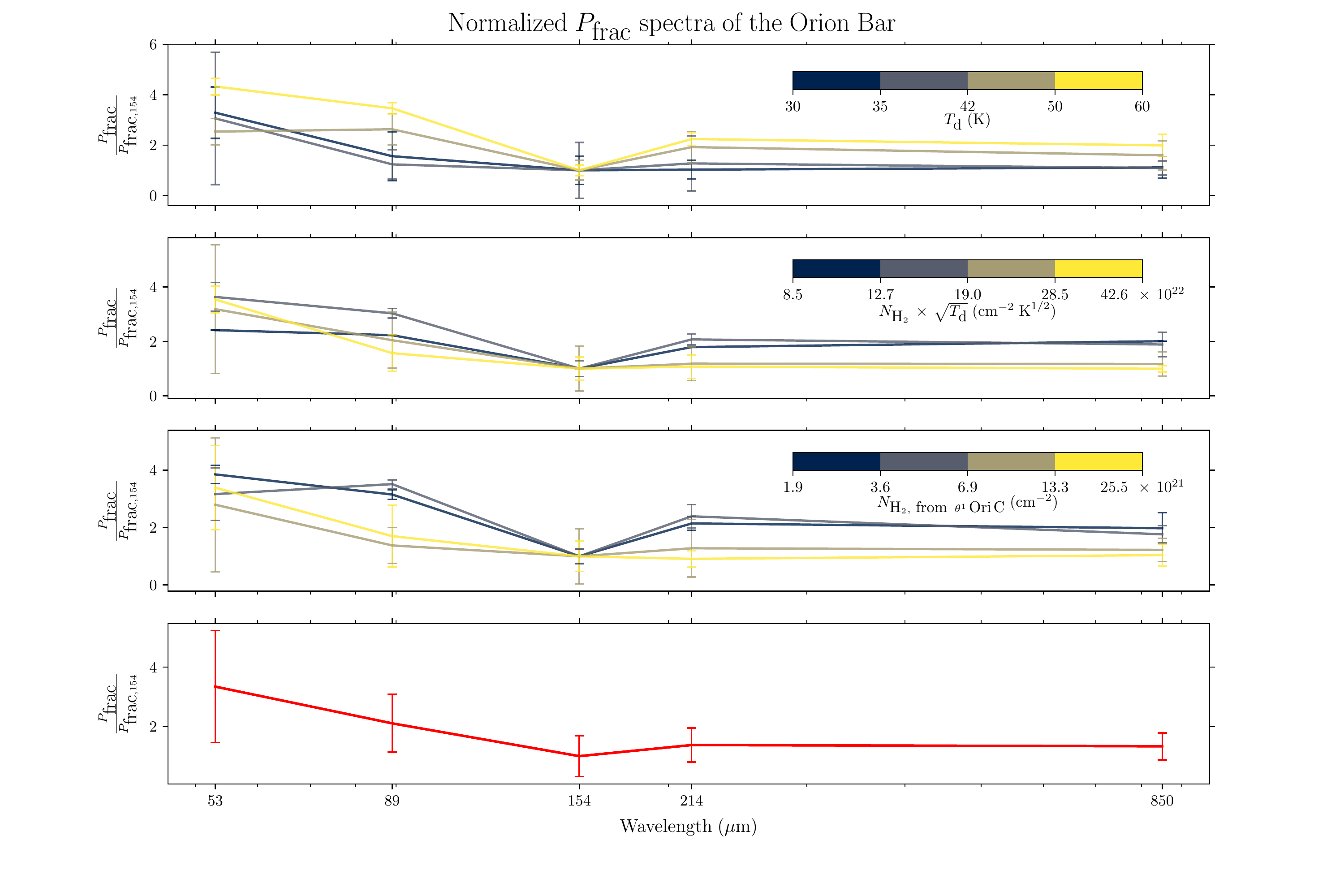}
\caption{\small Polarization fraction spectra calculated across the five wavelength of our observations (the SOFIA OTFMAP mode observations we present in this paper and the JCMT observations presented in \citealt{Pattle2017}). We normalize each spectra by the polarization fraction value at 154 $\mu$m. In each of the top three panels, we plot the normalized polarization fraction spectrum for four ranges of \Td, \NHtsqT, and \NHtc, indicated by the colorbars. The last panel shows the normalized polarization fraction spectrum averaging all data points. The error bars associated with the data points represent the standard deviation of each bin of data.}
\label{fig:pfrac_spectra}
\end{figure*}

Finally, the effect of RATD is also hard to calibrate because of the uncertain degree of mixing of carbon and silicate grains in aggregates. Indeed, as carbon grains are diamagnetic and unlikely to align \citep{Andersson2022}, they are not subject to RATD.  Because we have assumed a maximum efficiency of RATs when calculating the RATD parameters, this effect is not included here. 
The effects of “wrong” alignment of grains with respect to the magnetic field for large grains in the absence of internal alignment is also not included in this study \citep{Hoang2009}.
An even more important factor affecting the efficiency of RATD is the size of iron clusters embedded in dust, which significantly changes the magnetic susceptibility of dust grains, and in turn their inclination to experience fast alignment and potentially disruption \citep{ChauGiang2022}.

\subsection{Next Steps - What is needed to further test and verify grain alignment theories ?}
\label{sec:required_RAT}

The Bar is not optically thin along the direction of the radiation field, and the UV photons are quickly attenuated. This affects the radiation field spectrum responsible for the heating of the dust that produce most of the IR emission across the Bar. Therefore, a radiative transfer modeling of the Orion Bar can be useful to produce a spatially resolved SED model across the PDR. Synthetic observations of such polarization radiative transfer modeling could thus evaluate the potential contribution of \krat aligned grains to the polarized dust emission retrieved by HAWC+, and the RATD-survival of large silicates grains in supra-thermal rotation. While the radiative transfer code POLARIS \citep{Reissl2016} has been recently updated to include the calculation of the fractions of high-$J$ aligned grains as a function of grain size and environmental conditions \citep{ChauGiang2022}, the \krat and RATD mechanisms are not implemented yet. 
Such modeling would need to be run on a radiation-magneto-hydrodynamic simulation code that includes a proper gas cooling scheme, in order to reliably reproduce the magnetic field morphology, and the sub density structures, lying below the spatial resolution of our SOFIA observations. The magnetic field morphology that affects the polarized dust emission via depolarization effects may be hard to reproduce, and can make the comparisons between such models and dust polarization observations challenging. In theory, if deeper and higher angular resolution MIR to FIR dust polarization observations are made to resolve the typical dust evolution spatial scale, one can constrain the dust evolution scenario via a joint modeling of the grain alignment mechanisms and SED of the (polarized) dust emission.

Additionally, the polarization fraction spectrum can be a powerful tool to constrain grain alignment theories or dust characteristics. Indeed, it is known to vary with the local physical conditions (\ie dust temperature, grain alignment efficiency, dust grain characteristics such as their composition), that affects the polarized dust emission differently across wavelength \citep{Hildebrand1999,Vaillancourt2008,Vaillancourt2012,Fanciullo2022}. With optical thickness decreasing with wavelength, the polarization fraction is expected to increase with wavelength toward the dense interior of cores \citep{Hildebrand1999}. However, for less dense regions, the polarization fraction is expected to decrease with wavelength, as warm dust traced at short wavelength is more easily aligned due to the higher efficiency of radiative torques. The available polarization spectrum models \citep{Draine2009,Guillet2018,Hensley2022} predict the 50$-$200 $\mu$m spectra to be $\sim\,$flat toward irradiated regions, \ie regions close to the irradiation front in our case. Figure \ref{fig:pfrac_spectra} presents the normalized polarization fraction spectra of the Orion Bar, for four ranges of \Td, \NHtsqT, and \NHtc, and averaging all data points (selecting the pixels meting the SNR($I$)$\;\geq\;200$ \& SNR($P$)$\;\geq\;5$ \& $\Pf\;\leq\;30\,\%$ conditions at all wavelengths). We note here again that the radiation field interacting with the grains is a function of both the input radiation field and the density of the cloud, which make the \Td and \NHtc quantities probe the same area of the Bar (\ie the high \Td locations correspond to low \NHtc, and inversely; see Figure \ref{fig:Td_NH}).
Data from the SOFIA HAWC+ OTFMAP mode observations are used alongside the JCMT POL-2 observations presented in \citealt{WardThompson2017,Pattle2017}. The data have been smoothed to the lowest resolution of the observations we consider, \ie the 214 $\mu$m data, and regridded to Nyquist-sample the beam. 
% The resulting 8$^{\prime\prime}$ pixels may retrieve the emission emanating from various sub-regions where the population of aligned dust grains come from fairly different environmental conditions. This could hinder future studies attempting to fit such polarization spectra with dust models implementing specific environmental conditions. 
\citet{Michail2021} measured a falling 50$-$220 $\mu$m polarization fraction spectrum toward OMC-1, that they attributed to variation of dust grain population and grain alignment efficiency in the LOS. 
Given the uncertainties in the spectra of Figure \ref{fig:pfrac_spectra}, we find that the spectra are on average flat. 
However, there is a slight decreasing tendency---the spectra have either a minimum at 154 $\mu$m or at 850 $\mu$m, and appear tentatively falling between 53 and 154 $\mu$m (this latter trend has also been observed in starburst galaxies, see \citealt{LopezRodriguez2022b}).
We also notice that the apparent dip at 154 $\mu$m is more prominent at high temperature and low reddening (\ie the low density region toward the irradiated side of the Bar) within the covered range of physical conditions. However, the error bar on the the mean polarization of the Band A data is large, and as noted in Appendix \ref{app:c2n_vs_otfmap}, the polarization fractions values of the Band A data may suffer from systematics.
Future FIR observatories with multi-wavelength polarization capabilities will be able to put stronger constraints on the dust grain properties and grain alignment mechanisms at play in PDRs.

Finally, optical and NIR spectro-polarimetry observations of background stars represent a strong tool to constrain the size distribution of foreground aligned grains. The shape of the polarization fraction versus wavelength curve, \ie the Serkowski curve, is sensitive to the total-to-selective extinction $R_{\rm{V}}$, which is a good tracer of dust grain evolution \citep{Serkowski1975,Whittet1978,Andersson2007,Fanciullo2017,ChauGiang2020,Vaillancourt2020}. Such method can precisely constrain the maximum size of aligned grains, but it would require a good number of polarization detections of background stars as a function of the depth in the Bar, along the direction of the radiation field, to constrain the dust evolution throughout the PDR.
Such work shall also be enabled by future FIR polarization capabilities (see Appendix \ref{app:starlight}).

\section{Conclusions and summary}
\label{sec:ccl}

We present multiwavelength SOFIA HAWC+ polarimetric scan-pol dust polarization observations of the Orion Bar. Our goal is to characterize the origin of such polarized dust emission in such a highly illuminated region. In particular, we investigate the possibility that the reference direction for the grain alignment might change from the magnetic field ("\brat") to the radiation field k-vector ("\krat") in the regions of strongest illumination of the Bar. In addition, we explore the grain size parameter space in which aligned dust grains would be affected by the RAdiative Torques Disruption (RATD) mechanism, which fragments the largest aligned grains in efficient grain alignment conditions. The main results and conclusions of this study are as follows:

% To summarize, we suggest that the evolution of the grain alignment efficiency with the local conditions (using the available constrains on the dust temperature and gas density) cannot be explained by the RAT theory alone without considering dust evolution. The evolution of dust properties throughout the Bar can explain the slight decrease (increase) of alignment efficiency with dust temperature (reddening), which can be caused by RATD. However, we note that RATD, if occurring, does not totally hinder grain alignment in the Bar

\begin{enumerate}
\item We find overall consistent polarization position angles across all wavelengths, at 53, 89, 154, and 214 $\mu$m, within 15-25 $^{\circ}$. The linear polarization presents a chaotic morphology, precluding us from building precise predictions of the mechanism responsible for grain alignment in the Bar. However, the maps of polarization angles show that the radiation field direction is not the preferred grain alignment axis. We thus conclude that no evidence of the $k$-RAT grain alignment mechanism is found throughout the Orion Bar in our FIR dust polarization observations.

\item Using grey-body SED fits of FIR and sub-mm archival data, we derive the local environmental conditions that we use a proxy for the grain alignment conditions: the dust temperature, the reddening of the radiation field, and the gas pressure.
We compare them with the polarization quantities, \ie the polarization fraction with and without a correction for the depolarization caused by the disorganized magnetic field lines. 
While the grain alignment efficiency drops with increasing dust temperature, the dust polarization exhibits a high grain alignment efficiency toward the embedded side of the Bar, where the temperature is the lowest, and the reddening is the highest.
This evolution of the polarization degree as a function of the local environmental conditions cannot be explained by the RAT theory alone without considering dust evolution.
This suggests that the alignment properties of the large grains change from the cold to the hot side of the Bar, which can be explained by an evolution of dust properties across the Bar, possibly caused by RATD.

\item The level of grain alignment efficiency obtained from these FIR dust polarization observations across such a highly illuminated PDR suggests that RATD does not totally hinder grain alignment. The exact grain size distribution of the population of aligned grains responsible for the FIR polarization in the Bar remains to be constrained.

\item We calculate the grain size parameter space affected by \krat and RATD, as function of the depth into the Bar. 
% We conclude that \krat is not the dominant cause of dust polarization in the Orion Bar because
The typical grain sizes above which the alignment shifts from \brat to \krat (from 0.1 to 10 $\mu$m depending on the location within the Orion Bar and the grain characteristics) corresponds to grain sizes too large to be the dominant source of polarized dust emission (given this transition size is close to the expected maximum grain size $\leq\,0.5\,\mu$m), and that should be rotationally disrupted before they reach this typical size. However, the grains subject to \krat and RATD may not be the same population of aligned grains, \ie as they correspond to low versus high grain angular momentum states, respectively (low- or high-$J$). The evolution of the fraction of grains at high-$J$ in the typical environmental conditions encountered in PDRs remain to be explored.

\item Extrapolating our analytical calculations, we predict that \krat may be active toward the edge of the denser $n_{\rm{H}}\,=\,10^{6}-10^{7}$ cm$^{-3}$ sub-structures lying at the dissociation front.
\end{enumerate}

While the effective impact of RATD seems to remain moderate throughout the Bar, higher angular resolution multi-wavelength dust polarization observations will be required to truly quantify the effects of this mechanism. Our analysis tentatively suggests that this could be an active factor of dust grain evolution at the edge of highly illuminated PDRs, controlling the size distribution of the large aligned dust grains.

% \begin{acknowledgments}
\textit{Acknowledgments:}
We are grateful to Olivier Bern\'e, who provided
the SOFIA FORCAST photometric data. We thank Umit Kavak for his inputs about the literature of the Orion nebula. We are grateful to Alexander Tielens for the use of the SOFIA HAWC+ 09-0107 program he is the PI of.
We are grateful to Prof. Dan Clemens for providing the Orion Mimir data.
Based on observations made with the NASA/DLR Stratospheric
Observatory for Infrared Astronomy (SOFIA). SOFIA
is jointly operated by the Universities Space Research
Association, Inc. (USRA), under NASA contract NAS2-
97001, and the Deutsches SOFIA Institut (DSI) under DLR
contract 50 OK 0901 to the University of Stuttgart.
Financial support for this work was provided by NASA through awards
\#SOF 09-0037 issued by USRA.
This research was conducted in part using the Mimir instrument, jointly developed at Boston University and Lowell Observatory and supported by NASA, NSF, and the W.M. Keck Foundation
% \end{acknowledgments}

\textit{Facilities:} SOFIA.

\textit{Software:} APLpy, an open-source plotting package for Python hosted at \url{http://aplpy.github.com} \citep{Robitaille2012}.  CASA \citep{McMullin2007}.  Astropy \citep{Astropy2018}.

\bibliography{ms}

\begin{thebibliography}{}
\expandafter\ifx\csname natexlab\endcsname\relax\def\natexlab#1{#1}\fi

\bibitem[{{Abergel}(2010)}]{Abergel2010}
{Abergel}, A. 2010, {SDP\_aabergel\_3: Evolution of interstellar dust},
  Herschel Space Observatory Proposal, id.285

\bibitem[{{Alina} {et~al.}(2016){Alina}, {Montier}, {Ristorcelli}, {Bernard},
  {Levrier}, \& {Abdikamalov}}]{Alina2016}
{Alina}, D., {Montier}, L., {Ristorcelli}, I., {et~al.} 2016, \aap, 595, A57

\bibitem[{{Allers} {et~al.}(2005){Allers}, {Jaffe}, {Lacy}, {Draine}, \&
  {Richter}}]{Allers2005}
{Allers}, K.~N., {Jaffe}, D.~T., {Lacy}, J.~H., {Draine}, B.~T., \& {Richter},
  M.~J. 2005, \apj, 630, 368

\bibitem[{{Andersson} {et~al.}(2015){Andersson}, {Lazarian}, \&
  {Vaillancourt}}]{Andersson2015}
{Andersson}, B.-G., {Lazarian}, A., \& {Vaillancourt}, J.~E. 2015, \araa, 53,
  501

\bibitem[{{Andersson} \& {Potter}(2007)}]{Andersson2007}
{Andersson}, B.~G., \& {Potter}, S.~B. 2007, \apj, 665, 369

\bibitem[{{Andersson} {et~al.}(2013){Andersson}, {Piirola}, {De Buizer},
      {Clemens}, {Uomoto}, {Charcos-Llorens},  {Geballe}, {Lazarian},
  {Hoang}, \& {Vornanen}}]{Andersson2013}
{Andersson}, B.-G., {Piirola}, V., {De Buizer}, J., {et~al.} 2013, \apj, 775,
  84

\bibitem[{{Andersson} {et~al.}(2022){Andersson}, {Lopez-Rodriguez}, {Medan},
  {Soam}, {Hoang}, {Vaillancourt}, {Lazarian}, {Sandin}, {Mattsson}, \&
  {Tahani}}]{Andersson2022}
{Andersson}, B.~G., {Lopez-Rodriguez}, E., {Medan}, I., {et~al.} 2022, \apj,
  931, 80

\bibitem[{{Andr{\'e}}(2011)}]{Andre2011}
{Andr{\'e}}, P. 2011, {GT2\_pandre\_5: Completion of the Gould Belt and HOBYS
  surveys}, Herschel Space Observatory Proposal, id.1533

\bibitem[{{Andr{\'e}} {et~al.}(2010){Andr{\'e}}, {Men'shchikov}, {Bontemps},
  {K{\"o}nyves}, {Motte}, {Schneider}, {Didelon}, {Minier}, {Saraceno},
  {Ward-Thompson}, {di Francesco}, {White}, {Molinari}, {Testi}, {Abergel},
  {Griffin}, {Henning}, {Royer}, {Mer{\'\i}n}, {Vavrek}, {Attard},
  {Arzoumanian}, {Wilson}, {Ade}, {Aussel}, {Baluteau}, {Benedettini},
  {Bernard}, {Blommaert}, {Cambr{\'e}sy}, {Cox}, {di Giorgio}, {Hargrave},
  {Hennemann}, {Huang}, {Kirk}, {Krause}, {Launhardt}, {Leeks}, {Le Pennec},
  {Li}, {Martin}, {Maury}, {Olofsson}, {Omont}, {Peretto}, {Pezzuto}, {Prusti},
  {Roussel}, {Russeil}, {Sauvage}, {Sibthorpe}, {Sicilia-Aguilar}, {Spinoglio},
  {Waelkens}, {Woodcraft}, \& {Zavagno}}]{Andre2010}
{Andr{\'e}}, P., {Men'shchikov}, A., {Bontemps}, S., {et~al.} 2010, \aap, 518,
  L102

\bibitem[{{Andree-Labsch} {et~al.}(2017){Andree-Labsch}, {Ossenkopf-Okada}, \&
  {R{\"o}llig}}]{AndreeLabsch2017}
{Andree-Labsch}, S., {Ossenkopf-Okada}, V., \& {R{\"o}llig}, M. 2017, \aap,
  598, A2

\bibitem[{{Arab} {et~al.}(2012){Arab}, {Abergel}, {Habart}, {Bernard-Salas},
  {Ayasso}, {Dassas}, {Martin}, \& {White}}]{Arab2012}
{Arab}, H., {Abergel}, A., {Habart}, E., {et~al.} 2012, \aap, 541, A19

\bibitem[{{Astropy Collaboration} {et~al.}(2018){Astropy Collaboration},
  {Price-Whelan}, {Sip{\H o}cz}, {G{\"u}nther}, {Lim}, {Crawford}, {Conseil},
  {Shupe}, {Craig}, {Dencheva}, {Ginsburg}, {VanderPlas}, {Bradley},
  {P{\'e}rez-Su{\'a}rez}, {de Val-Borro}, {Aldcroft}, {Cruz}, {Robitaille},
  {Tollerud}, {Ardelean}, {Babej}, {Bach}, {Bachetti}, {Bakanov}, {Bamford},
  {Barentsen}, {Barmby}, {Baumbach}, {Berry}, {Biscani}, {Boquien}, {Bostroem},
  {Bouma}, {Brammer}, {Bray}, {Breytenbach}, {Buddelmeijer}, {Burke},
  {Calderone}, {Cano Rodr{\'{\i}}guez}, {Cara}, {Cardoso}, {Cheedella},
  {Copin}, {Corrales}, {Crichton}, {D'Avella}, {Deil}, {Depagne}, {Dietrich},
  {Donath}, {Droettboom}, {Earl}, {Erben}, {Fabbro}, {Ferreira}, {Finethy},
  {Fox}, {Garrison}, {Gibbons}, {Goldstein}, {Gommers}, {Greco}, {Greenfield},
  {Groener}, {Grollier}, {Hagen}, {Hirst}, {Homeier}, {Horton}, {Hosseinzadeh},
  {Hu}, {Hunkeler}, {Ivezi{\'c}}, {Jain}, {Jenness}, {Kanarek}, {Kendrew},
  {Kern}, {Kerzendorf}, {Khvalko}, {King}, {Kirkby}, {Kulkarni}, {Kumar},
  {Lee}, {Lenz}, {Littlefair}, {Ma}, {Macleod}, {Mastropietro}, {McCully},
  {Montagnac}, {Morris}, {Mueller}, {Mumford}, {Muna}, {Murphy}, {Nelson},
  {Nguyen}, {Ninan}, {N{\"o}the}, {Ogaz}, {Oh}, {Parejko}, {Parley}, {Pascual},
  {Patil}, {Patil}, {Plunkett}, {Prochaska}, {Rastogi}, {Reddy Janga},
  {Sabater}, {Sakurikar}, {Seifert}, {Sherbert}, {Sherwood-Taylor}, {Shih},
  {Sick}, {Silbiger}, {Singanamalla}, {Singer}, {Sladen}, {Sooley},
  {Sornarajah}, {Streicher}, {Teuben}, {Thomas}, {Tremblay}, {Turner},
  {Terr{\'o}n}, {van Kerkwijk}, {de la Vega}, {Watkins}, {Weaver}, {Whitmore},
  {Woillez}, {Zabalza}, \& {Astropy Contributors}}]{Astropy2018}
{Astropy Collaboration}, {Price-Whelan}, A.~M., {Sip{\H o}cz}, B.~M., {et~al.}
  2018, \aj, 156, 123

\bibitem[{{Bakes} \& {Tielens}(1994)}]{Bakes1994}
{Bakes}, E.~L.~O., \& {Tielens}, A.~G.~G.~M. 1994, \apj, 427, 822

\bibitem[{{Bendo} {et~al.}(2013){Bendo}, {Griffin}, {Bock}, {Conversi},
  {Dowell}, {Lim}, {Lu}, {North}, {Papageorgiou}, {Pearson}, {Pohlen},
  {Polehampton}, {Schulz}, {Shupe}, {Sibthorpe}, {Spencer}, {Swinyard},
  {Valtchanov}, \& {Xu}}]{Bendo2013}
{Bendo}, G.~J., {Griffin}, M.~J., {Bock}, J.~J., {et~al.} 2013, \mnras, 433,
  3062

\bibitem[{{Bradley}(1994)}]{Bradley1994}
{Bradley}, J.~P. 1994, Science, 265, 925

\bibitem[{{Bron} {et~al.}(2014){Bron}, {Le Bourlot}, \& {Le Petit}}]{Bron2014}
{Bron}, E., {Le Bourlot}, J., \& {Le Petit}, F. 2014, \aap, 569, A100

\bibitem[{{Chau Giang} {et~al.}(2022){Chau Giang}, {Hoang}, {Kim}, \&
  {Tram}}]{ChauGiang2022}
{Chau Giang}, N., {Hoang}, T., {Kim}, J.-G., \& {Tram}, L.~N. 2022, arXiv
  e-prints, arXiv:2210.01036

\bibitem[{{Chrysostomou} {et~al.}(1994){Chrysostomou}, {Hough}, {Burton}, \&
  {Tamura}}]{Chrysostomou1994}
{Chrysostomou}, A., {Hough}, J.~H., {Burton}, M.~G., \& {Tamura}, M. 1994,
  \mnras, 268, 325

\bibitem[{{Chuss} {et~al.}(2019){Chuss}, {Andersson}, {Bally}, {Dotson},
  {Dowell}, {Guerra}, {Harper}, {Houde}, {Jones}, {Lazarian}, {Lopez
  Rodriguez}, {Michail}, {Morris}, {Novak}, {Siah}, {Staguhn}, {Vaillancourt},
  {Volpert}, {Werner}, {Wollack}, {Benford}, {Berthoud}, {Cox}, {Crutcher},
  {Dale}, {Fissel}, {Goldsmith}, {Hamilton}, {Hanany}, {Henning}, {Looney},
  {Moseley}, {Santos}, {Stephens}, {Tassis}, {Trinh}, {Van Camp},
  {Ward-Thompson}, \& {HAWC + Science Team}}]{Chuss2019}
{Chuss}, D.~T., {Andersson}, B.~G., {Bally}, J., {et~al.} 2019, \apj, 872, 187

\bibitem[{{Clemens} {et~al.}(2007){Clemens}, {Sarcia}, {Grabau}, {Tollestrup},
  {Buie}, {Dunham}, \& {Taylor}}]{Clemens2007}
{Clemens}, D.~P., {Sarcia}, D., {Grabau}, A., {et~al.} 2007, \pasp, 119, 1385

\bibitem[{{Compi{\`e}gne} {et~al.}(2011){Compi{\`e}gne}, {Verstraete}, {Jones},
  {Bernard}, {Boulanger}, {Flagey}, {Le Bourlot}, {Paradis}, \&
  {Ysard}}]{Compiegne2011}
{Compi{\`e}gne}, M., {Verstraete}, L., {Jones}, A., {et~al.} 2011, \aap, 525,
  A103

\bibitem[{{Cortes} {et~al.}(2021){Cortes}, {Le Gouellec}, {Hull}, {Girart},
  {Louvet}, {Fomalont}, {Kameno}, {Moellenbrock}, {Nagai}, {Nakanishi}, \&
  {Villard}}]{Cortes2021}
{Cortes}, P.~C., {Le Gouellec}, V. J.~M., {Hull}, C. L.~H., {et~al.} 2021,
  \apj, 907, 94

\bibitem[{{Desert} {et~al.}(1990){Desert}, {Boulanger}, \&
  {Puget}}]{Desert1990}
{Desert}, F.~X., {Boulanger}, F., \& {Puget}, J.~L. 1990, \aap, 237, 215

\bibitem[{{Dicker} {et~al.}(2009){Dicker}, {Mason}, {Korngut}, {Cotton},
  {Compi{\`e}gne}, {Devlin}, {Martin}, {Ade}, {Benford}, {Irwin}, {Maddalena},
  {McMullin}, {Shepherd}, {Sievers}, {Staguhn}, \& {Tucker}}]{Dicker2009}
{Dicker}, S.~R., {Mason}, B.~S., {Korngut}, P.~M., {et~al.} 2009, \apj, 705,
  226

\bibitem[{{Dolginov} \& {Mitrofanov}(1976)}]{Dolginov1976}
{Dolginov}, A.~Z., \& {Mitrofanov}, I.~G. 1976, \apss, 43, 291

\bibitem[{{Dowell} {et~al.}(2010){Dowell}, {Cook}, {Harper}, {Lin}, {Looney},
  {Novak}, {Stephens}, {Berthoud}, {Chuss}, {Crutcher}, {Dotson}, {Hildebrand},
  {Houde}, {Jones}, {Krejny}, {Lazarian}, {Moseley}, {Tassis}, {Vaillancourt},
  \& {Werner}}]{Dowell2010}
{Dowell}, C.~D., {Cook}, B.~T., {Harper}, D.~A., {et~al.} 2010, in Society of
  Photo-Optical Instrumentation Engineers (SPIE) Conference Series, Vol. 7735,
  Ground-based and Airborne Instrumentation for Astronomy III, ed. I.~S.
  {McLean}, S.~K. {Ramsay}, \& H.~{Takami}, 77356H

\bibitem[{{Draine}(2011)}]{Draine2011}
{Draine}, B.~T. 2011, {Physics of the Interstellar and Intergalactic Medium}

\bibitem[{{Draine} \& {Fraisse}(2009)}]{Draine2009}
{Draine}, B.~T., \& {Fraisse}, A.~A. 2009, \apj, 696, 1

\bibitem[{{Draine} \& {Hensley}(2021)}]{Draine2021}
{Draine}, B.~T., \& {Hensley}, B.~S. 2021, \apj, 909, 94

\bibitem[{{Draine} \& {Lazarian}(1998)}]{Draine1998}
{Draine}, B.~T., \& {Lazarian}, A. 1998, \apj, 508, 157

\bibitem[{{Draine} \& {Weingartner}(1996)}]{Draine1996}
{Draine}, B.~T., \& {Weingartner}, J.~C. 1996, \apj, 470, 551

\bibitem[{{Draine} \& {Weingartner}(1997)}]{Draine1997}
---. 1997, \apj, 480, 633

\bibitem[{{Einstein} \& {de Haas}(1915)}]{Einstein1915}
{Einstein}, A., \& {de Haas}, W.~J. 1915, Koninklijke Nederlandse Akademie van
  Wetenschappen Proceedings Series B Physical Sciences, 18, 696

\bibitem[{{Fanciullo} {et~al.}(2017){Fanciullo}, {Guillet}, {Boulanger}, \&
  {Jones}}]{Fanciullo2017}
{Fanciullo}, L., {Guillet}, V., {Boulanger}, F., \& {Jones}, A.~P. 2017, \aap,
  602, A7

\bibitem[{{Fanciullo} {et~al.}(2022){Fanciullo}, {Kemper}, {Pattle}, {Koch},
  {Sadavoy}, {Coud{\'e}}, {Soam}, {Hoang}, {Onaka}, {Le Gouellec},
  {Arzoumanian}, {Berry}, {Eswaraiah}, {Chung}, {Furuya}, {Hull}, {Hwang},
  {Johnstone}, {Kang}, {Kim}, {Kirchschlager}, {K{\"o}nyves}, {Kwon}, {Kwon},
  {Lai}, {Lee}, {Liu}, {Lyo}, {Stephens}, {Tamura}, {Tang}, {Ward-Thompson},
  {Whitworth}, \& {Shinnaga}}]{Fanciullo2022}
{Fanciullo}, L., {Kemper}, F., {Pattle}, K., {et~al.} 2022, \mnras, 512, 1985

\bibitem[{{Fissel} {et~al.}(2016){Fissel}, {Ade}, {Angil{\`e}}, {Ashton},
  {Benton}, {Devlin}, {Dober}, {Fukui}, {Galitzki}, {Gandilo}, {Klein},
  {Korotkov}, {Li}, {Martin}, {Matthews}, {Moncelsi}, {Nakamura},
  {Netterfield}, {Novak}, {Pascale}, {Poidevin}, {Santos}, {Savini}, {Scott},
  {Shariff}, {Diego Soler}, {Thomas}, {Tucker}, {Tucker}, \&
  {Ward-Thompson}}]{Fissel2016}
{Fissel}, L.~M., {Ade}, P.~A.~R., {Angil{\`e}}, F.~E., {et~al.} 2016, \apj,
  824, 134

\bibitem[{{Gaia Collaboration} {et~al.}(2022){Gaia Collaboration}, {Vallenari},
  {Brown}, {Prusti}, {de Bruijne}, {Arenou}, {Babusiaux}, {Biermann},
  {Creevey}, {Ducourant}, \& et~al.}]{GaiaDR32022}
{Gaia Collaboration}, {Vallenari}, A., {Brown}, A.~G.~A., {et~al.} 2022, arXiv
  e-prints, arXiv:2208.00211

\bibitem[{{Genzel} \& {Stutzki}(1989)}]{Genzel1989}
{Genzel}, R., \& {Stutzki}, J. 1989, \araa, 27, 41

\bibitem[{{Giang} {et~al.}(2020){Giang}, {Hoang}, \& {Tram}}]{ChauGiang2020}
{Giang}, N.~C., {Hoang}, T., \& {Tram}, L.~N. 2020, \apj, 888, 93

\bibitem[{{Goicoechea} \& {Le Bourlot}(2007)}]{Goicoechea2007}
{Goicoechea}, J.~R., \& {Le Bourlot}, J. 2007, \aap, 467, 1

\bibitem[{{Goicoechea} {et~al.}(2011){Goicoechea}, {Joblin}, {Contursi},
  {Bern{\'e}}, {Cernicharo}, {Gerin}, {Le Bourlot}, {Bergin}, {Bell}, \&
  {R{\"o}llig}}]{Goicoechea2011}
{Goicoechea}, J.~R., {Joblin}, C., {Contursi}, A., {et~al.} 2011, \aap, 530,
  L16

\bibitem[{{Goicoechea} {et~al.}(2016){Goicoechea}, {Pety}, {Cuadrado},
  {Cernicharo}, {Chapillon}, {Fuente}, {Gerin}, {Joblin}, {Marcelino}, \&
  {Pilleri}}]{Goicoechea2016}
{Goicoechea}, J.~R., {Pety}, J., {Cuadrado}, S., {et~al.} 2016, \nat, 537, 207

\bibitem[{{Goicoechea} {et~al.}(2017){Goicoechea}, {Cuadrado}, {Pety}, {Bron},
  {Black}, {Cernicharo}, {Chapillon}, {Fuente}, \& {Gerin}}]{Goicoechea2017}
{Goicoechea}, J.~R., {Cuadrado}, S., {Pety}, J., {et~al.} 2017, \aap, 601, L9

\bibitem[{{Goodman} \& {Whittet}(1995)}]{Goodman1995a}
{Goodman}, A.~A., \& {Whittet}, D.~C.~B. 1995, \apjl, 455, L181

\bibitem[{{Gordon} {et~al.}(2018){Gordon}, {Lopez-Rodriguez}, {Andersson},
  {Clarke}, {Coude}, {Moullet}, {Richards}, {Shuping}, {Vacca}, \&
  {Yorke}}]{Gordon2018}
{Gordon}, M.~S., {Lopez-Rodriguez}, E., {Andersson}, B.~G., {et~al.} 2018,
  arXiv e-prints, arXiv:1811.03100

\bibitem[{{Gorti} \& {Hollenbach}(2002)}]{Gorti2002}
{Gorti}, U., \& {Hollenbach}, D. 2002, \apj, 573, 215

\bibitem[{{G{\"u}del} {et~al.}(2008){G{\"u}del}, {Briggs}, {Montmerle},
  {Audard}, {Rebull}, \& {Skinner}}]{Gudel2008}
{G{\"u}del}, M., {Briggs}, K.~R., {Montmerle}, T., {et~al.} 2008, Science, 319,
  309

\bibitem[{{Guerra} {et~al.}(2021){Guerra}, {Chuss}, {Dowell}, {Houde},
  {Michail}, {Siah}, \& {Wollack}}]{Guerra2021}
{Guerra}, J.~A., {Chuss}, D.~T., {Dowell}, C.~D., {et~al.} 2021, \apj, 908, 98

\bibitem[{{Guillet} {et~al.}(2018){Guillet}, {Fanciullo}, {Verstraete},
  {Boulanger}, {Jones}, {Miville-Desch{\^e}nes}, {Ysard}, {Levrier}, \&
  {Alves}}]{Guillet2018}
{Guillet}, V., {Fanciullo}, L., {Verstraete}, L., {et~al.} 2018, \aap, 610, A16

\bibitem[{{Habart} {et~al.}(2022){Habart}, {Le Gal}, {Alvarez}, {Peeters},
  {Bern{\'e}}, {Wolfire}, {Goicoechea}, {Schirmer}, {Bron}, \&
  {R{\"o}llig}}]{Habart2022}
{Habart}, E., {Le Gal}, R., {Alvarez}, C., {et~al.} 2022, arXiv e-prints,
  arXiv:2206.08245

\bibitem[{{Harper} {et~al.}(2018){Harper}, {Runyan}, {Dowell}, {Wirth},
  {Amato}, {Ames}, {Amiri}, {Banks}, {Bartels}, {Benford}, {Berthoud},
  {Buchanan}, {Casey}, {Chapman}, {Chuss}, {Cook}, {Derro}, {Dotson}, {Evans},
  {Fixsen}, {Gatley}, {Guerra}, {Halpern}, {Hamilton}, {Hamlin}, {Hansen},
  {Heimsath}, {Hermida}, {Hilton}, {Hirsch}, {Hollister}, {Hostetter}, {Irwin},
  {Jhabvala}, {Jhabvala}, {Kastner}, {Kov{\'a}cs}, {Lin}, {Loewenstein},
  {Looney}, {Lopez-Rodriguez}, {Maher}, {Michail}, {Miller}, {Moseley},
  {Novak}, {Pernic}, {Rennick}, {Rhody}, {Sandberg}, {Sandford}, {Santos},
  {Shafer}, {Sharp}, {Shirron}, {Siah}, {Silverberg}, {Sparr}, {Spotz},
  {Staguhn}, {Toorian}, {Towey}, {Tuttle}, {Vaillancourt}, {Voellmer},
  {Volpert}, {Wang}, \& {Wollack}}]{Harper2018}
{Harper}, D.~A., {Runyan}, M.~C., {Dowell}, C.~D., {et~al.} 2018, Journal of
  Astronomical Instrumentation, 7, 1840008

\bibitem[{{Hensley} \& {Draine}(2022)}]{Hensley2022}
{Hensley}, B.~S., \& {Draine}, B.~T. 2022, arXiv e-prints, arXiv:2208.12365

\bibitem[{{Herrmann} {et~al.}(1997){Herrmann}, {Madden}, {Nikola}, {Poglitsch},
  {Timmermann}, {Geis}, {Townes}, \& {Stacey}}]{Herrmann1997}
{Herrmann}, F., {Madden}, S.~C., {Nikola}, T., {et~al.} 1997, \apj, 481, 343

\bibitem[{{Hildebrand} {et~al.}(1999){Hildebrand}, {Dotson}, {Dowell},
  {Schleuning}, \& {Vaillancourt}}]{Hildebrand1999}
{Hildebrand}, R.~H., {Dotson}, J.~L., {Dowell}, C.~D., {Schleuning}, D.~A., \&
  {Vaillancourt}, J.~E. 1999, \apj, 516, 834

\bibitem[{{Hildebrand} \& {Dragovan}(1995)}]{Hildebrand1995}
{Hildebrand}, R.~H., \& {Dragovan}, M. 1995, \apj, 450, 663

\bibitem[{{Hiltner}(1949)}]{Hiltner1949}
{Hiltner}, W.~A. 1949, Science, 109, 165

\bibitem[{{Hoang}(2019)}]{Hoang2019c}
{Hoang}, T. 2019, \apj, 876, 13

\bibitem[{{Hoang} {et~al.}(2018){Hoang}, {Cho}, \& {Lazarian}}]{Hoang2018}
{Hoang}, T., {Cho}, J., \& {Lazarian}, A. 2018, \apj, 852, 129

\bibitem[{{Hoang} \& {Lazarian}(2008)}]{Hoang2008}
{Hoang}, T., \& {Lazarian}, A. 2008, \mnras, 388, 117

\bibitem[{{Hoang} \& {Lazarian}(2009{\natexlab{a}})}]{Hoang2009}
---. 2009{\natexlab{a}}, \apj, 697, 1316

\bibitem[{{Hoang} \& {Lazarian}(2009{\natexlab{b}})}]{Hoang2009a}
---. 2009{\natexlab{b}}, \apj, 695, 1457

\bibitem[{{Hoang} \& {Lazarian}(2016)}]{Hoang2016}
---. 2016, \apj, 831, 159

\bibitem[{{Hoang} {et~al.}(2015){Hoang}, {Lazarian}, \&
  {Andersson}}]{Hoang2015}
{Hoang}, T., {Lazarian}, A., \& {Andersson}, B.~G. 2015, \mnras, 448, 1178

\bibitem[{{Hoang} {et~al.}(2019){Hoang}, {Tram}, {Lee}, \&
  {Ahn}}]{Hoang2019NatAs}
{Hoang}, T., {Tram}, L.~N., {Lee}, H., \& {Ahn}, S.-H. 2019, Nature Astronomy,
  3, 766

\bibitem[{{Hoang} {et~al.}(2021){Hoang}, {Tram}, {Lee}, {Diep}, \&
  {Ngoc}}]{Hoang2020b}
{Hoang}, T., {Tram}, L.~N., {Lee}, H., {Diep}, P.~N., \& {Ngoc}, N.~B. 2021,
  \apj, 908, 218

\bibitem[{{Hoang} {et~al.}(2022){Hoang}, {Tram}, {Minh Phan}, {Giang},
  {Phuong}, \& {Dieu}}]{Hoang2022b}
{Hoang}, T., {Tram}, L.~N., {Minh Phan}, V.~H., {et~al.} 2022, \aj, 164, 248

\bibitem[{{Hogerheijde} {et~al.}(1995){Hogerheijde}, {Jansen}, \& {van
  Dishoeck}}]{Hogerheijde1995}
{Hogerheijde}, M.~R., {Jansen}, D.~J., \& {van Dishoeck}, E.~F. 1995, \aap,
  294, 792

\bibitem[{{Hopkins} {et~al.}(2022){Hopkins}, {Rosen}, {Squire}, {Panopoulou},
  {Soliman}, {Seligman}, \& {Steinwandel}}]{Hopkins2022}
{Hopkins}, P.~F., {Rosen}, A.~L., {Squire}, J., {et~al.} 2022, \mnras, 517,
  1491

\bibitem[{{Hough} {et~al.}(1986){Hough}, {Axon}, {Burton}, {Gatley}, {Sato},
  {Bailey}, {McCaughrean}, {McLean}, {Nagata}, {Allen}, {Garden}, {Hasegawa},
  {Hayashi}, {Kaifu}, {Morimoto}, \& {Walther}}]{Hough1986}
{Hough}, J.~H., {Axon}, D.~J., {Burton}, M.~G., {et~al.} 1986, \mnras, 222, 629

\bibitem[{{Hull} \& {Plambeck}(2015)}]{Hull2015b}
{Hull}, C.~L.~H., \& {Plambeck}, R.~L. 2015, Journal of Astronomical
  Instrumentation, 4, 1550005

\bibitem[{{Hull} {et~al.}(2014){Hull}, {Plambeck}, {Kwon}, {Bower},
  {Carpenter}, {Crutcher}, {Fiege}, {Franzmann}, {Hakobian}, {Heiles}, {Houde},
  {Hughes}, {Lamb}, {Looney}, {Marrone}, {Matthews}, {Pillai}, {Pound},
  {Rahman}, {Sandell}, {Stephens}, {Tobin}, {Vaillancourt}, {Volgenau}, \&
  {Wright}}]{Hull2014}
{Hull}, C.~L.~H., {Plambeck}, R.~L., {Kwon}, W., {et~al.} 2014, \apjs, 213, 13

\bibitem[{{Hull} {et~al.}(2022){Hull}, {Yang}, {Cort{\'e}s}, {Dent}, {Kral},
  {Li}, {Le Gouellec}, {Hughes}, {Milli}, {Teague}, \& {Wyatt}}]{Hull2022}
{Hull}, C. L.~H., {Yang}, H., {Cort{\'e}s}, P.~C., {et~al.} 2022, \apj, 930, 49

\bibitem[{{Jones} {et~al.}(1990){Jones}, {Duley}, \& {Williams}}]{Jones1990}
{Jones}, A.~P., {Duley}, W.~W., \& {Williams}, D.~A. 1990, \qjras, 31, 567

\bibitem[{{Jones} {et~al.}(2013){Jones}, {Fanciullo}, {K{\"o}hler},
  {Verstraete}, {Guillet}, {Bocchio}, \& {Ysard}}]{Jones2013}
{Jones}, A.~P., {Fanciullo}, L., {K{\"o}hler}, M., {et~al.} 2013, \aap, 558,
  A62

\bibitem[{{Jones} \& {Habart}(2015)}]{JonesA2015}
{Jones}, A.~P., \& {Habart}, E. 2015, \aap, 581, A92

\bibitem[{{Jones} {et~al.}(2017){Jones}, {K{\"o}hler}, {Ysard}, {Bocchio}, \&
  {Verstraete}}]{Jones2017}
{Jones}, A.~P., {K{\"o}hler}, M., {Ysard}, N., {Bocchio}, M., \& {Verstraete},
  L. 2017, \aap, 602, A46

\bibitem[{{Jones} \& {Spitzer}(1967)}]{Jones1967}
{Jones}, R.~V., \& {Spitzer}, Lyman, J. 1967, \apj, 147, 943

\bibitem[{{Jones} {et~al.}(1992){Jones}, {Klebe}, \& {Dickey}}]{Jones1992}
{Jones}, T.~J., {Klebe}, D., \& {Dickey}, J.~M. 1992, \apj, 389, 602

\bibitem[{{Kavak} {et~al.}(2022){Kavak}, {Goicoechea}, {Pabst}, {Bally}, {van
  der Tak}, \& {Tielens}}]{Kavak2022}
{Kavak}, {\"U}., {Goicoechea}, J.~R., {Pabst}, C.~H.~M., {et~al.} 2022, \aap,
  660, A109

\bibitem[{{Kneller} \& {Luborsky}(1963)}]{Kneller1963}
{Kneller}, E.~F., \& {Luborsky}, F.~E. 1963, Journal of Applied Physics, 34,
  656

\bibitem[{{K{\"o}hler} {et~al.}(2015){K{\"o}hler}, {Ysard}, \&
  {Jones}}]{Kohler2015}
{K{\"o}hler}, M., {Ysard}, N., \& {Jones}, A.~P. 2015, \aap, 579, A15

\bibitem[{{Koumpia} {et~al.}(2015){Koumpia}, {Harvey}, {Ossenkopf}, {van der
  Tak}, {Mookerjea}, {Fuente}, \& {Kramer}}]{Koumpia2015}
{Koumpia}, E., {Harvey}, P.~M., {Ossenkopf}, V., {et~al.} 2015, \aap, 580, A68

\bibitem[{{Kounkel} {et~al.}(2017){Kounkel}, {Hartmann}, {Loinard},
  {Ortiz-Le{\'o}n}, {Mioduszewski}, {Rodr{\'{\i}}guez}, {Dzib}, {Torres},
  {Pech}, {Galli}, {Rivera}, {Boden}, {Evans}, {Brice{\~n}o}, \&
  {Tobin}}]{Kounkel2017}
{Kounkel}, M., {Hartmann}, L., {Loinard}, L., {et~al.} 2017, \apj, 834, 142

\bibitem[{{Kounkel} {et~al.}(2018){Kounkel}, {Covey}, {Su{\'a}rez},
  {Rom{\'a}n-Z{\'u}{\~n}iga}, {Hernandez}, {Stassun}, {Jaehnig}, {Feigelson},
  {Pe{\~n}a Ram{\'\i}rez}, {Roman-Lopes}, {Da Rio}, {Stringfellow}, {Kim},
  {Borissova}, {Fern{\'a}ndez-Trincado}, {Burgasser},
  {Garc{\'\i}a-Hern{\'a}ndez}, {Zamora}, {Pan}, \& {Nitschelm}}]{Kounkel2018}
{Kounkel}, M., {Covey}, K., {Su{\'a}rez}, G., {et~al.} 2018, \aj, 156, 84

\bibitem[{{Kov{\'a}cs}(2006)}]{Kovacs2006PhD}
{Kov{\'a}cs}, A. 2006, PhD thesis, California Institute of Technology

\bibitem[{{Kov{\'a}cs}(2008)}]{Kovacs2008}
{Kov{\'a}cs}, A. 2008, in Society of Photo-Optical Instrumentation Engineers
  (SPIE) Conference Series, Vol. 7020, Millimeter and Submillimeter Detectors
  and Instrumentation for Astronomy IV, ed. W.~D. {Duncan}, W.~S. {Holland},
  S.~{Withington}, \& J.~{Zmuidzinas}, 70201S

\bibitem[{{Lazarian} \& {Draine}(1999)}]{Lazarian1999a}
{Lazarian}, A., \& {Draine}, B.~T. 1999, \apjl, 520, L67

\bibitem[{{Lazarian} \& {Efroimsky}(1999)}]{Lazarian1999c}
{Lazarian}, A., \& {Efroimsky}, M. 1999, \mnras, 303, 673

\bibitem[{{Lazarian} \& {Hoang}(2007)}]{LazarianHoang2007}
{Lazarian}, A., \& {Hoang}, T. 2007, \mnras, 378, 910

\bibitem[{{Lazarian} \& {Hoang}(2021)}]{Lazarian2021}
---. 2021, \apj, 908, 12

\bibitem[{{Le Bourlot} {et~al.}(2012){Le Bourlot}, {Le Petit}, {Pinto},
  {Roueff}, \& {Roy}}]{LeBourlot2012}
{Le Bourlot}, J., {Le Petit}, F., {Pinto}, C., {Roueff}, E., \& {Roy}, F. 2012,
  \aap, 541, A76

\bibitem[{{Le Gouellec} {et~al.}(2020){Le Gouellec}, {Maury}, {Guillet},
  {Hull}, {Girart}, {Verliat}, {Mignon-Risse}, {Valdivia}, {Hennebelle},
  {Gonz{\'a}lez}, \& {Louvet}}]{LeGouellec2020}
{Le Gouellec}, V.~J.~M., {Maury}, A.~J., {Guillet}, V., {et~al.} 2020, \aap,
  644, A11

\bibitem[{{Li} {et~al.}(2022){Li}, {Lopez-Rodriguez}, {Ajeddig}, {Andr{\'e}},
  {McKee}, {Rho}, \& {Klein}}]{LiPS2022}
{Li}, P.~S., {Lopez-Rodriguez}, E., {Ajeddig}, H., {et~al.} 2022, \mnras, 510,
  6085

\bibitem[{{Lis} \& {Schilke}(2003)}]{Lis2003}
{Lis}, D.~C., \& {Schilke}, P. 2003, \apjl, 597, L145

\bibitem[{{Lopez-Rodriguez} {et~al.}(2022{\natexlab{a}}){Lopez-Rodriguez},
  {Clarke}, {Shenoy}, {Vacca}, {Coude}, {Arneson}, {Ashton}, {Eftekharzadeh},
  {Beck}, {Beckman}, {Borlaff}, {Clark}, {Dale}, {Martin-Alvarez}, {Ntormousi},
  {Reach}, {Roman-Duval}, {Tassis}, {Harper}, \& {Marcum}}]{LopezRodriguez2022}
{Lopez-Rodriguez}, E., {Clarke}, M., {Shenoy}, S., {et~al.} 2022{\natexlab{a}},
  \apj, 936, 65

\bibitem[{{Lopez-Rodriguez} {et~al.}(2022{\natexlab{b}}){Lopez-Rodriguez},
  {Mao}, {Beck}, {Borlaff}, {Ntormousi}, {Tassis}, {Dale}, {Roman-Duval},
  {Subramanian}, {Martin-Alvarez}, {Marcum}, {Clark}, {Reach}, {Harper}, \&
  {Zweibel}}]{LopezRodriguez2022b}
{Lopez-Rodriguez}, E., {Mao}, S.~A., {Beck}, R., {et~al.} 2022{\natexlab{b}},
  \apj, 936, 92

\bibitem[{{Mairs} {et~al.}(2016){Mairs}, {Johnstone}, {Kirk}, {Buckle},
  {Berry}, {Broekhoven-Fiene}, {Currie}, {Fich}, {Graves}, {Hatchell},
  {Jenness}, {Mottram}, {Nutter}, {Pattle}, {Pineda}, {Salji}, {Di Francesco},
  {Hogerheijde}, {Ward-Thompson}, {Bastien}, {Bresnahan}, {Butner}, {Chen},
  {Chrysostomou}, {Coud{\'e}}, {Davis}, {Drabek-Maunder}, {Duarte-Cabral},
  {Fiege}, {Friberg}, {Friesen}, {Fuller}, {Greaves}, {Gregson}, {Holland},
  {Joncas}, {Kirk}, {Knee}, {Marsh}, {Matthews}, {Moriarty-Schieven}, {Mowat},
  {Rawlings}, {Richer}, {Robertson}, {Rosolowsky}, {Rumble}, {Sadavoy},
  {Thomas}, {Tothill}, {Viti}, {White}, {Wouterloot}, {Yates}, \&
  {Zhu}}]{Mairs2016}
{Mairs}, S., {Johnstone}, D., {Kirk}, H., {et~al.} 2016, \mnras, 461, 4022

\bibitem[{{Marconi} {et~al.}(1998){Marconi}, {Testi}, {Natta}, \&
  {Walmsley}}]{Marconi1998}
{Marconi}, A., {Testi}, L., {Natta}, A., \& {Walmsley}, C.~M. 1998, \aap, 330,
  696

\bibitem[{{Martin}(1995)}]{Martin1995}
{Martin}, P.~G. 1995, \apjl, 445, L63

\bibitem[{{Mathis} {et~al.}(1977){Mathis}, {Rumpl}, \&
  {Nordsieck}}]{Mathis1977}
{Mathis}, J.~S., {Rumpl}, W., \& {Nordsieck}, K.~H. 1977, \apj, 217, 425

\bibitem[{{Mathis} \& {Whiffen}(1989)}]{Mathis1989}
{Mathis}, J.~S., \& {Whiffen}, G. 1989, \apj, 341, 808

\bibitem[{{McMullin} {et~al.}(2007){McMullin}, {Waters}, {Schiebel}, {Young},
  \& {Golap}}]{McMullin2007}
{McMullin}, J.~P., {Waters}, B., {Schiebel}, D., {Young}, W., \& {Golap}, K.
  2007, in Astronomical Society of the Pacific Conference Series, Vol. 376,
  Astronomical Data Analysis Software and Systems XVI, ed. R.~A. {Shaw},
  F.~{Hill}, \& D.~J. {Bell}, 127

\bibitem[{{Michail} {et~al.}(2021){Michail}, {Ashton}, {Berthoud}, {Chuss},
  {Dowell}, {Guerra}, {Harper}, {Novak}, {Santos}, {Siah}, {Sukay}, {Taylor},
  {Tram}, {Vaillancourt}, \& {Wollack}}]{Michail2021}
{Michail}, J.~M., {Ashton}, P.~C., {Berthoud}, M.~G., {et~al.} 2021, \apj, 907,
  46

\bibitem[{{Morrish}(2001)}]{Morrish2001}
{Morrish}, A.~H. 2001, {The Physical Principles of Magnetism}

\bibitem[{{Murga} {et~al.}(2022){Murga}, {Kirsanova}, {Wiebe}, \&
  {Boley}}]{Murga2022}
{Murga}, M.~S., {Kirsanova}, M.~S., {Wiebe}, D.~S., \& {Boley}, P.~A. 2022,
  \mnras, 509, 800

\bibitem[{{Novak} {et~al.}(1997){Novak}, {Dotson}, {Dowell}, {Goldsmith},
  {Hildebrand}, {Platt}, \& {Schleuning}}]{Novak1997}
{Novak}, G., {Dotson}, J.~L., {Dowell}, C.~D., {et~al.} 1997, \apj, 487, 320

\bibitem[{{O'dell}(2001)}]{ODell2001}
{O'dell}, C.~R. 2001, \araa, 39, 99

\bibitem[{{Ossenkopf} {et~al.}(2013){Ossenkopf}, {R{\"o}llig}, {Neufeld},
  {Pilleri}, {Lis}, {Fuente}, {van der Tak}, \& {Bergin}}]{Ossenkopf2013}
{Ossenkopf}, V., {R{\"o}llig}, M., {Neufeld}, D.~A., {et~al.} 2013, \aap, 550,
  A57

\bibitem[{{Pabst} {et~al.}(2019){Pabst}, {Higgins}, {Goicoechea}, {Teyssier},
  {Berne}, {Chambers}, {Wolfire}, {Suri}, {Guesten}, {Stutzki}, {Graf},
  {Risacher}, \& {Tielens}}]{Pabst2019}
{Pabst}, C., {Higgins}, R., {Goicoechea}, J.~R., {et~al.} 2019, \nat, 565, 618

\bibitem[{{Pabst} {et~al.}(2020){Pabst}, {Goicoechea}, {Teyssier}, {Bern{\'e}},
  {Higgins}, {Chambers}, {Kabanovic}, {G{\"u}sten}, {Stutzki}, \&
  {Tielens}}]{Pabst2020}
{Pabst}, C.~H.~M., {Goicoechea}, J.~R., {Teyssier}, D., {et~al.} 2020, \aap,
  639, A2

\bibitem[{{Parikka} {et~al.}(2017){Parikka}, {Habart}, {Bernard-Salas},
  {Goicoechea}, {Abergel}, {Pilleri}, {Dartois}, {Joblin}, {Gerin}, \&
  {Godard}}]{Parikka2017}
{Parikka}, A., {Habart}, E., {Bernard-Salas}, J., {et~al.} 2017, \aap, 599, A20

\bibitem[{{Pattle} {et~al.}(2017){Pattle}, {Ward-Thompson}, {Berry},
  {Hatchell}, {Chen}, {Pon}, {Koch}, {Kwon}, {Kim}, {Bastien}, {Cho},
  {Coud{\'e}}, {Di Francesco}, {Fuller}, {Furuya}, {Graves}, {Johnstone},
  {Kirk}, {Kwon}, {Lee}, {Matthews}, {Mottram}, {Parsons}, {Sadavoy},
  {Shinnaga}, {Soam}, {Hasegawa}, {Lai}, {Qiu}, \& {Friberg}}]{Pattle2017}
{Pattle}, K., {Ward-Thompson}, D., {Berry}, D., {et~al.} 2017, \apj, 846, 122

\bibitem[{{Pellegrini} {et~al.}(2009){Pellegrini}, {Baldwin}, {Ferland},
  {Shaw}, \& {Heathcote}}]{Pellegrini2009}
{Pellegrini}, E.~W., {Baldwin}, J.~A., {Ferland}, G.~J., {Shaw}, G., \&
  {Heathcote}, S. 2009, \apj, 693, 285

\bibitem[{{Planck Collaboration} {et~al.}(2020){Planck Collaboration},
  {Aghanim}, {Akrami}, {Alves}, {Ashdown}, {Aumont}, {Baccigalupi},
  {Ballardini}, {Banday}, {Barreiro}, {Bartolo}, {Basak}, {Benabed}, {Bernard},
  {Bersanelli}, {Bielewicz}, {Bock}, {Bond}, {Borrill}, {Bouchet}, {Boulanger},
  {Bracco}, {Bucher}, {Burigana}, {Calabrese}, {Cardoso}, {Carron}, {Chary},
  {Chiang}, {Colombo}, {Combet}, {Crill}, {Cuttaia}, {de Bernardis}, {de
  Zotti}, {Delabrouille}, {Delouis}, {Di Valentino}, {Dickinson}, {Diego},
  {Dor{\'e}}, {Douspis}, {Ducout}, {Dupac}, {Efstathiou}, {Elsner},
  {En{\ss}lin}, {Eriksen}, {Falgarone}, {Fantaye}, {Fernandez-Cobos},
  {Ferri{\`e}re}, {Finelli}, {Forastieri}, {Frailis}, {Fraisse}, {Franceschi},
  {Frolov}, {Galeotta}, {Galli}, {Ganga}, {G{\'e}nova-Santos}, {Gerbino},
  {Ghosh}, {Gonz{\'a}lez-Nuevo}, {G{\'o}rski}, {Gratton}, {Green}, {Gruppuso},
  {Gudmundsson}, {Guillet}, {Handley}, {Hansen}, {Helou}, {Herranz}, {Hivon},
  {Huang}, {Jaffe}, {Jones}, {Keih{\"a}nen}, {Keskitalo}, {Kiiveri}, {Kim},
  {Krachmalnicoff}, {Kunz}, {Kurki-Suonio}, {Lagache}, {Lamarre}, {Lasenby},
  {Lattanzi}, {Lawrence}, {Le Jeune}, {Levrier}, {Liguori}, {Lilje},
  {Lindholm}, {L{\'o}pez-Caniego}, {Lubin}, {Ma}, {Mac{\'\i}as-P{\'e}rez},
  {Maggio}, {Maino}, {Mandolesi}, {Mangilli}, {Marcos-Caballero}, {Maris},
  {Martin}, {Mart{\'\i}nez-Gonz{\'a}lez}, {Matarrese}, {Mauri}, {McEwen},
  {Melchiorri}, {Mennella}, {Migliaccio}, {Miville-Desch{\^e}nes}, {Molinari},
  {Moneti}, {Montier}, {Morgante}, {Moss}, {Natoli}, {Pagano}, {Paoletti},
  {Patanchon}, {Perrotta}, {Pettorino}, {Piacentini}, {Polastri}, {Polenta},
  {Puget}, {Rachen}, {Reinecke}, {Remazeilles}, {Renzi}, {Ristorcelli},
  {Rocha}, {Rosset}, {Roudier}, {Rubi{\~n}o-Mart{\'\i}n}, {Ruiz-Granados},
  {Salvati}, {Sandri}, {Savelainen}, {Scott}, {Sirignano}, {Sunyaev},
  {Suur-Uski}, {Tauber}, {Tavagnacco}, {Tenti}, {Toffolatti}, {Tomasi},
  {Trombetti}, {Valiviita}, {Vansyngel}, {Van Tent}, {Vielva}, {Villa},
  {Vittorio}, {Wandelt}, {Wehus}, {Zacchei}, \& {Zonca}}]{Planck2018XII}
{Planck Collaboration}, {Aghanim}, N., {Akrami}, Y., {et~al.} 2020, \aap, 641,
  A12

\bibitem[{{Purcell}(1979)}]{Purcell1979}
{Purcell}, E.~M. 1979, \apj, 231, 404

\bibitem[{{Rao} {et~al.}(1998){Rao}, {Crutcher}, {Plambeck}, \&
  {Wright}}]{Rao1998}
{Rao}, R., {Crutcher}, R.~M., {Plambeck}, R.~L., \& {Wright}, M.~C.~H. 1998,
  \apjl, 502, L75+

\bibitem[{{Reissl} {et~al.}(2020){Reissl}, {Guillet}, {Brauer}, {Levrier},
  {Boulanger}, \& {Klessen}}]{Reissl2020}
{Reissl}, S., {Guillet}, V., {Brauer}, R., {et~al.} 2020, \aap, 640, A118

\bibitem[{{Reissl} {et~al.}(2022){Reissl}, {Meehan}, \& {Klessen}}]{Reissl2022}
{Reissl}, S., {Meehan}, P., \& {Klessen}, R.~S. 2022, arXiv e-prints,
  arXiv:2201.03694

\bibitem[{{Reissl} {et~al.}(2016){Reissl}, {Wolf}, \& {Brauer}}]{Reissl2016}
{Reissl}, S., {Wolf}, S., \& {Brauer}, R. 2016, \aap, 593, A87

\bibitem[{{Robitaille} \& {Bressert}(2012)}]{Robitaille2012}
{Robitaille}, T., \& {Bressert}, E. 2012, {APLpy: Astronomical Plotting Library
  in Python}, Astrophysics Source Code Library, ascl:1208.017

\bibitem[{{Salgado} {et~al.}(2016){Salgado}, {Bern{\'e}}, {Adams}, {Herter},
  {Keller}, \& {Tielens}}]{Salgado2016}
{Salgado}, F., {Bern{\'e}}, O., {Adams}, J.~D., {et~al.} 2016, \apj, 830, 118

\bibitem[{{Schirmer} {et~al.}(2020){Schirmer}, {Abergel}, {Verstraete},
  {Ysard}, {Juvela}, {Jones}, \& {Habart}}]{Schirmer2020}
{Schirmer}, T., {Abergel}, A., {Verstraete}, L., {et~al.} 2020, \aap, 639, A144

\bibitem[{{Schirmer} {et~al.}(2022){Schirmer}, {Ysard}, {Habart}, {Jones},
  {Abergel}, \& {Verstraete}}]{Schirmer2022}
{Schirmer}, T., {Ysard}, N., {Habart}, E., {et~al.} 2022, \aap, 666, A49

\bibitem[{{Schleuning}(1998)}]{Schleuning1998}
{Schleuning}, D.~A. 1998, \apj, 493, 811

\bibitem[{{Serkowski} {et~al.}(1975){Serkowski}, {Mathewson}, \&
  {Ford}}]{Serkowski1975}
{Serkowski}, K., {Mathewson}, D.~S., \& {Ford}, V.~L. 1975, \apj, 196, 261

\bibitem[{{Sim{\'o}n-D{\'\i}az} {et~al.}(2006){Sim{\'o}n-D{\'\i}az}, {Herrero},
  {Esteban}, \& {Najarro}}]{SimonDiaz2006}
{Sim{\'o}n-D{\'\i}az}, S., {Herrero}, A., {Esteban}, C., \& {Najarro}, F. 2006,
  \aap, 448, 351

\bibitem[{{Soam} {et~al.}(2021){Soam}, {Andersson}, {Acosta-Pulido},
  {L{\'o}pez}, {Vaillancourt}, {Widicus Weaver}, {Piirola}, \&
  {Gordon}}]{Soam2021a}
{Soam}, A., {Andersson}, B.~G., {Acosta-Pulido}, J., {et~al.} 2021, \apj, 907,
  93

\bibitem[{{Tang} {et~al.}(2010){Tang}, {Ho}, {Koch}, \& {Rao}}]{Tang2010}
{Tang}, Y.-W., {Ho}, P.~T.~P., {Koch}, P.~M., \& {Rao}, R. 2010, \apj, 717,
  1262

\bibitem[{{Tazaki} {et~al.}(2017){Tazaki}, {Lazarian}, \&
  {Nomura}}]{Tazaki2017}
{Tazaki}, R., {Lazarian}, A., \& {Nomura}, H. 2017, \apj, 839, 56

\bibitem[{{Tielens} \& {Hollenbach}(1985{\natexlab{a}})}]{Tielens1985a}
{Tielens}, A.~G.~G.~M., \& {Hollenbach}, D. 1985{\natexlab{a}}, \apj, 291, 722

\bibitem[{{Tielens} \& {Hollenbach}(1985{\natexlab{b}})}]{Tielens1985b}
---. 1985{\natexlab{b}}, \apj, 291, 747

\bibitem[{{Tielens} {et~al.}(1993){Tielens}, {Meixner}, {van der Werf},
  {Bregman}, {Tauber}, {Stutzki}, \& {Rank}}]{Tielens1993}
{Tielens}, A.~G.~G.~M., {Meixner}, M.~M., {van der Werf}, P.~P., {et~al.} 1993,
  Science, 262, 86

\bibitem[{{Tram} {et~al.}(2021{\natexlab{a}}){Tram}, {Hoang}, {Lee}, {Santos},
  {Soam}, {Lesaffre}, {Gusdorf}, \& {Reach}}]{Tram2021}
{Tram}, L.~N., {Hoang}, T., {Lee}, H., {et~al.} 2021{\natexlab{a}}, \apj, 906,
  115

\bibitem[{{Tram} {et~al.}(2021{\natexlab{b}}){Tram}, {Lee}, {Hoang}, {Michail},
  {Chuss}, {Nickerson}, {Rangwala}, \& {Reach}}]{Tram2021b}
{Tram}, L.~N., {Lee}, H., {Hoang}, T., {et~al.} 2021{\natexlab{b}}, \apj, 908,
  159

\bibitem[{{Tram} {et~al.}(2021{\natexlab{c}}){Tram}, {Hoang},
  {Lopez-Rodriguez}, {Coud{\'e}}, {Soam}, {Andersson}, {Lee}, {Bonne}, {Vacca},
  \& {Lee}}]{Tram2021c}
{Tram}, L.~N., {Hoang}, T., {Lopez-Rodriguez}, E., {et~al.} 2021{\natexlab{c}},
  \apj, 923, 130

\bibitem[{{Tremblin} {et~al.}(2012){Tremblin}, {Audit}, {Minier}, {Schmidt}, \&
  {Schneider}}]{Tremblin2012}
{Tremblin}, P., {Audit}, E., {Minier}, V., {Schmidt}, W., \& {Schneider}, N.
  2012, \aap, 546, A33

\bibitem[{{Vaillancourt}(2006)}]{Vaillancourt2006}
{Vaillancourt}, J.~E. 2006, \pasp, 118, 1340

\bibitem[{{Vaillancourt} {et~al.}(2020){Vaillancourt}, {Andersson}, {Clemens},
  {Piirola}, {Hoang}, {Becklin}, \& {Caputo}}]{Vaillancourt2020}
{Vaillancourt}, J.~E., {Andersson}, B.~G., {Clemens}, D.~P., {et~al.} 2020,
  \apj, 905, 157

\bibitem[{{Vaillancourt} \& {Matthews}(2012)}]{Vaillancourt2012}
{Vaillancourt}, J.~E., \& {Matthews}, B.~C. 2012, \apjs, 201, 13

\bibitem[{{Vaillancourt} {et~al.}(2007){Vaillancourt}, {Chuss}, {Crutcher},
  {Dotson}, {Dowell}, {Harper}, {Hildebrand}, {Jones}, {Lazarian}, {Novak}, \&
  {Werner}}]{Vaillancourt2007}
{Vaillancourt}, J.~E., {Chuss}, D.~T., {Crutcher}, R.~M., {et~al.} 2007, in
  \procspie, Vol. 6678, Infrared Spaceborne Remote Sensing and Instrumentation
  XV, 66780D

\bibitem[{{Vaillancourt} {et~al.}(2008){Vaillancourt}, {Dowell}, {Hildebrand},
  {Kirby}, {Krejny}, {Li}, {Novak}, {Houde}, {Shinnaga}, \&
  {Attard}}]{Vaillancourt2008}
{Vaillancourt}, J.~E., {Dowell}, C.~D., {Hildebrand}, R.~H., {et~al.} 2008,
  \apjl, 679, L25

\bibitem[{{Van De Putte} {et~al.}(2020){Van De Putte}, {Gordon}, {Roman-Duval},
  {Williams}, {Baes}, {Tchernyshyov}, {Lawton}, \& {Arab}}]{VanDePutte2020}
{Van De Putte}, D., {Gordon}, K.~D., {Roman-Duval}, J., {et~al.} 2020, \apj,
  888, 22

\bibitem[{{Ward-Thompson} {et~al.}(2017){Ward-Thompson}, {Pattle}, {Bastien},
  {Furuya}, {Kwon}, {Lai}, {Qiu}, {Berry}, {Choi}, {Coud{\'e}}, {Di Francesco},
  {Hoang}, {Franzmann}, {Friberg}, {Graves}, {Greaves}, {Houde}, {Johnstone},
  {Kirk}, {Koch}, {Kwon}, {Lee}, {Li}, {Matthews}, {Mottram}, {Parsons}, {Pon},
  {Rao}, {Rawlings}, {Shinnaga}, {Sadavoy}, {van Loo}, {Aso}, {Byun},
  {Eswaraiah}, {Chen}, {Chen}, {Chen}, {Ching}, {Cho}, {Chrysostomou}, {Chung},
  {Doi}, {Drabek-Maunder}, {Eyres}, {Fiege}, {Friesen}, {Fuller}, {Gledhill},
  {Griffin}, {Gu}, {Hasegawa}, {Hatchell}, {Hayashi}, {Holland}, {Inoue},
  {Inutsuka}, {Iwasaki}, {Jeong}, {Kang}, {Kang}, {Kang}, {Kawabata}, {Kemper},
  {Kim}, {Kim}, {Kim}, {Kim}, {Kim}, {Kim}, {Lacaille}, {Lee}, {Lee}, {Li},
  {Li}, {Liu}, {Liu}, {Liu}, {Liu}, {Lyo}, {Mairs}, {Matsumura},
  {Moriarty-Schieven}, {Nakamura}, {Nakanishi}, {Ohashi}, {Onaka}, {Peretto},
  {Pyo}, {Qian}, {Retter}, {Richer}, {Rigby}, {Robitaille}, {Savini}, {Scaife},
  {Soam}, {Tamura}, {Tang}, {Tomisaka}, {Wang}, {Wang}, {Whitworth}, {Yen},
  {Yoo}, {Yuan}, {Zhang}, {Zhang}, {Zhou}, {Zhu}, {Andr{\'e}}, {Dowell},
  {Falle}, \& {Tsukamoto}}]{WardThompson2017}
{Ward-Thompson}, D., {Pattle}, K., {Bastien}, P., {et~al.} 2017, \apj, 842, 66

\bibitem[{{Wardle} \& {Kronberg}(1974)}]{WardleKronberg1974}
{Wardle}, J.~F.~C., \& {Kronberg}, P.~P. 1974, \apj, 194, 249

\bibitem[{{Weingartner} \& {Draine}(2001)}]{Weingartner2001b}
{Weingartner}, J.~C., \& {Draine}, B.~T. 2001, \apjs, 134, 263

\bibitem[{{Whittet} \& {van Breda}(1978)}]{Whittet1978}
{Whittet}, D.~C.~B., \& {van Breda}, I.~G. 1978, \aap, 66, 57

\bibitem[{{Wolfire} {et~al.}(2022){Wolfire}, {Vallini}, \&
  {Chevance}}]{Wolfire2022}
{Wolfire}, M.~G., {Vallini}, L., \& {Chevance}, M. 2022, \araa, 60, 247

\bibitem[{{Zubko} {et~al.}(2004){Zubko}, {Dwek}, \& {Arendt}}]{Zubko2004}
{Zubko}, V., {Dwek}, E., \& {Arendt}, R.~G. 2004, \apjs, 152, 211

\end{thebibliography}
\bibliographystyle{apj}

\newpage
\clearpage
\newpage

% \begin{appendix}
\appendix
\addcontentsline{toc}{section}{Appendix}
\renewcommand{\thesection}{\Alph{section}}

\section{\normalfont{A. Chop-nod and OTFMAP modes comparisons}}
\label{app:c2n_vs_otfmap}

\begin{figure*}[!tbh]
\centering
\subfigure{
\includegraphics[scale=0.44,clip,trim= 0.5cm 1.5cm 0cm 0cm]{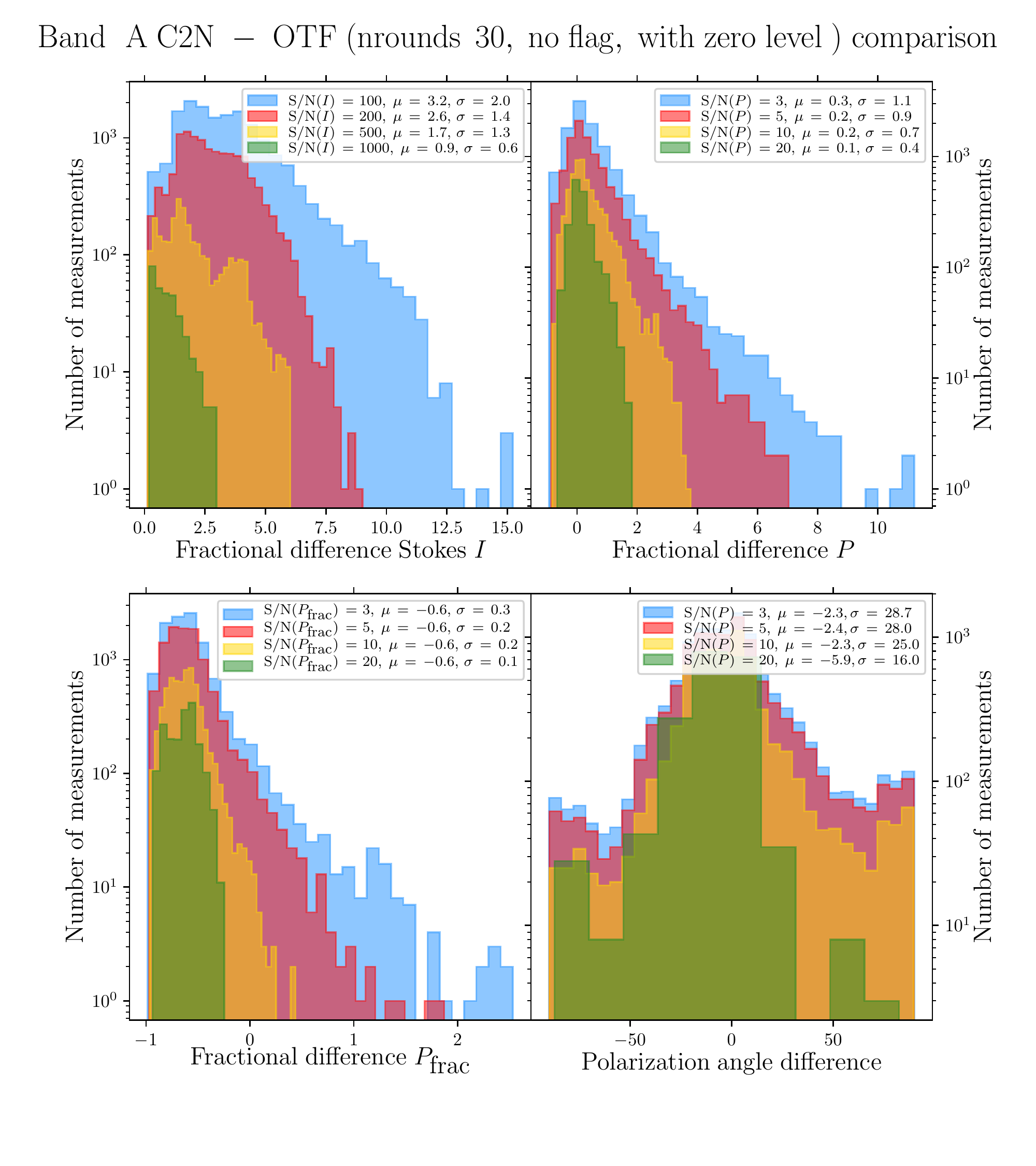}}
\subfigure{
\includegraphics[scale=0.44,clip,trim= 0.5cm 1.5cm 0cm 0cm]{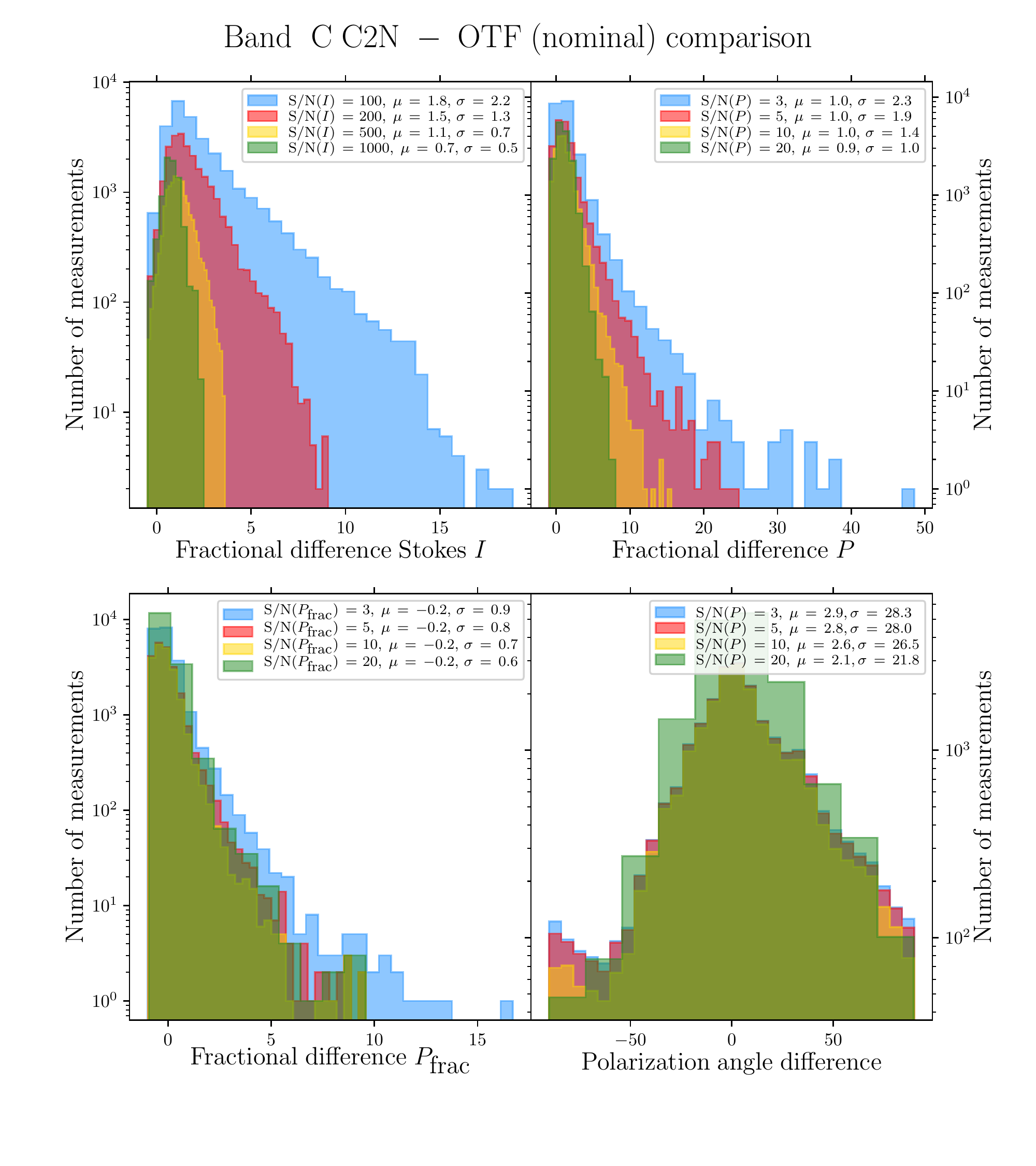}}
\subfigure{
\includegraphics[scale=0.44,clip,trim= 0.5cm 1.5cm 0cm 0cm]{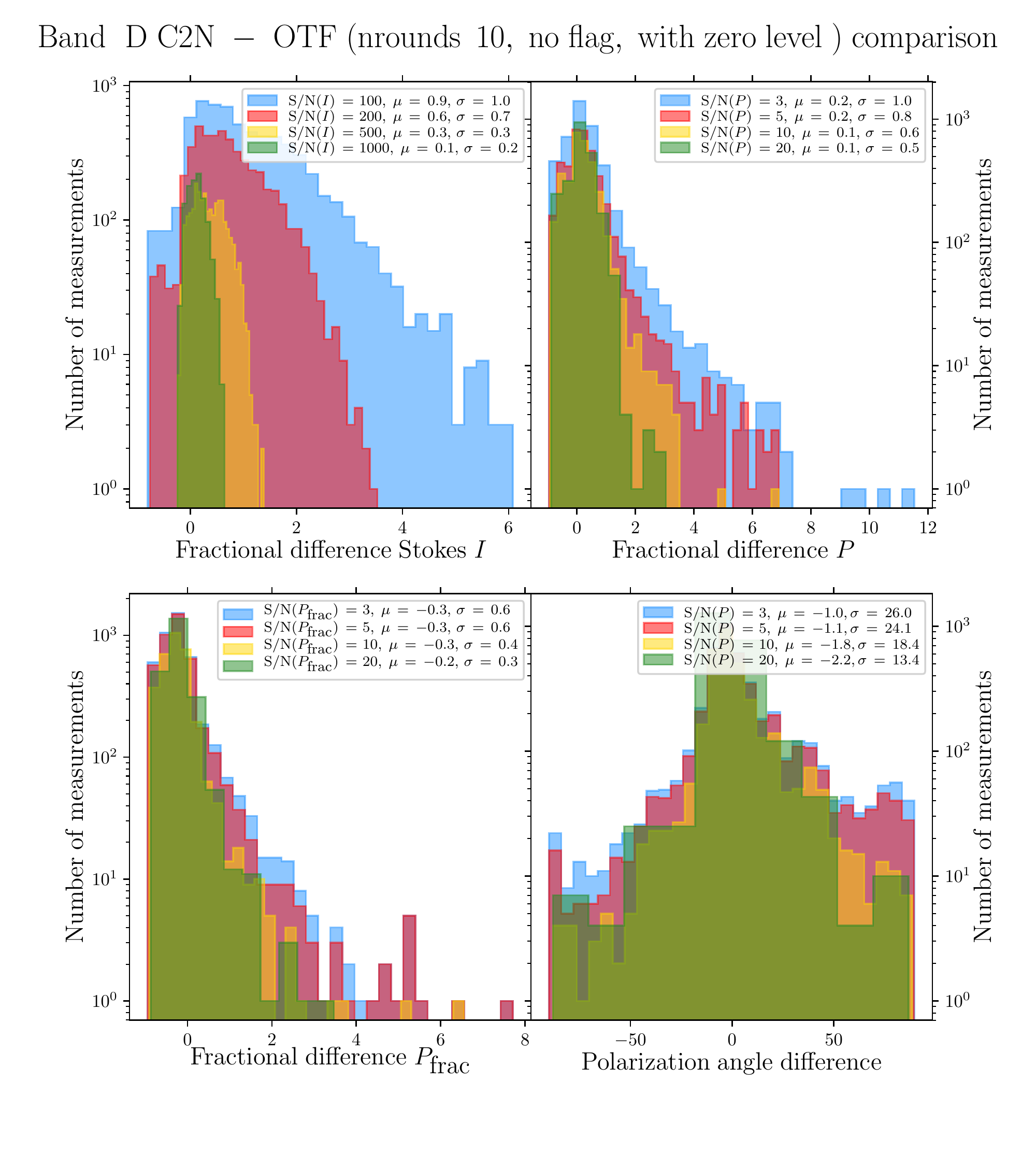}}
\subfigure{
\includegraphics[scale=0.44,clip,trim= 0.5cm 1.5cm 0cm 0cm]{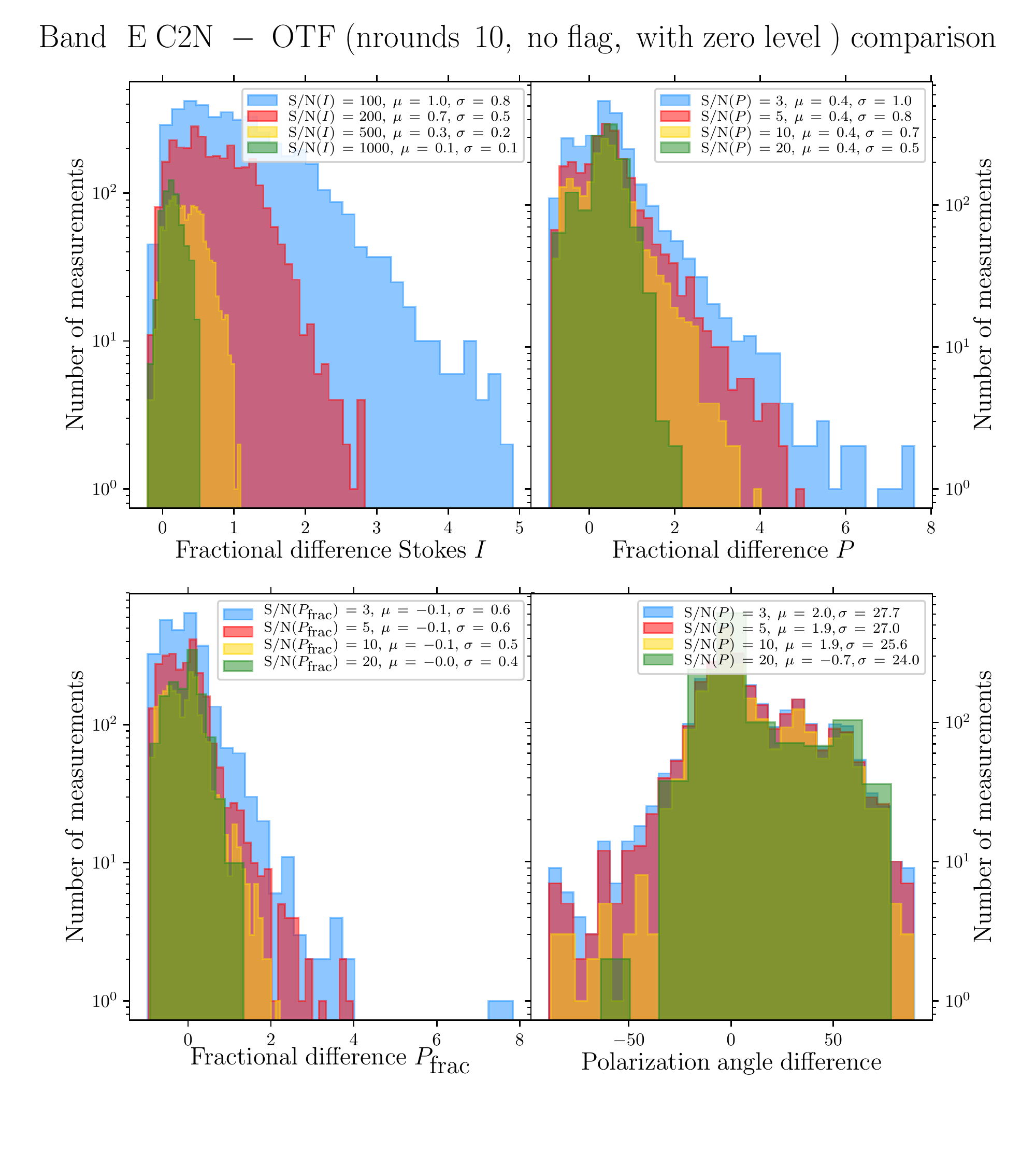}}
\caption{\small Comparisons of the the dust polarization quantities between the C2N and OTFMAP mode, for the Band A (\textit{top-left} panel), C (\textit{top-right} panel), D (\textit{bottom-left} panel), and E observations (\textit{bottom-right} panel). In each panel, distributions of the fractional difference of Stokes $I$, $P$, $\Pf$, and $\phi$ are shown. Each color corresponds to the SNR criteria specified in the legend along with the mean ($\mu$) and the standard deviation ($\sigma$) of the corresponding distribution.
}
\label{fig:c2n_otf_comp}
\end{figure*}

\begin{figure*}[!tbh]
\centering
\subfigure{
\includegraphics[scale=0.44,clip,trim= 0.5cm 1.5cm 0cm 0cm]{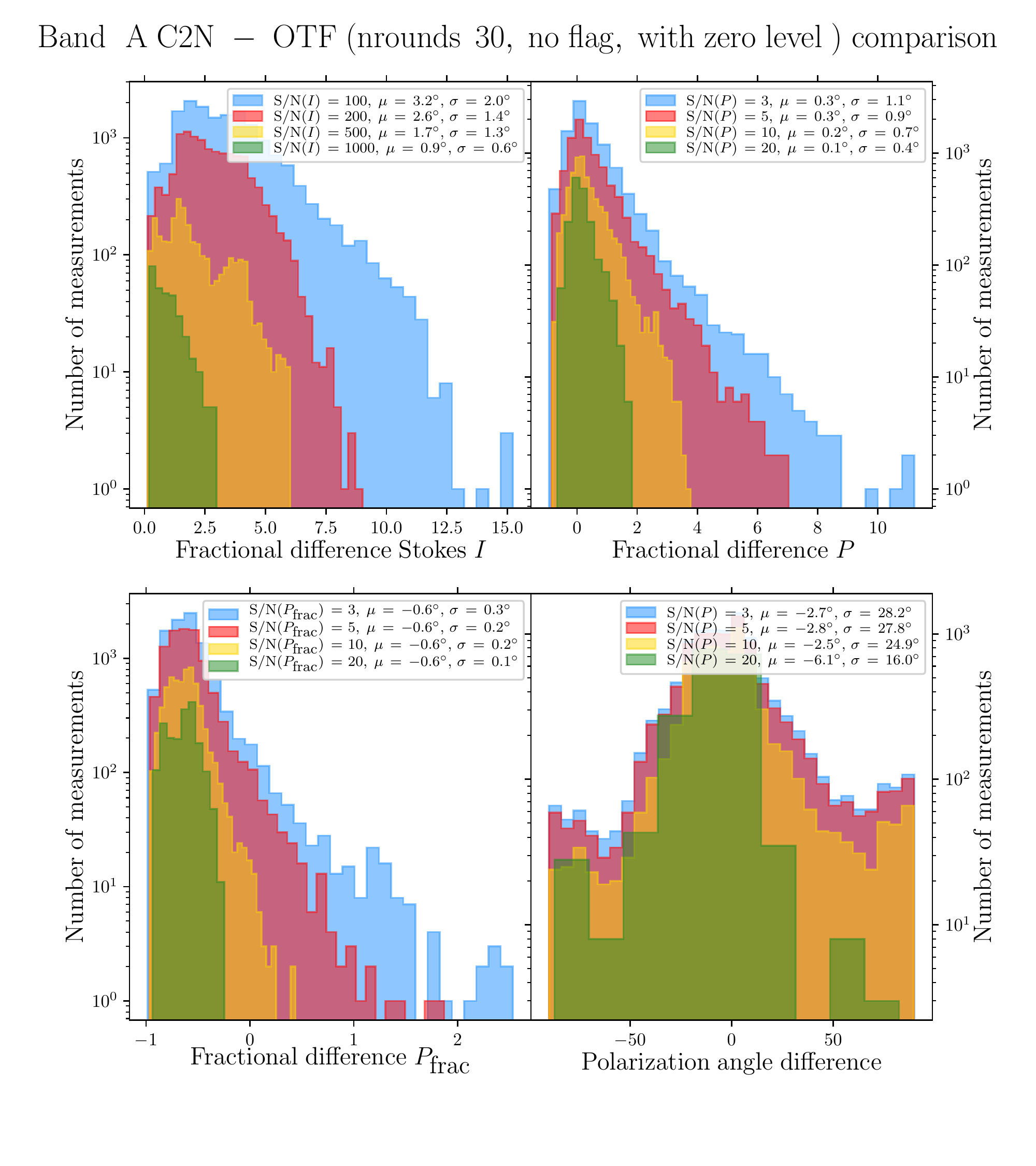}}
\subfigure{
\includegraphics[scale=0.44,clip,trim= 0.5cm 1.5cm 0cm 0cm]{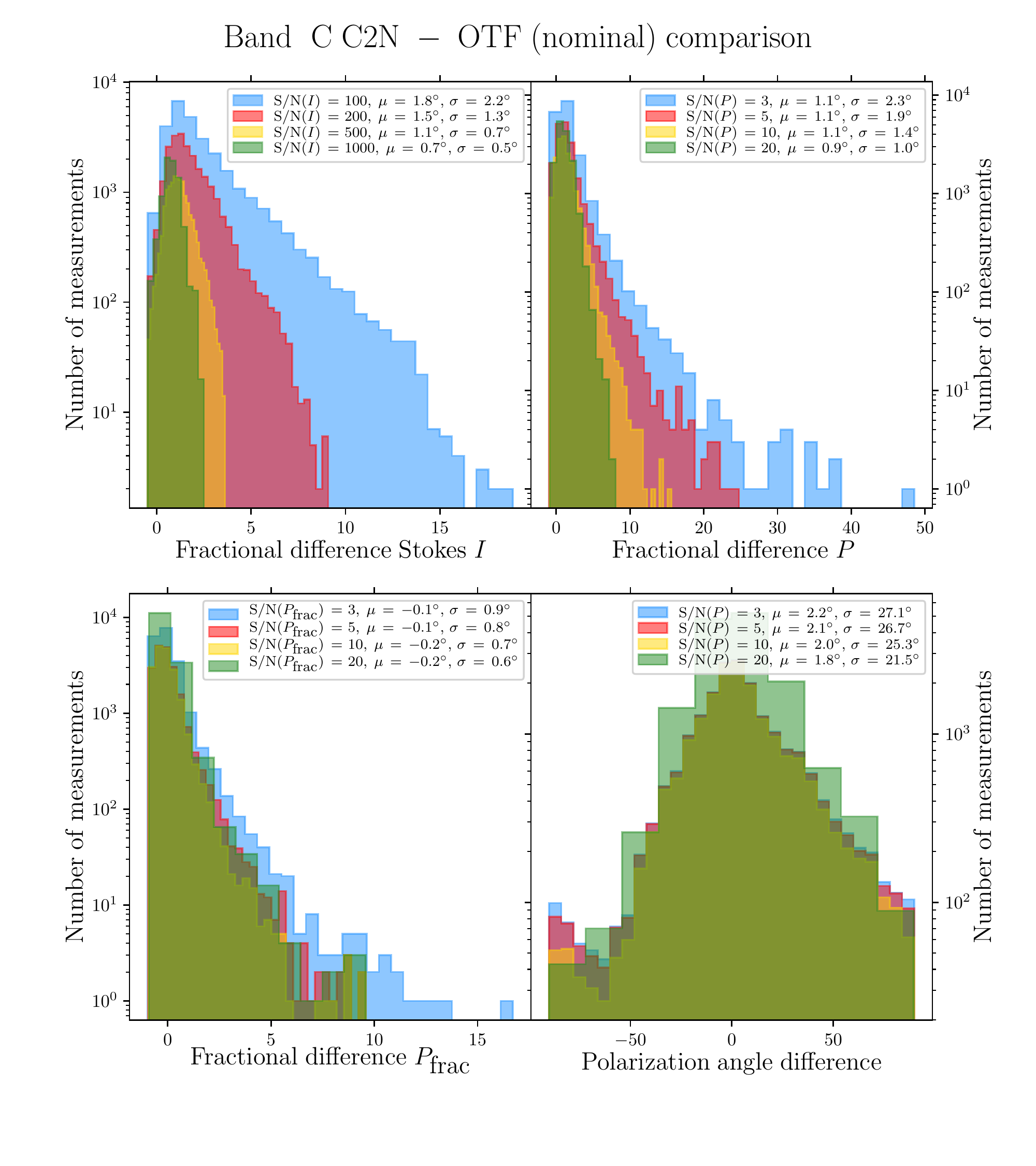}}
\subfigure{
\includegraphics[scale=0.44,clip,trim= 0.5cm 1.5cm 0cm 0cm]{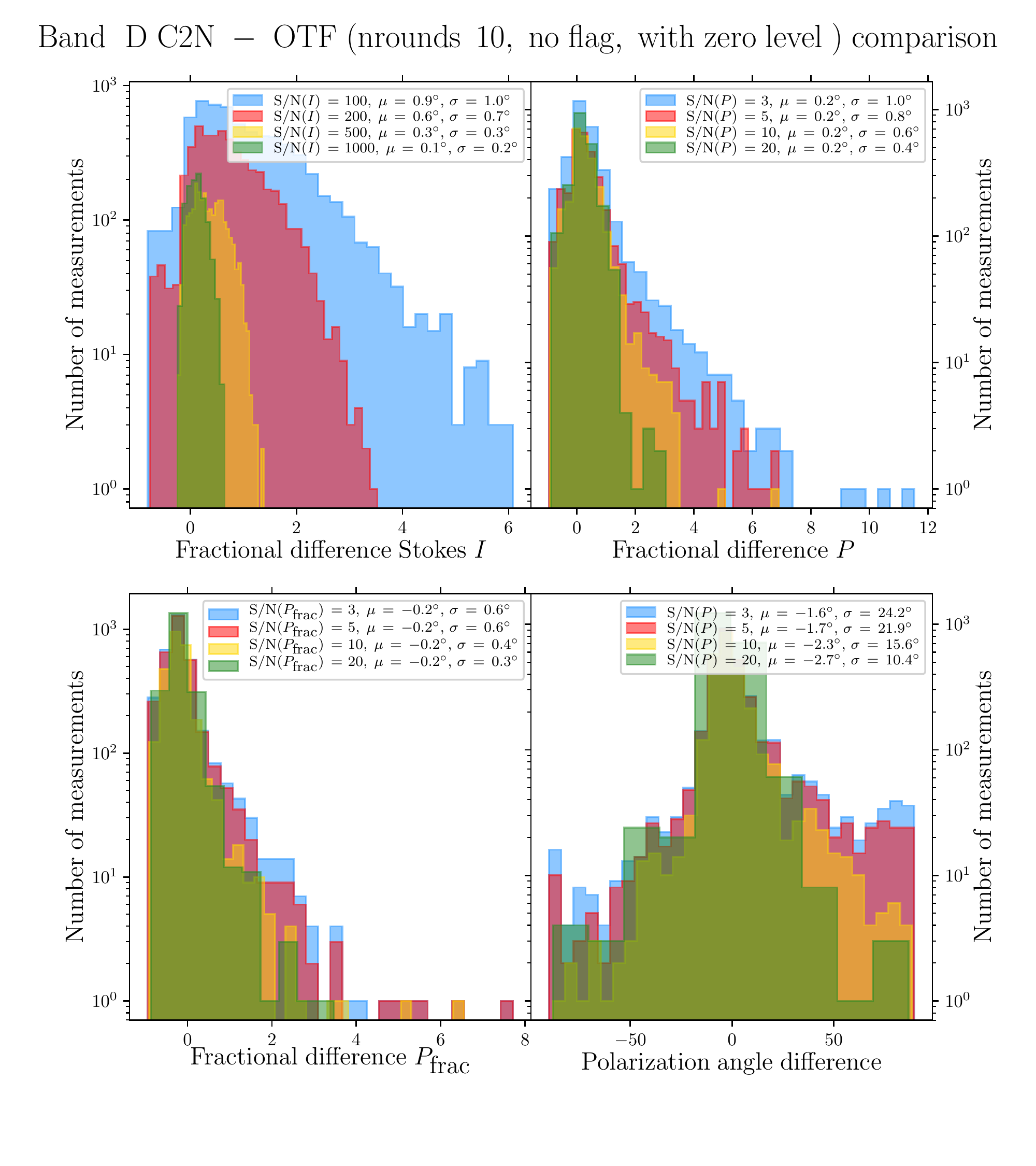}}
\subfigure{
\includegraphics[scale=0.44,clip,trim= 0.5cm 1.5cm 0cm 0cm]{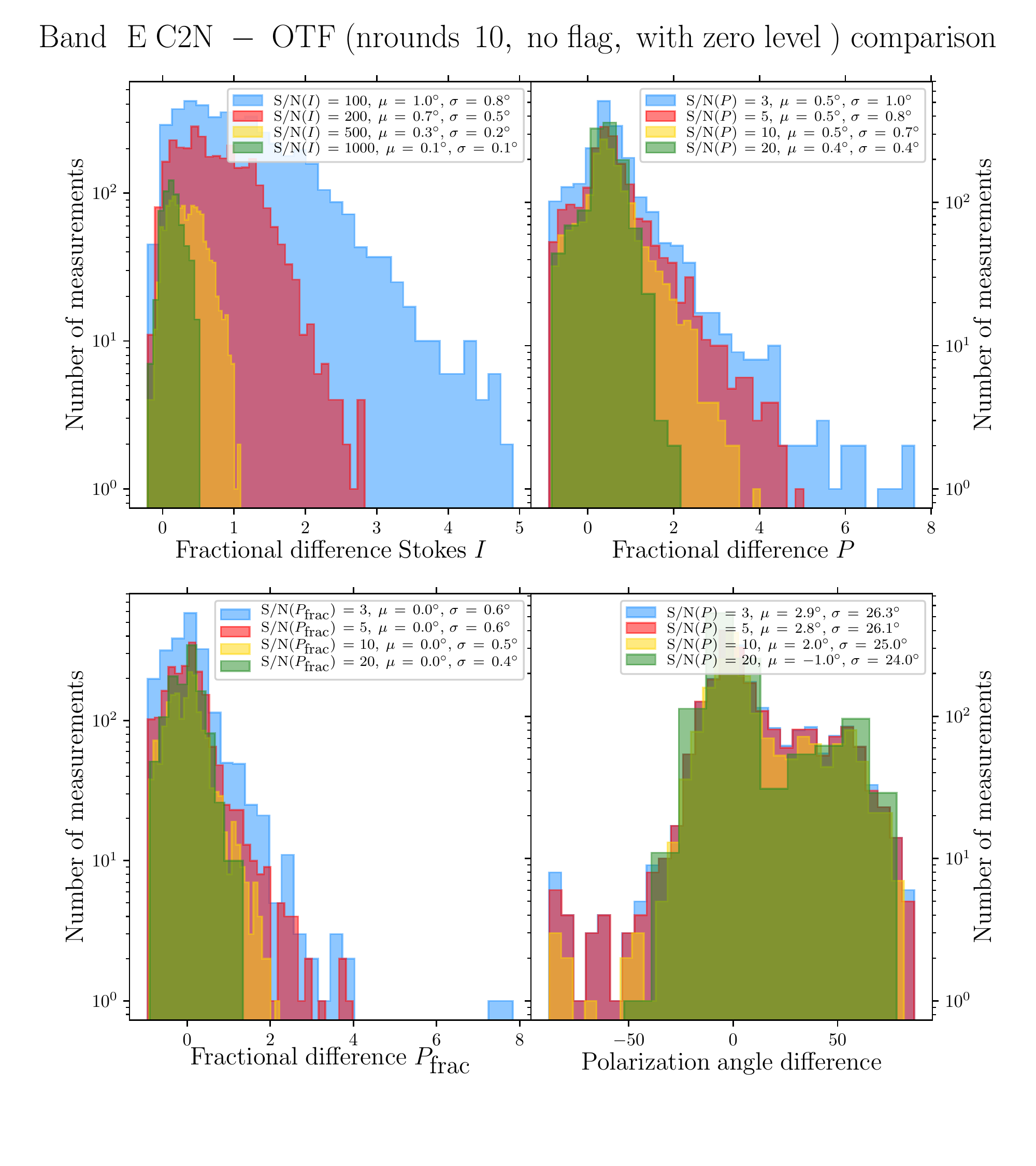}}
\caption{\small Same as Figure \ref{fig:c2n_otf_comp} but using the masks of Figure 10 in \citet{Chuss2019} for the polarization quantities. These masks flag all pixels whose off-beam polarization affects the polarization angles more than 10$^{\circ}$.
}
\label{fig:c2n_otf_comp_offbeam}
\end{figure*}

Figure \ref{fig:c2n_otf_comp} presents quantitative comparisons between the polarization quantities Stokes $I$, $P$, $\Pf$, and $\phi$, obtained with OTFMAP and the C2N HAWC+ polarimetric mode, at the four wavelengths used in this paper. The same comparisons are shown in Figure \ref{fig:c2n_otf_comp_offbeam}, but these new histograms take into account the off-beam contamination corrections \citep{Novak1997} performed in the Section 3.4 of \citet{Chuss2019}.
% We plot the distributions of fractional difference (doing C2N-OTFMAP/OTFMAP) for Stokes $I$, $P$, and $\Pf$, and the distributions of polarization angle difference.

In all four bands, we analyze how does the fractional difference distribution evolve with increasing SNR criteria. The fractional difference distributions of $P$ and Stokes $I$ narrower for increasing SNR criteria (from 100 to 1000 for Stokes $I$, and from 3 to 20 for $P$). 
For the fractional difference distributions of $\phi$ and $\Pf$, the increase of the SNR criteria also decreases the dispersion of the distributions, but this decrease is peculiarly significant on the Band A data.
This suggests that the quality of the Band A OTFMAP observations is lower.
From the shape of the distributions, one can check if any distributions present a clear departure from zero with a statistical significance, and with sufficiently high SNR criteria. This would suggest a systematic over(under)-estimation of a given polarization quantity. Among all the distributions, only the fractional difference distribution of polarization fraction values in Band A data present a statistical significant offset from zero with a mean value $|\mu|\,\geq\,2\sigma$ (where $\sigma$ is the standard deviation) for all SNR criteria.
Finally, the distributions presented in Figure \ref{fig:c2n_otf_comp_offbeam} are on average slightly narrower with the additional flag on the C2N data that select out pixels with too high off-beam contamination .

\section{\normalfont{B. Additional dust polarization observations plots}}
\label{app:add_pol_plots}

\begin{figure*}[!tbh]
\centering
\subfigure{
\includegraphics[scale=0.87,clip,trim= 0.2cm 5.35cm 10cm 4.5cm]{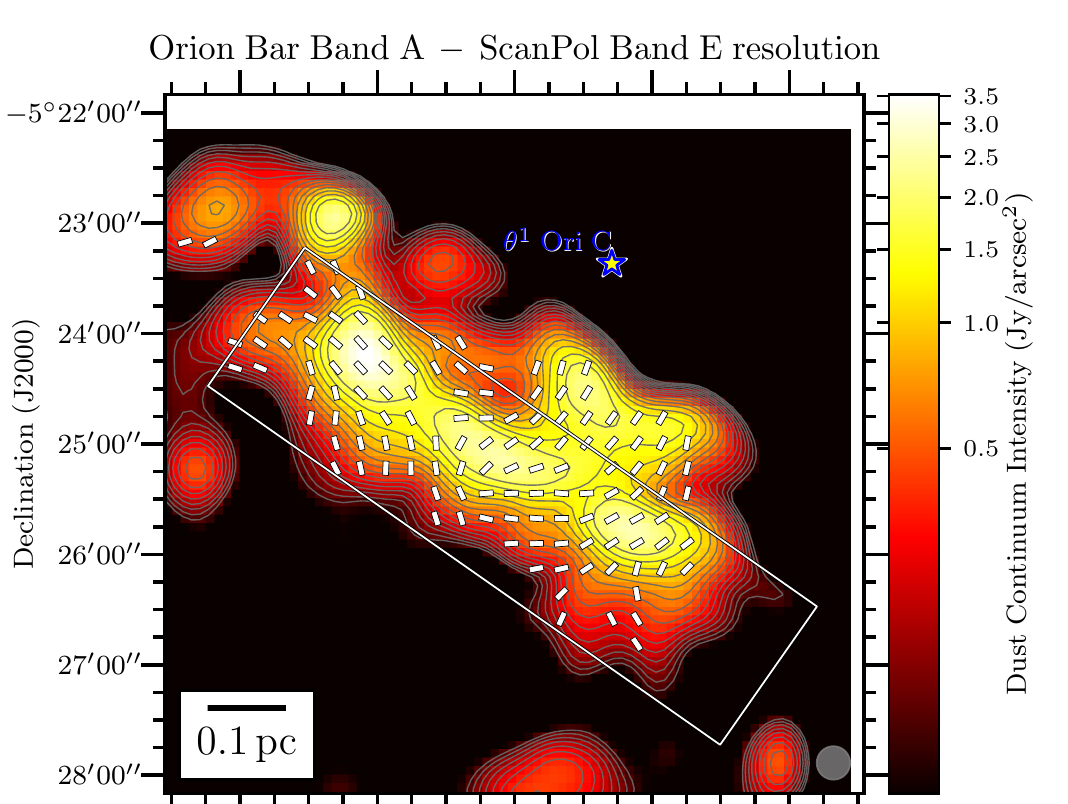}}
\subfigure{
\includegraphics[scale=0.87,clip,trim= 2cm 5.34cm 10cm 4.5cm]{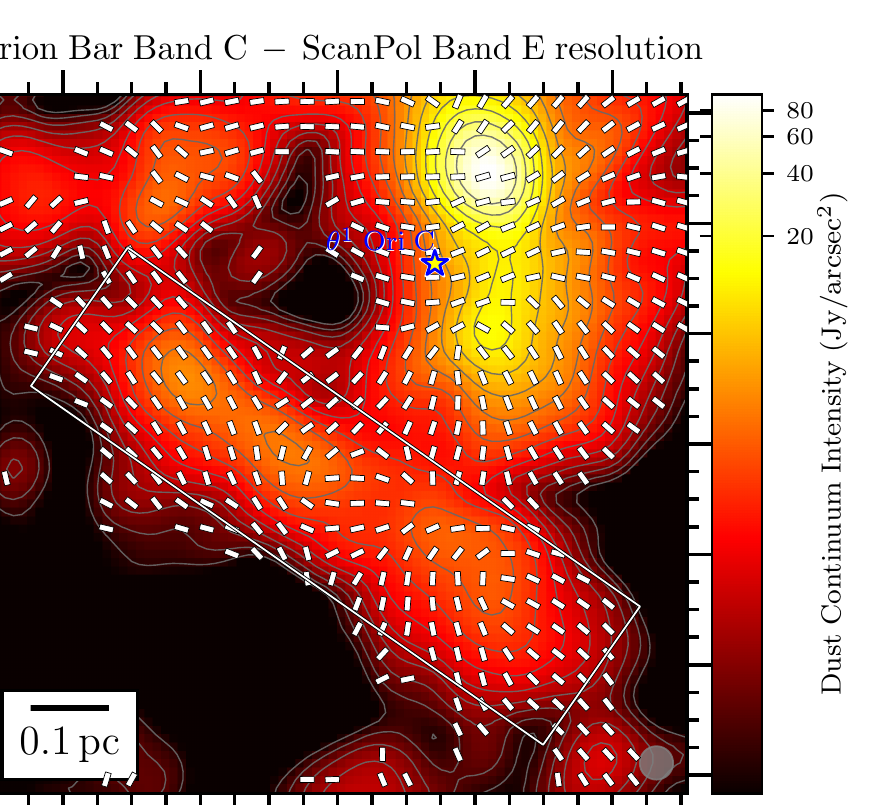}}
\subfigure{
\includegraphics[scale=0.87,clip,trim= 0.2cm 4.5cm 10cm 4.5cm]{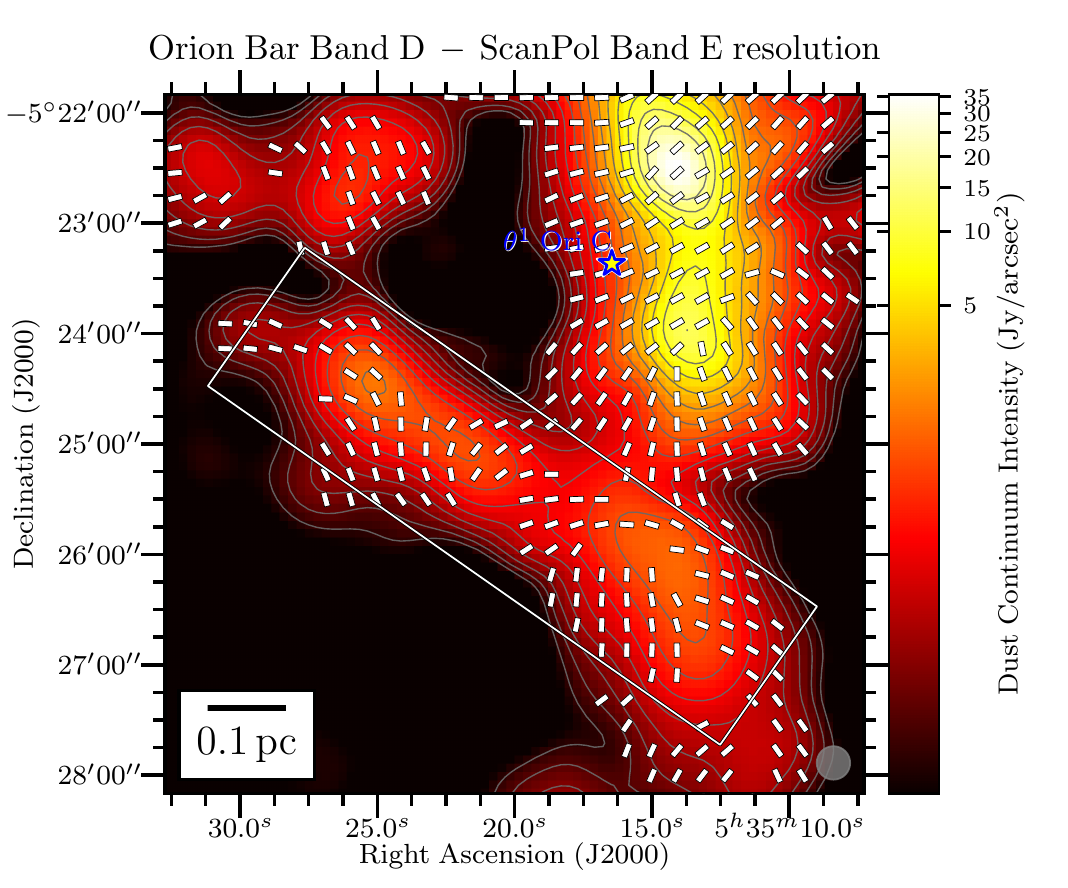}}
\subfigure{
\includegraphics[scale=0.87,clip,trim= 2cm 4.5cm 10cm 4.5cm]{plots_obs/scanpol/BandE/F0686_HA_POL_0802093_HAWEHWPE_PMP_169-172cont_pola.pdf}}
\caption{\small Same as Figure \ref{fig:obs_bar_4bands}, but with the Band A, C, and D observations smoothed and regridded at the Band E resolution and pixel size.}
\label{fig:obs_bar_4bands_resE}
\end{figure*}

\begin{figure*}[!tbh]
\centering
\subfigure{
\includegraphics[scale=0.625,clip,trim= 0.2cm 3.5cm 7cm 2.5cm]{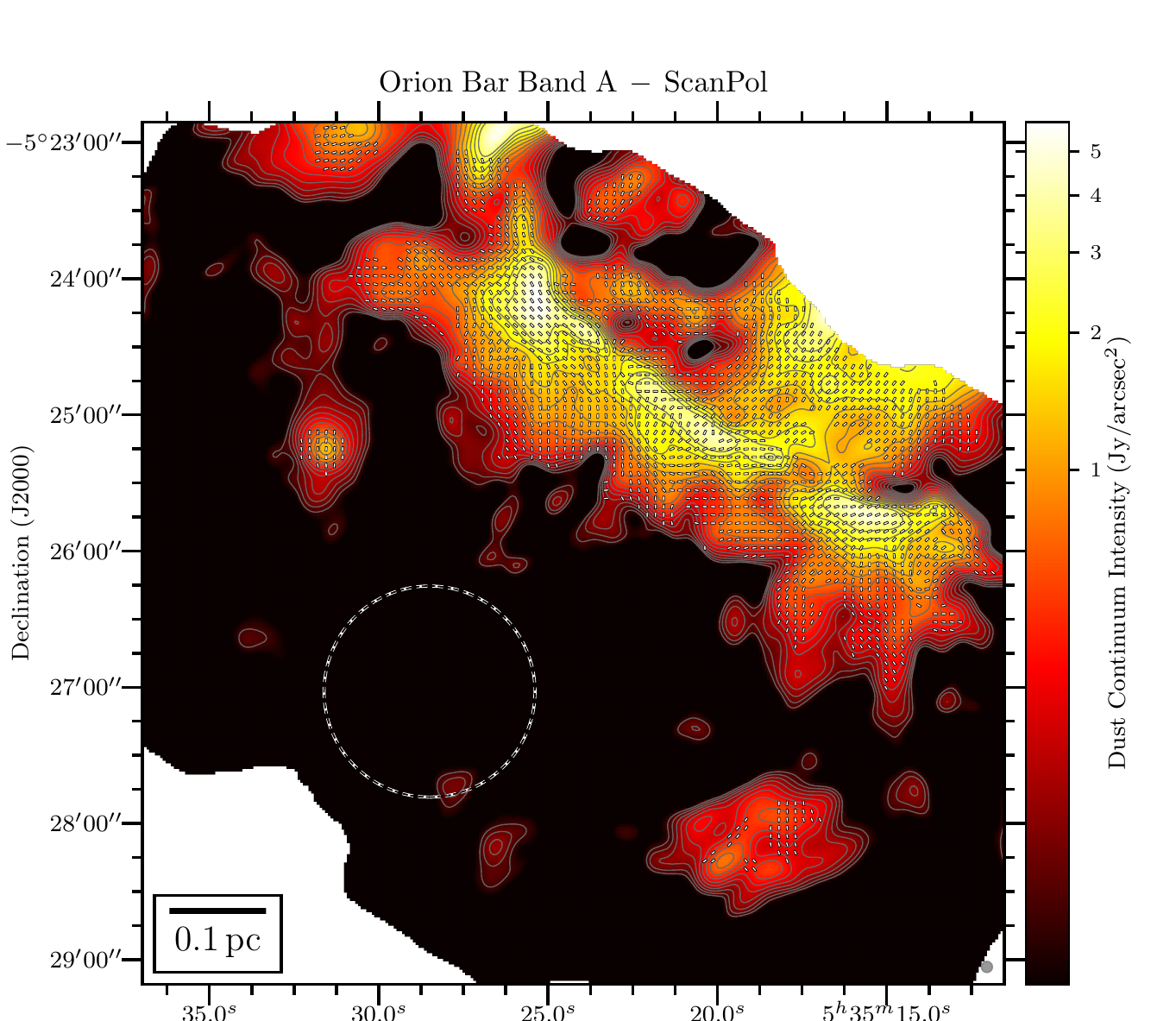}}
\subfigure{
\includegraphics[scale=0.625,clip,trim= 0.2cm 3.5cm 7cm 2.5cm]{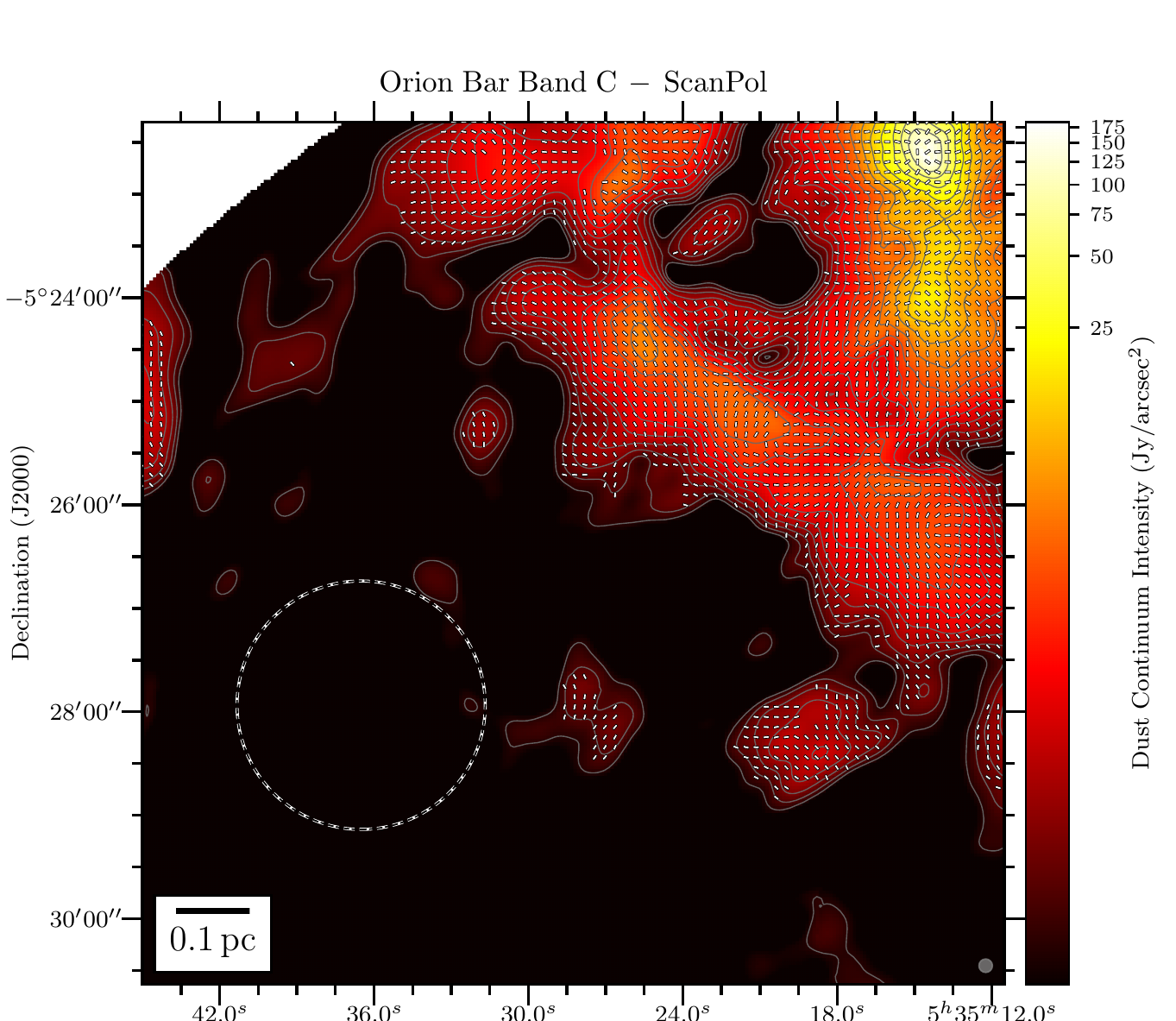}}
\subfigure{
\includegraphics[scale=0.625,clip,trim= 0.2cm 3.5cm 7cm 2.5cm]{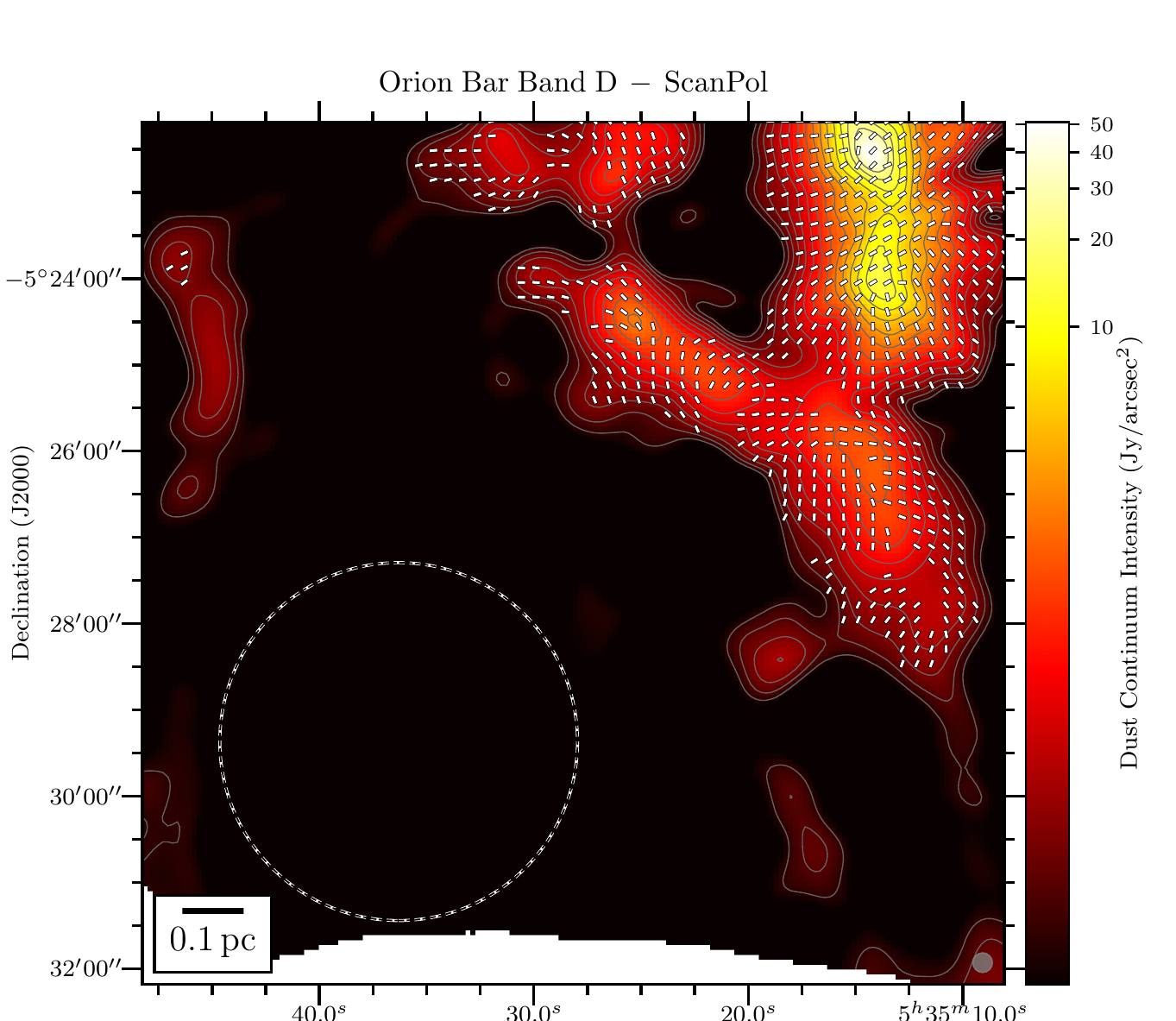}}
\subfigure{
\includegraphics[scale=0.625,clip,trim= 0.2cm 3.5cm 7cm 2.5cm]{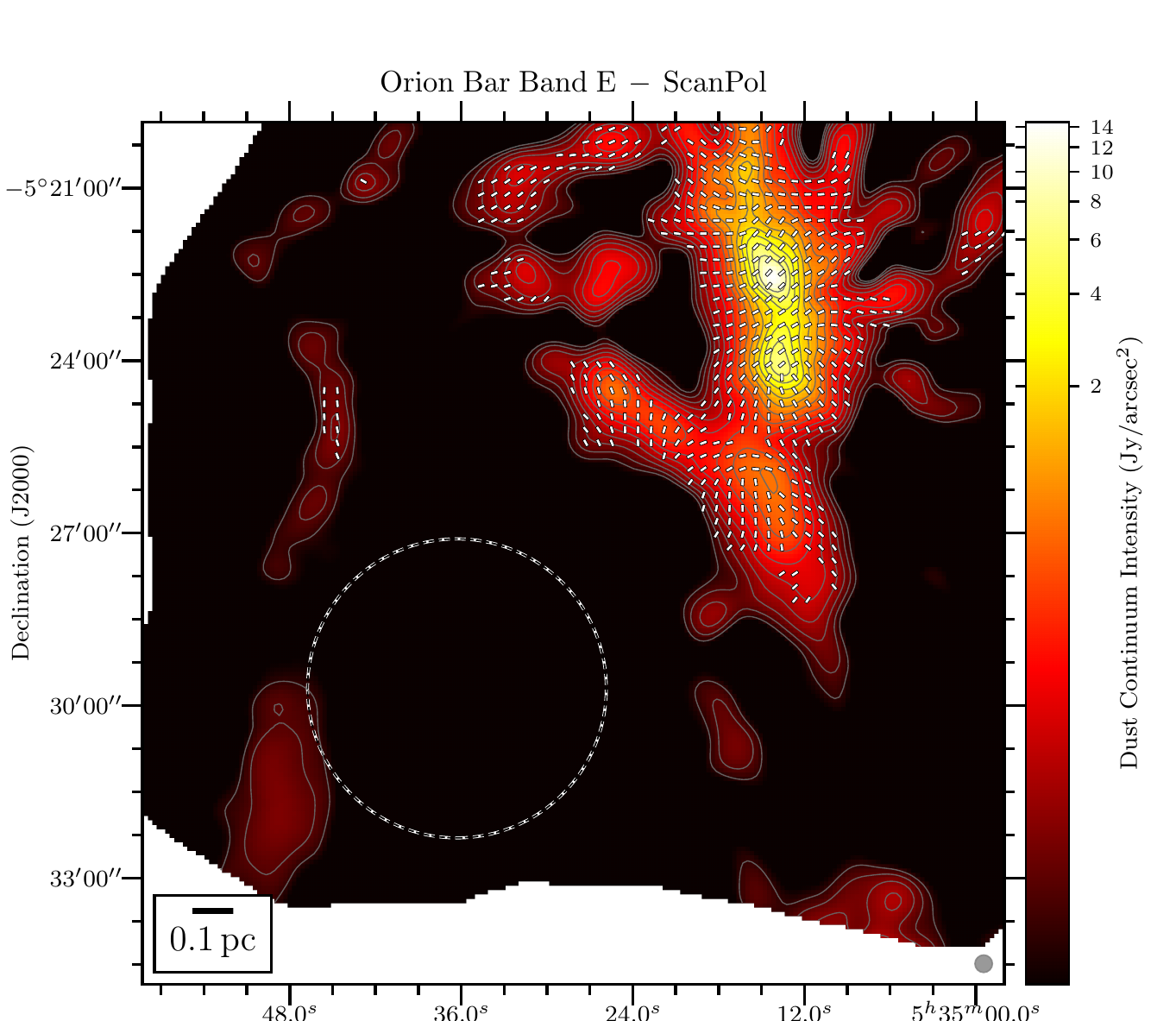}}
% \vspace{-0.1cm}
\caption{\small Same as Figure \ref{fig:obs_bar_4bands}. We show with a dashed gray circle the region taken to perform the zero level background correction following \citet{LiPS2022,LopezRodriguez2022}.}
\label{fig:obs_bar_4bands_zero_lvl_zone}
\vspace{0.2cm}
\end{figure*}

Figure \ref{fig:obs_bar_4bands_resE} presents the four HAWC+ observations shown in Figure \ref{fig:obs_bar_4bands}, but smoothed and regridded to the Band E angular resolution and pixel size, using a flux conserving algorithm. These maps are the one used throughout Section \ref{sec:results} and \ref{sec:disc} of this paper. We also show in Figure \ref{fig:obs_bar_4bands_zero_lvl_zone} the region taken to perform the zero level background correction in each band (see Section \ref{sec:obs}). The location of this region is consistent for Band C, D, and E, which correspond to low average emission in the 100, 160, and 250 $\mu$m Herschel maps. For Band A we were limited by the field of view of the observations.

\section{\normalfont{C. Can mechanical torques contribute to the alignment of dust grains?}}
\label{app:METs}

\begin{figure*}[!tbh]
\centering
\subfigure{
\includegraphics[scale=0.63,clip,trim= 0.3cm 0cm 1cm 0.5cm]{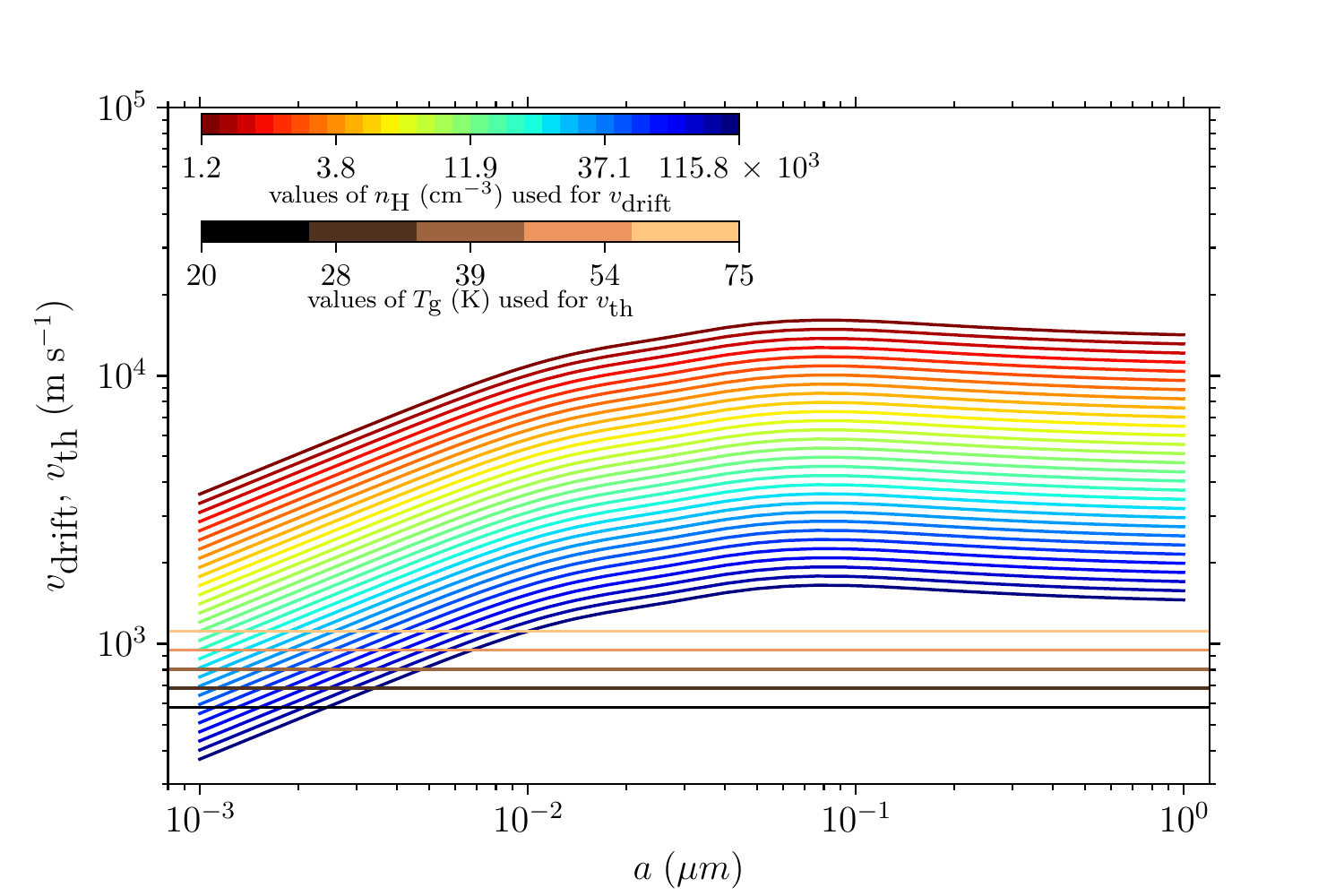}}
\subfigure{
\includegraphics[scale=0.63,clip,trim= 0.3cm 0cm 1cm 0.5cm]{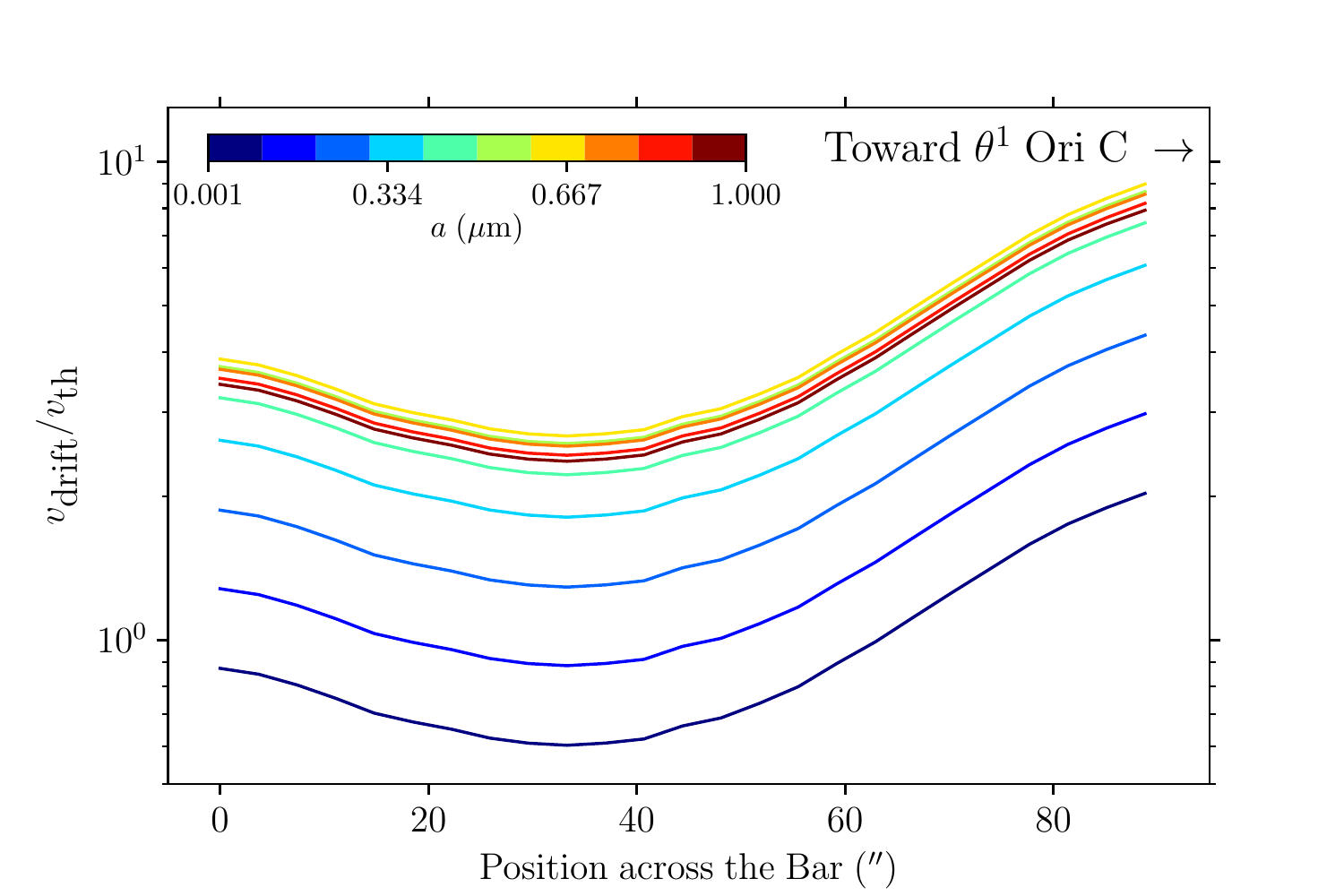}}
\caption{\small Gas-dust drift and thermal velocities derived with the environmental conditions of the Orion Bar. \textit{Left panel:} The gas-dust drift (thermal) velocities are shown with rainbow colors (tints of orange) for a range of gas volume density (gas temperature) values encountered in the Orion Bar. We estimate the gas-dust drift using the Equation 19 of \citet{Schirmer2022}. \textit{Right panel:} Evolution of gas-dust drift to thermal velocity ratio as a function of the position across the Orion Bar (we used the map of $n_\textrm{H}$ and $T_\textrm{d}$ we derived from \citet{Chuss2019}) for several grain sizes, indicated by the different colors.}
\label{fig:vdrift}
\end{figure*}

We now discuss the potential contribution that Mechanical Torques (METs) can have on the alignment of grains within the Orion Bar. Similarly to RATs, METs consist in the torque applied to dust grains induced by an anisotropic flow of colliding gas particles \citep{LazarianHoang2007}. Numerically tested in \citet{Hoang2018,Reissl2022}, the MET mechanism can contribute to the alignment of grains with the magnetic field (or even with the gas flow direction for high enough velocity), if internal alignment is ensured. There is, as of yet, no direct observational evidence for METs, given that is it highly degenerate to measure separately the orientation of magnetic field lines and gas-dust drift velocity. However, we can estimate the drift velocities of aggregates in the Orion Bar driven by the radiation pressure using a one-dimensional approach. We follow the method developed in Section 6.3 of \citet{Schirmer2022}, which consists in deriving the asymptotic drift velocity $v_{\textrm{drift}}$ of a grain subject to a radiative pressure force (induced by radiation field spectrum of \ThOriC), a drag force induced by gaseous collisions, and a gravitational force (assuming the local gravitational potential is governed by the Trapezium cluster). The left panel of Figure \ref{fig:vdrift} shows the drift velocities obtained for the range of gas volume density values encountered in the Orion Bar (using the density values derived above from \citealt{Chuss2019}; see Section \ref{sec:Pf_vs_Td_NH2}) as a function of grain sizes. We also show the thermal velocities $v_{\textrm{th}}$ (\ie the typical velocity of gas particles, with $v_{\textrm{th}}=\sqrt{2 k_{\textrm{B}} T_{\textrm{g}}/m_{\textrm{H}}}$) for the range of temperature values of Bar (assuming here $T_{\textrm{d}}\,\approx\,T_{\textrm{g}}$). We compute the ratio $v_{\textrm{drift}}/v_{\textrm{th}}$ for every pixel of the Bar, and show in the right panel of Figure \ref{fig:vdrift} the average evolution of the $v_{\textrm{drift}}/v_{\textrm{th}}$ ratio as a function of the position along the minor axis of the Bar, directed toward \ThOriC, for different grain sizes. On the irradiated side of the Bar, we reach $v_{\textrm{drift}}/v_{\textrm{th}}$ values of $\sim\,3-8$ for grains of $0.01-1\,\mu$m in size, and values of $\sim\,0.6-3$ for grains of $\leq\,0.01\,\mu$m in size.

These calculations are, however, subject to several caveats. We assume that the drag force is totally anisotropic and entirely due to gas collisions. The plasma drag and Lorentz forces are thus not taken into account, even though they are expected to play an important role in the development of instabilities, that decouple gas from dust (see the work of \citep{Hopkins2022} for HII regions). Additionally, because of its spatial resolution, the map of gas volume density used in our study does not reflect the presence of high density clumps ($n_{\textrm{H}}\,=\,10^{6}-10^{7} \;\textrm{cm}^{-3}$; see \citealt{Lis2003,AndreeLabsch2017,Habart2022}). Such high density layers should be efficient at reprocessing the radiation field from the Trapezium cluster. Indeed, the subsequent modification and attenuation of the spectrum of the heating source (taken as the blackbody of \ThOriC in the equation) can significantly decrease the efficiency of the radiative pressure force, which in turn can decrease the gas-dust drift velocity. Therefore, the drift velocities derived here are most likely overestimates, especially for the values passed the dissociation front.

\begin{figure}[!tbh]
\centering
\includegraphics[scale=0.65,clip,trim= 0.1cm 0.1cm 1cm 0.5cm]{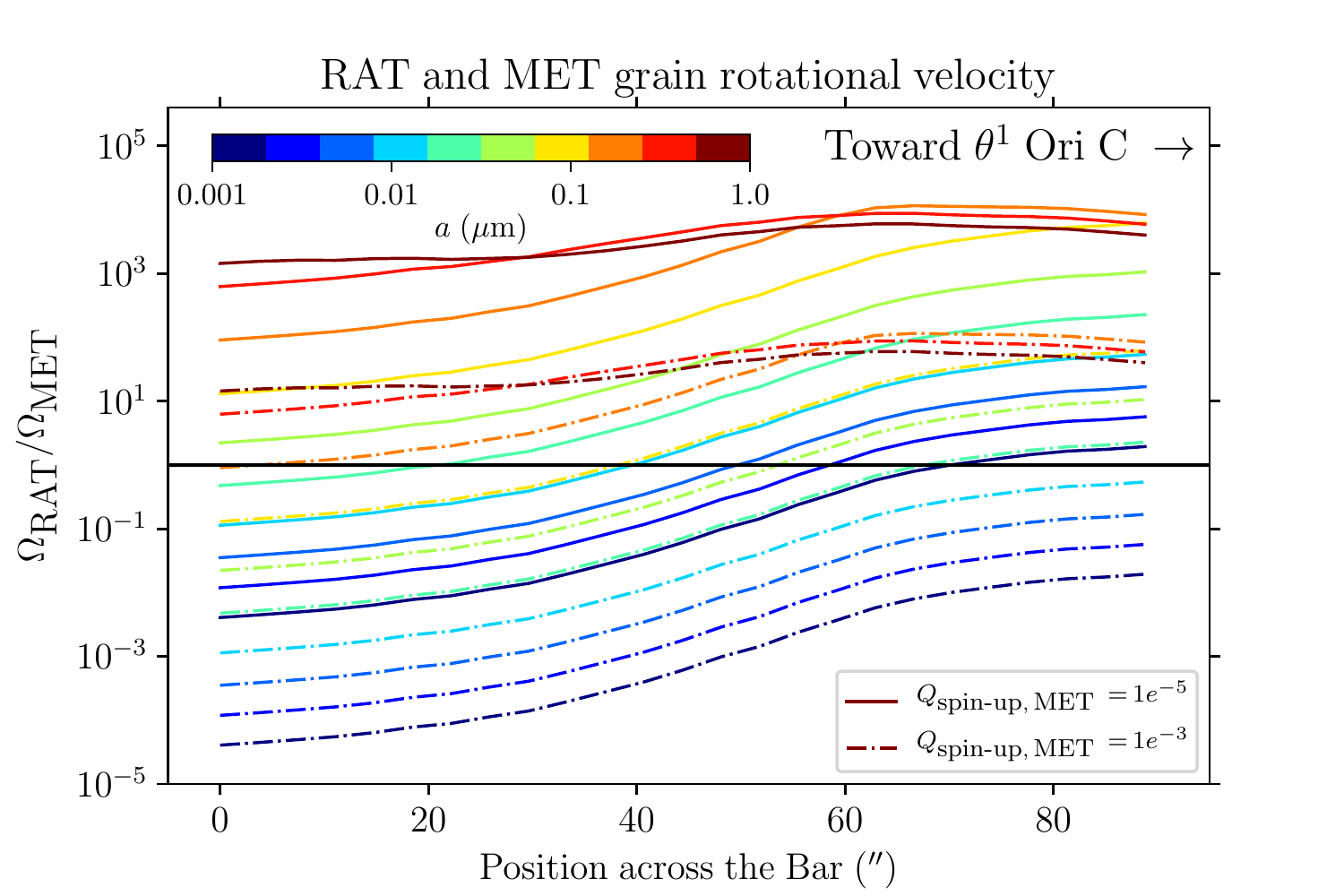}
\caption{\small Evolution of grain rotational velocity ratio $\Omega_{\textrm{RAT}}/\Omega_{\textrm{MET}}$ as a function of the position across the Orion Bar. The color of the lines correspond to different grain sizes. Solid (dot-dashed) lines correspond to grain rotational velocity ratio calculated for a MET spin-up efficiency of $Q_{\textrm{spin-up},\,\textrm{MET}}\,=\,10^{-5}$ ($10^{-3}$). The horizontal black line denotes the location where $\Omega_{\textrm{RAT}}\,=\,\Omega_{\textrm{MET}}$.
}
\label{fig:omega_RAT/MET}
\end{figure}

In order to determine which of the two alignment mechanisms between RATs and METs dominate within the Orion Bar, we show in Figure \ref{fig:omega_RAT/MET} the ratio of the respective grain rotational velocities $\Omega_{\textrm{RAT}}/\Omega_{\textrm{MET}}$ as a function of the position across the Orion Bar, for different grain sizes. We use Equations of Section 3.2 and 3.4 of \citet{Hoang2022b}, with our derivation of the gas-dust drift velocities presented in Figure \ref{fig:vdrift}. We treat the spin-up efficiency of METs as a free parameter, and use explore the range $Q_{\textrm{spin-up},\,\textrm{MET}}\,=\,10^{-5}-10^{-3}$, given the numerical calculations of \citet{Hoang2018,Reissl2022}. We find that the deeper into the Orion Bar, the more important the contribution of METs is, due to the strong attenuation of the radiation field. However, this is where we trust our derivation of the gas-dust drift velocities the least, for the reasons discussed above. Close to the irradiated edge of the Bar, RATs dominate if we assume a low value of MET spin-up efficiency, \ie $10^{-5}$, and METs can contribute to the alignment of grains $\leq\,0.01\,\mu$m in size if we assume $Q_{\textrm{spin-up}}\,=\,10^{-3}$. In summary, the degeneracies generated by the estimation of the gas-dust drift velocities and the MET spin-up efficiencies make prediction of the role played by MET challenging. Staying conservative with those degeneracies, we do not predict that METs are dominant in the Orion Bar PDR.

\section{\normalfont{D. Starlight Polarization}}
\label{app:starlight}

\begin{figure*}[!tbh]
\centering
\subfigure{
\includegraphics[scale=0.848,clip,trim= 0.2cm 5.5cm 13.4cm 6cm]{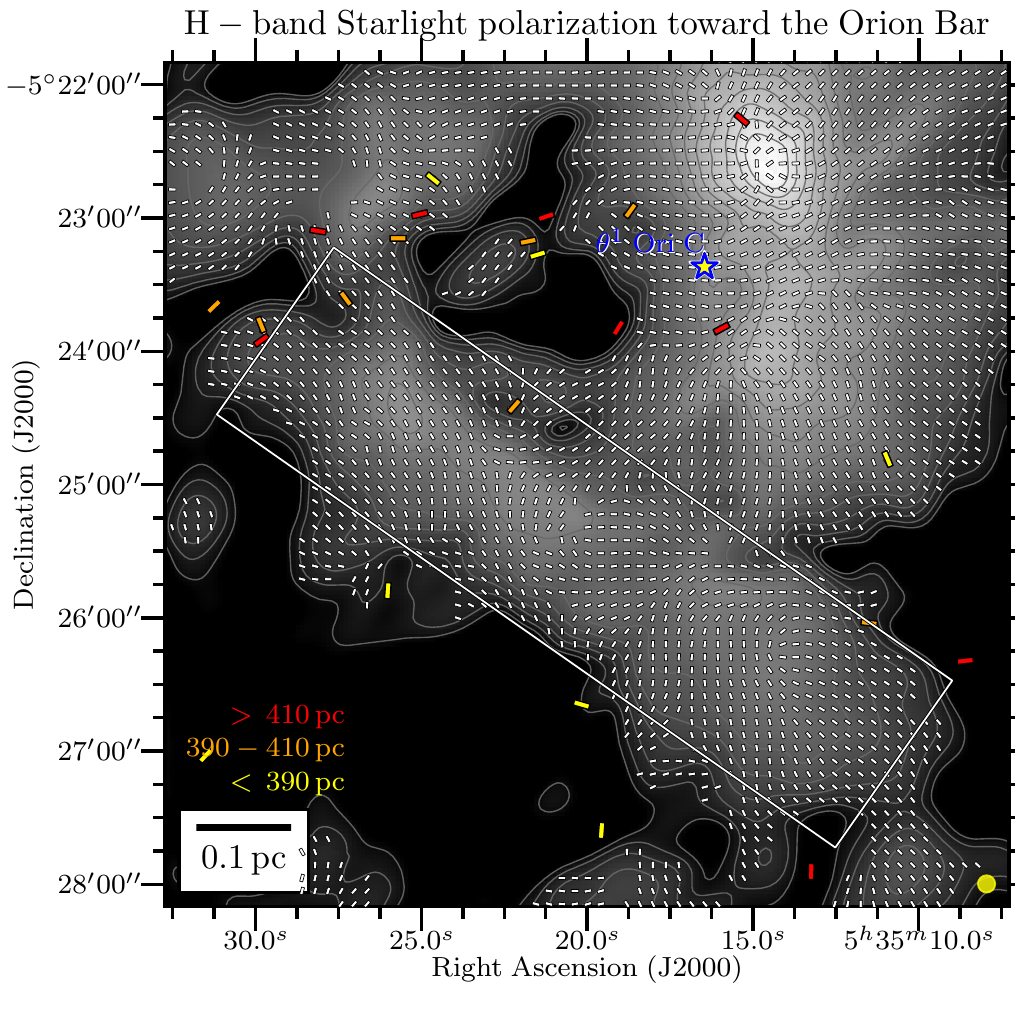}}
\subfigure{
\includegraphics[scale=0.848,clip,trim= 1.8cm 5.5cm 11.5cm 6cm]{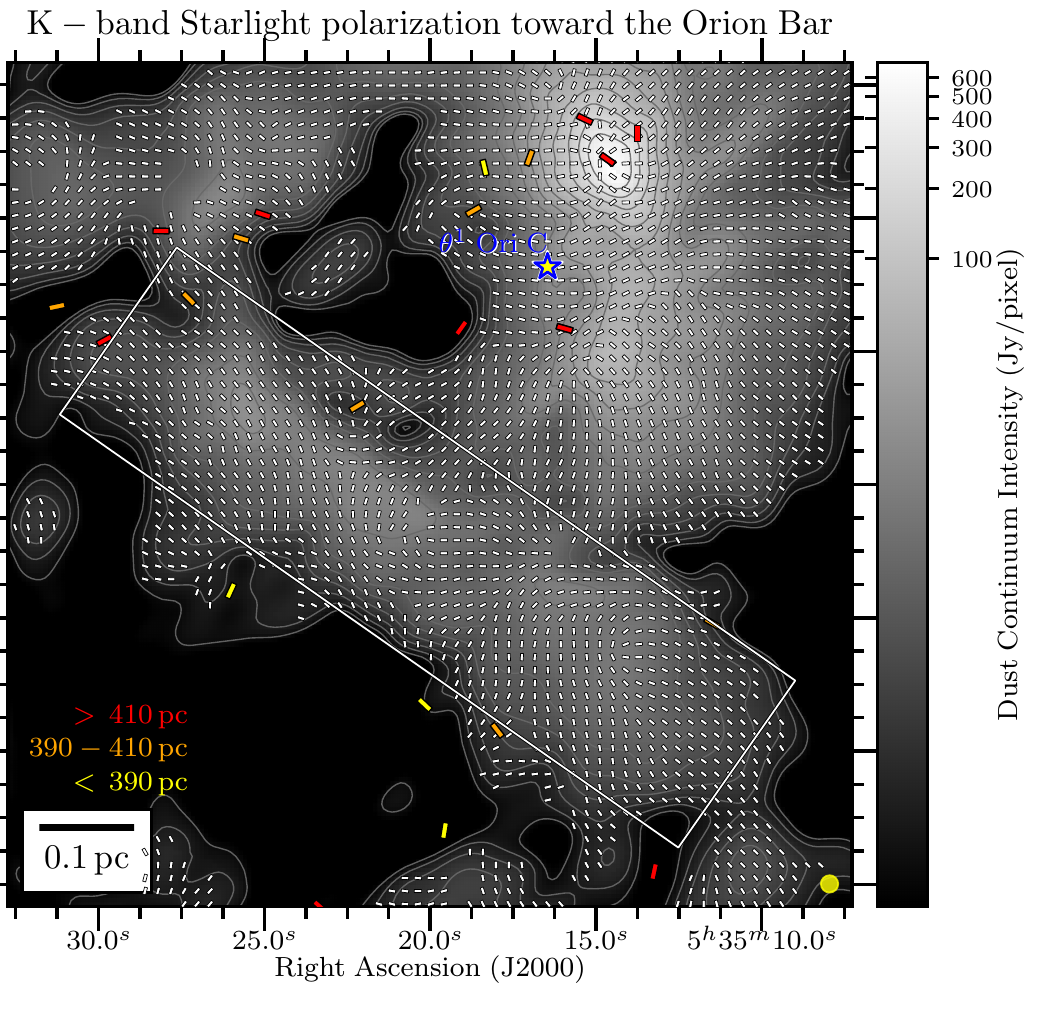}}
\caption{\small Maps of K- and H- band starlight polarization and 89 $\mu$m dust polarization in emission. Like in Figure \ref{fig:obs_bar_4bands}, top-right panel, the colorscale is the total intensity and the white line segments, are the B-field vectors, obtained from the 89 $\mu$m HAWC+ polarimetric observations. H-band and K-band starlight polarization vectors (MIMIR; \citealt{Clemens2007}) are shown by the colored line segments, in the left and right panels, respectively. The colors of the starlight polarization vectors indicate the distance range in which the star is thought to be, using the Gaia DR3 \citep{GaiaDR32022}. Starlight polarization vectors are plotted is $\Pf/\sigma_{\Pf}\,>\,2$.
}
\label{fig:starlight_pola}
\end{figure*}

Figure \ref{fig:starlight_pola} presents detections of H-band (1.65 $\mu$m) and K-band (2.2 $\mu$m) starlight polarization detection toward the Orion nebula, on top of the HAWC+ Band C dust polarization map. The starlight polarization data (Clemens et al. in preparation) was taken with the Mimir instrument \citep{Clemens2007} on the the 1.83m Perkins telescope, located outside Flagstaff, AZ. We retrieve the distance of those stars with Gaia DR3 \citep{GaiaDR32022}, and separate them into three sub groups, \ie foreground ($d$ < 390 pc), within the cloud (410 < $d$ < 390 pc), and background ($d$ > 410 pc). Our goal is to analyze the evolution of the starlight polarization angle as a function of distance, in order to constrain the 3D magnetic field topology, and to quantify the contribution of the background OMC-1 cloud in the dust polarized emission, in the LOSs of the Bar. Unfortunately, not enough stars have been detected within our region of interest to compute statistics. However, we highlight that such comparisons between polarization in extinction and polarization seen in emission is promising, toward such environments that are expected to experience intense gradients in grain alignment conditions.
Figure \ref{fig:starlight_pola} shows that such work can be done toward the Orion Veil nebula with future, more sensitive FIR polarization observation.

% \end{appendix}

\end{document}